\documentclass[onecolumn,final]{elsart3p}

\usepackage{amssymb}
\usepackage{amsfonts}
\usepackage{amsmath}
\usepackage{graphicx}
\usepackage{bm,bbm}
\usepackage{comment}
\usepackage{rotating}
\usepackage{color}
\usepackage{multirow}
\usepackage{numprint}
\usepackage{overpic}

\usepackage[numbers,sort]{natbib}

\usepackage{float}
\usepackage{mdframed}
\usepackage{outlines}
\usepackage{soul}

\usepackage{url}

\usepackage{savesym}
\savesymbol{AND}
\usepackage{algpseudocode}
\usepackage{algorithm}

\theoremstyle{plain}
\newtheorem{theorem}{Theorem}[section]

\newtheorem{remark}{Remark}[section]

\theoremstyle{remark}

\newcommand{\EOD}{\end{document}}

\newcommand{\DG}{DG{}}

\definecolor{MyDarkGreen}{rgb}{0,0.45,0}

\renewcommand{\cv}{\bm{c}}

\newcommand{\ev}{\bm{e}}

\newcommand{\gv}{\bm{g}}

\newcommand{\nv}{\bm{n}}

\newcommand{\uv}{\bm{u}}
\newcommand{\vv}{\bm{v}}

\newcommand{\xv}{\bm{x}}

\newcommand{\Av}{\bm{A}}
\newcommand{\Bv}{\bm{B}}

\newcommand{\Ev}{\bm{E}}
\newcommand{\Fv}{\bm{F}}
\newcommand{\Gv}{\bm{G}}

\newcommand{\Jv}{\bm{J}}

\newcommand{\Ov}{\bm{O}}
\newcommand{\Pv}{\bm{P}}

\newcommand{\Rv}{\bm{R}}
\newcommand{\Sv}{\bm{W}}

\newcommand{\Uv}{\bm{U}}
\newcommand{\Vv}{\bm{V}}

\newcommand{\UvC}{\bm{u}}

\newcommand{\as}{a}
\newcommand{\bs}{b}
\newcommand{\cs}{c}
\newcommand{\ds}{d}

\newcommand{\fs}{f}

\newcommand{\hs}{h}

\newcommand{\ks}{k}

\newcommand{\ms}{m}
\newcommand{\ns}{n}

\newcommand{\qs}{q}

\renewcommand{\ss}{s}
\newcommand{\ts}{t}
\newcommand{\us}{u}
\newcommand{\vs}{v}

\newcommand{\xs}{x}
\newcommand{\ys}{y}
\newcommand{\zs}{z}

\newcommand{\As}{A}
\newcommand{\Bs}{B}
\newcommand{\Cs}{C}

\newcommand{\Fs}{F}

\newcommand{\Hs}{H}
\newcommand{\Is}{I}

\newcommand{\Ls}{L}

\newcommand{\Ys}{Y}

\renewcommand{\fs}{f^s}

\newcommand{\xiv}{{\bm\xi}}

\newcommand{\Phiv}{{\bm\Phi}}
\newcommand{\Psiv}{{\bm\Psi}}

\newcommand{\Cvs}  {\mathbf{C}^s}

\newcommand{\C}{C}
\newcommand{\CS}{C^s}

\newcommand{\PT}[1]{\ensuremath{\dfrac{\partial #1}{\partial t}}}

\renewcommand{\as}{\alpha^s}
\renewcommand{\us}{u^s}

\newcommand{\GRADX}{\nabla_{\xv}}
\newcommand{\GRADV}{\nabla_{\vv}}

\newcommand{\matAx}{\mathbbm{A}_x}
\newcommand{\matAy}{\mathbbm{A}_y}
\newcommand{\matAz}{\mathbbm{A}_z}
\newcommand{\matAb}{\mathbbm{A}_{\beta}}

\newcommand{\matRb}{\mathbf{R}_{\beta}}

\newcommand{\matDb}{\mathbf{D}_{\beta}}

\newcommand{\matA}{\mathbbm{A}}

\newcommand{\matF}{\mathbbm{F}}

\newcommand{\dxv}{d\xv}
\newcommand{\dvv}{d\vv}

\newcommand{\dS}{dS}

\newcommand{\Nn}{\ensuremath{N_{v_x}}}
\newcommand{\Nm}{\ensuremath{N_{v_y}}}
\newcommand{\Np}{\ensuremath{N_{v_z}}}

\newcommand{\ABS}[1]{|#1|}

\newcommand{\REAL}{\mathbbm{R}}

\newcommand{\vphi}{\bm\varphi}
\newcommand{\varphiIl}{\varphi^{I,l}}
\newcommand{\varphiIlp}{\varphi^{I,l'}}

\renewcommand{\ms}{m^s}
\renewcommand{\qs}{q^s}

\newcommand{\dx}{\,dx}
\newcommand{\dy}{\,dy}
\newcommand{\dz}{\,dz}

\newcommand{\dt}{dt}
\renewcommand{\ds}{ds}

\newcommand{\PS}[1]{\mathbb{P}_{#1}}

\newcommand{\Diffx}[1]{\delta_x\big[ #1 \big]}
\newcommand{\Diffy}[1]{\delta_y\big[ #1 \big]}
\newcommand{\Diffz}[1]{\delta_z\big[ #1 \big]}

\newcommand{\Diffv}[2]{\delta_{#1}\big[ #2 \big]}

\def\trait #1 #2 #3 {\vrule width #1pt height #2pt depth #3pt}
\def\fin{\hfill
        \trait .3 5 0
        \trait 5 .3 0
        \kern-5pt
        \trait 5 5 -4.7
        \trait 0.3 5 0
\medskip}
\newcommand{\ENDPROOF}{\fin}

\usepackage{scalerel,stackengine}
\stackMath
\newcommand\reallywidehat[1]{%
\savestack{\tmpbox}{\stretchto{%
  \scaleto{%
    \scalerel*[\widthof{\ensuremath{#1}}]{\kern-.6pt\bigwedge\kern-.6pt}%
    {\rule[-\textheight/2]{1ex}{\textheight}}
  }{\textheight}%
}{0.5ex}}%
\stackon[1pt]{#1}{\tmpbox}%
}

\newcommand{\NFLX}[1]{\reallywidehat{#1}}        
\newcommand{\JUMP}[1]{\big[\hspace{-1mm}\big[#1\big]\hspace{-1mm}\big]}

\newcommand{\JMPF}{\mathcal{J}_{\F}}

\renewcommand{\Hs} {\mathcal{H}}

\renewcommand{\Fs} {\mathcal{F}}

\newcommand{\Ox}{\Omega_x}
\renewcommand{\Ov}{\Omega_\vs}

\newcommand{\fsH}{f^{s,H}}
\newcommand{\fsN}{f^{s,N}}
\newcommand{\fsNz}{f^{s,N}_{0}}

\newcounter{numbs}
\newcounter{numbi}
\newcounter{numbii}

\newcommand{\PGRAPH}[1]{\noindent\textbf{#1}}

\newcommand{\BegBoxQuote}[1]{
   \vspace{\baselineskip}
   \begin{center}
   \begin{minipage}{#1\textwidth}
   \begin{mdframed}
}
\newcommand{\EndBoxQuote}{
   \end{mdframed}
   \end{minipage}
   \end{center}
}

\newcommand{\ocs}{\omega_{cs}}
\newcommand{\oce}{\omega_{ce}}
\newcommand{\ope}{\omega_{pe}}

\newcommand{\asx}{\as_x}
\newcommand{\asy}{\as_y}
\newcommand{\asz}{\as_z}

\newcommand{\usx}{\us_x}
\newcommand{\usy}{\us_y}
\newcommand{\usz}{\us_z}

\newcommand{\vsx}{\vs_x}
\newcommand{\vsy}{\vs_y}
\newcommand{\vsz}{\vs_z}

\newcommand{\Ex}{E_x}
\newcommand{\Ey}{E_y}
\newcommand{\Ez}{E_z}
\newcommand{\ExN}{E_x^N}
\newcommand{\EyN}{E_y^N}
\newcommand{\EzN}{E_z^N}

\newcommand{\Bx}{B_x}
\newcommand{\By}{B_y}
\newcommand{\Bz}{B_z}
\newcommand{\BxN}{B_x^N}
\newcommand{\ByN}{B_y^N}
\newcommand{\BzN}{B_z^N}

\newcommand{\vpsi}{\psi}

\newcommand{\cL}{\mathcal{L}}
\newcommand{\cN}{\mathcal{N}}

\newcommand{\Il} {I,l}
\newcommand{\Ilp}{I,l'}

\newcommand{\pxI}{\partial I_x}
\newcommand{\pyI}{\partial I_y}
\newcommand{\pzI}{\partial I_z}

\newcommand{\F}{\mathsf{f}}
\newcommand{\Fip}{\mathsf{f}_{i+\frac{1}{2},j,k}}
\newcommand{\Fjp}{\mathsf{f}_{i,j+\frac{1}{2},k}}
\newcommand{\Fkp}{\mathsf{f}_{i,j,k+\frac{1}{2}}}

\newcommand{\Fim}{\mathsf{f}_{i-\frac{1}{2},j,k}}
\newcommand{\Fjm}{\mathsf{f}_{i,j-\frac{1}{2},k}}
\newcommand{\Fkm}{\mathsf{f}_{i,j,k-\frac{1}{2}}}

\newcommand{\Fipm}{\mathsf{f}_{i\pm\frac{1}{2},j,k}}
\newcommand{\Fjpm}{\mathsf{f}_{i,j\pm\frac{1}{2},k}}
\newcommand{\Fkpm}{\mathsf{f}_{i,j,k\pm\frac{1}{2}}}

\newcommand{\SPAN}{\mathsf{span}}

\newcommand{\Icp}{I^{+}}

\newcommand{\matAxp}{\matAx^{+}}
\newcommand{\matAxm}{\matAx^{-}}
\newcommand{\xpp}{x^{+}_{i+\frac{1}{2}}}
\newcommand{\xmp}{x^{-}_{i+\frac{1}{2}}}

\newcommand{\vecE}  {\vec{\Ev}}
\newcommand{\vecEN}  {\vec{\Ev}^N}
\newcommand{\vecENz}{\vec{\Ev}^N_0}
\newcommand{\vecEIl}{\vec{\Ev}^{\Il}}
\newcommand{\ExIl}  {\Ex^{\Il}}
\newcommand{\EyIl}  {\Ey^{\Il}}
\newcommand{\EzIl}  {\Ez^{\Il}}

\newcommand{\vecB}  {\vec{\Bv}}
\newcommand{\vecBN}  {\vec{\Bv}^N}
\newcommand{\vecBNz}{\vec{\Bv}^N_0}
\newcommand{\vecBIl}{\vec{\Bv}^{\Il}}
\newcommand{\BxIl}  {\Bx^{\Il}}
\newcommand{\ByIl}  {\By^{\Il}}
\newcommand{\BzIl}  {\Bz^{\Il}}

\newcommand{\BsE}{\Bs_{\Ev}}
\newcommand{\BsB}{\Bs_{\Bv}}

\newcommand{\KIN}{{\bf kin}}
\newcommand{\TOT}{{\bf tot}}
\newcommand{\UPW}{\textbf{upw}}

\newcommand{\ELE}{\Ev,\Bv}
\newcommand{\JMP}{\textbf{jump}}

\newcommand{\vecJ}{\vec{\Jv}}
\newcommand{\vecJN}{\vec{\Jv}^N}

\newcommand{\FF}{\mathbb{F}}
\newcommand{\Fvx}{\mathbf{F}_x}
\newcommand{\Fvy}{\mathbf{F}_y}
\newcommand{\Fvz}{\mathbf{F}_z}

\newcommand{\LTWO}{L^2}

\newcommand{\Nl}{N_l}

\newcommand{\lx}{l_x}
\newcommand{\ly}{l_y}
\newcommand{\lz}{l_z}

\newcommand{\xsI}{\xs_{i,j,k}}
\newcommand{\ysI}{\ys_{i,j,k}}
\newcommand{\zsI}{\zs_{i,j,k}}

\newcommand{\FFE}{\FF_{\vecE}}
\newcommand{\FFB}{\FF_{\vecB}}

\newcommand{\calEN}{\calE^N}

\newcommand{\fip}{\F_{i+\frac{1}{2},j,k}}
\newcommand{\fim}{\F_{i-\frac{1}{2},j,k}}

\newcommand{\fjp}{\F_{i,j+\frac{1}{2},k}}
\newcommand{\fjm}{\F_{i,j-\frac{1}{2},k}}

\newcommand{\fkp}{\F_{i,j,k+\frac{1}{2}}}
\newcommand{\fkm}{\F_{i,j,k-\frac{1}{2}}}

\newcommand{\Uvp}{\Uv^+}
\newcommand{\Uvm}{\Uv^-}

\newcommand{\csVM}{\frac{\qs}{\ms}\frac{\oce}{\ope}}

\newcommand{\fse}{f^s}
\newcommand{\Bve}{\Bv}
\newcommand{\Eve}{\Ev}

\newcommand{\rhoN}{\rho^N}
\newcommand{\rhoNs}{\rho^{s,N}}
\newcommand{\vecJNs}{\Jv^{s,N}}

\newcommand{\calH}{\mathcal{H}}
\newcommand{\calI}{\mathcal{I}}

\newcommand{\calV}{\mathcal{V}}

\newcommand{\calVN} {\calV^N} \newcommand{\calHN} {\calH^N}
\newcommand{\calHpN}{\mathcal{\widetilde{H}}^N}
\newcommand{\Fb}{{\F}_{\beta}}
\newcommand{\FFb}{{\matF}_{\beta}}

\newcommand{\NDG}{N_{\DG}}

\newcommand{\Vlasov}{f}
\newcommand{\Maxwell}{\Ev\times\Bv}

\newcommand{\phiRK} {\varphi_{RK}}
\newcommand{\phiRKh}{\widehat{\varphi}_{RK}}

\newcommand{\calE}{\mathcal{E}}

\newcommand{\vz}{{\bm\zeta}}

\newcommand{\ysb}{\overline{\ys}}

\newcommand{\CP}[1]{\textcolor{black}{#1}}
 \newcommand{\cecilia}[1]{\textcolor{red}{#1}}

\begin{document}
%
\begin{frontmatter} 
  \title{The multi-dimensional Hermite-discontinuous Galerkin method
    \\ for the Vlasov-Maxwell equations}

  \author[LANL] {O.~Koshkarov,}
  \author[LANL] {G.~Manzini,}
  \author[LANL] {G.~L.~Delzanno,}
  \author[CA]   {C. Pagliantini,}
  \author[SSI]  {V.~Roytershteyn}
  
  \address[LANL]{T-5 Applied Mathematics and Plasma Physics Group, Los
    Alamos National Laboratory, Los Alamos, NM 87545, USA}

    \address[CA]{Department of Mathematics and Computer Science, Eindhoven University of Technology, Eindhoven, The Netherlands}

  \address[SSI]{Space Science Institute, 4750 Walnut St, Suite 205,
    Boulder, CO 80301}
  
  \begin{abstract}
    We discuss the development, analysis, implementation, and
    numerical assessment of a spectral method for the numerical
    simulation of the three-dimensional
    Vlasov-Maxwell equations.
    The method is based on a spectral expansion of the velocity space 
    with the asymmetrically
    weighted Hermite functions.
    The resulting system of time-dependent nonlinear equations is
    discretized by the discontinuous Galerkin (DG) method in space and by the
    method of lines for the time integration using explicit
    Runge-Kutta integrators.
    The resulting code, called Spectral Plasma Solver (SPS-DG), is successfully applied
    to standard plasma physics benchmarks to demonstrate its accuracy, robustness, and
    parallel scalability.
  \end{abstract}
  
  \begin{keyword}
    3-D Vlasov-Maxwell equations,
    AW Hermite discretization,
    Discontinuous Galerkin method
  \end{keyword}
  
\end{frontmatter}

\renewcommand{\arraystretch}{1.}
\raggedbottom

\section{Introduction}
Efficient and accurate coupling of the microscopic physics into the
macroscopic system-scale dynamics (also known as ‘fluid-kinetic’
coupling) is arguably the most important and yet still unresolved
problem of computational plasma physics. 
It impacts a wide variety of systems, including the solar corona, the
Earth's magnetosphere, all the way to laboratory experiments such as
those for magnetic and inertial fusion energy. 
The crux of the matter is the large scale separation involved in
plasma dynamics. 
For instance, for the Earth's magnetosphere, the local plasma Debye
length (an important parameter governing the microscopic physics) can
easily be less than $\sim 1$ m, while the system scale is $\sim10^9$
m. In what follows we will use the terms microscopic or kinetic and
macroscopic or fluid interchangeably.

Recognizing the inability to solve this enormous scale separation at
the microscopic level with available (and foreseeable) supercomputers,
the next step is to develop methods that treat fluid-kinetic coupling
in some approximated form and this is under active development.
In one approach, the microscopic physics is treated only locally in
selected regions of physical space by embedding a kinetic solver
within a large-scale fluid framework~\cite{kolobov2012,sugiyama07,daldorff14,toth16,chen17,ho18}. 
Another approach seeks ``augmented'' fluid models, with better
closures of the fluid equations that improve the representation of the
kinetic physics~\cite{liang15,liang18}. A different line of investigations targets large-scale simulations by averaging out certain scales of the system, as in the gyrokinetics approach~\cite{Brizard2007} for magnetic fusion energy applications, or by treating electrons as a fluid (hence removing electron kinetic scales) while the ions are still treated kinetically, as in the hybrid approach~\cite{Lipatov:2002,Winske:1985,homa14,lin17,Palmroth2018}.

An alternative method for fluid-kinetic coupling is based on a
spectral expansion of the plasma distribution function at the kinetic
level. 
With a suitable choice of the spectral basis, fluid-kinetic coupling
is an intrinsic property of spectral methods: the low-order
coefficients of the expansion are akin to a fluid description of the
plasma, while the kinetic physics is captured by retaining additional
terms in the expansion~\cite{Camporeale-Delzanno-Lapenta-Daughton:2006,Vencels-Delzanno-Johnson-BoPeng-Laure-Markidis:2015}. 
\CP{(A proof-of-principle demonstration of  how fluid-kinetic coupling can be exploited in spectral methods can be found in Ref. \cite{Vencels-Delzanno-Johnson-BoPeng-Laure-Markidis:2015}.)}
One can therefore recognize that spectral methods might offer the
optimal way to treat microscopic physics in large-scale simulations
since one can envision adapting the spectral expansion in time and in
space to minimize/optimize the number of degrees of freedom for a
given accuracy. 
It is also important to notice that spectral methods enclose the two
approaches for fluid-kinetic coupling described in the previous
paragraph: they can be seen as improved fluid models that treat
kinetic physics in a reduced way but also as methods where the kinetic
physics can be treated only locally, where necessary. 
It is also worth emphasizing that spectral methods for the solution of
the kinetic equations are important in their own right, beyond
fluid-kinetic coupling. 
These methods date back to the
seventies~\cite{Armstrong:1970,Gajewski-Zacharias:1977,Klimas:1983},
where their application was limited to
one-dimensional electrostatic problems, and their development
continues to this day~\cite{%
  Holloway:1996,%
  Schumer-Holloway:1998,%
  cai2013,%
  Camporeale-Delzanno-Bergen-Moulton:2015,%
  Delzanno2015,%
  Vencels-Delzanno-Johnson-BoPeng-Laure-Markidis:2015,%
  parker_dellar_2015,
  Manzini-Delzanno-Markidis-Vencels:2016,%
  Loureiro16,
  Manzini-Funaro-Delzanno:2017,%
  cai2018,%
  Fatone-Funaro-Manzini:2019,%
  Fatone-Funaro-Manzini:2019b,%
  Pezzi_2019,%
  Di2019%
}.
\CP{Currently, some drawbacks of spectral methods for the kinetic equations are their inability to enforce the positivity of the distribution function and, for some expansions (e.g. the asymmetrically-weighted (AW) Hermite representation), the lack of a numerical stability theorem. The optimization of the expansion basis is also an important open problem. In addition, we also note that on simple one-dimensional electrostatic problems spectral methods based on the AW-Hermite presentation were shown to be orders of magnitude faster/more accurate than Particle-In-Cell (PIC) methods~\cite{Camporeale-Delzanno-Bergen-Moulton:2015}. References~\cite{vr19,gld19} also present some physics applications of the AW-Hermite spectral method that were too computationally expensive for PIC.}

To the best of our knowledge, the only implementation of a spectral method that
treats the full three-dimensional Vlasov-Maxwell equations is discussed in
Refs.~\cite{Delzanno2015,Vencels-Delzanno-Manzini-Markidis-BoPeng-Roytershteyn:2015,vr18} and led to the development of the Spectral Plasma Solver (SPS)
code~\cite{Vencels-Delzanno-Manzini-Markidis-BoPeng-Roytershteyn:2015,vr18}.
In those works, the physical space is discretized with a Fourier
expansion, while the velocity space is discretized with an (asymmetrically weighted) Hermite
expansion. 
The resulting numerical method features the conservation of total
mass, momentum, and energy in a finite time
step~\cite{Delzanno2015}. 
This approach is highly accurate and is
particularly well suited for problems involving periodic boundary
conditions and wave-like perturbations. 
For example, it has been successfully applied to studies of the
turbulent cascade in magnetized plasmas~\cite{vr18,vr19}. 
However, a Fourier decomposition in physical space leads to multiple
convolutions in the transformed equations, which result from nonlinear terms in
the original equations. 
From a practical point of view, convolutions are handled using
the pseudo-spectral method, which requires computing many Fast Fourier
Transforms (FFTs) at every time step.
In a parallel code, the FFTs involve global communication operations,
which limit the code scalability and overall performance.
To overcome this problem, in this paper we present a spectral method
based on a finite element, discontinuous-Galerkin (\DG{}) discretization
in physical space, coupled with a Hermite representation of velocity
space.

The \DG{} method was initially introduced for solving the neutron transport
equation~\cite{Reed-Hill:1973,Lasaint-Raviart:1974} and was later extended
to the numerical
approximation of nonlinear conservation laws and hyperbolic system of
partial differential equations~\cite{Cockburn-Shu:1989,Cockburn-Shu:1991,Cockburn-Lin-Shu:1991}.
We refer the reader
to~\cite{Cockburn-Karniadakis-Shu:2000} for an historical overview and
to~\cite{Shu:2009,Hesthaven-Warburton:2007-book} for a general
presentation.
Relevant to our work are the papers for the Vlasov-Poisson
system~\cite{Ayuso-Carrillo-Shu:2011,Ayuso-Carrillo-Shu:2012,Ayuso-Hajan:2012},
the Boltzmann-Poisson system~\cite{Cheng-Gamba-Majorana:2008,%
  Cheng-Gamba-Majorana-Shu:2009,%
  Cheng-Gamba-Majorana-Shu:2011,%
  Cheng-Gamba-Proft:2012,%
  Besse-Berthelin-Brenier-Bertrand:2009,%
  Morales-Gamba:2018%
}, the Maxwell equations~\cite{Cockburn-Li-Shu:2004}, and the
Vlasov-Maxwell
system~\cite{Cheng-Gamba-Li-Morrison:2014,Cheng-Christlieb-Zhong:2014,Juno-Hakim-TenBarge-Shi-Dorland:2018,ho18}.
Applications of the DG algorithm to reduced plasma models also exist, e.g. \cite{loverich2005,loverich2011,srinivasan2011}.
Unlike previous approaches, in this work we couple the \DG{} discretization of the spatial
terms of the Vlasov and Maxwell equations with a
spectral representation of the velocity space.
The formulation of our method is obtained by testing the conservative
form of the partial differential equations against elements of a finite
dimensional space of globally discontinuous functions, whose
restriction to any element of the computational mesh is a polynomial
of a maximum assigned degree.
Therefore, the approximation has an intrinsic \emph{local conservative
  nature}.
Moreover, the accuracy of the method is determined by the degree of
the local polynomials, so that increasing arbitrarily such parameter
makes it possible to obtain numerical approximations of
\emph{arbitrary order of accuracy}. Consistent
upwind numerical flux functions provide the exchange of information
between adjacent cells so that
the \DG{} method is characterized by an extreme \emph{locality in data
  and communication}. Importantly, this yields a method which is much 
better suited for implementations on high performance parallel architectures
than the equivalent method based on a Fourier discretization in physical 
space which involves FFTs and global communications.
Upwind numerical fluxes are chosen from numerical stability considerations.
Importantly, the elemental polynomial basis can be built independently
in each mesh element, so \emph{different approximation degrees} can be
used in different elements as in the $hp$ refinement strategy, and (in
the modal setting) almost independently of the geometric shape of the
element, thus providing a significant \emph{mesh flexibility} in the
application.
Indeed, the \DG{} method, which was originally developed using
discontinuous polynomials on triangles or quadrilaterals, can be
extended to more general unstructured meshes with polygonal (2D) and
polytopal (3D) cells~\cite{Cangiani-Dong-Georgoulis-Houston:2017}.
Finally, even though this topic is not pursued in the present work, it is worth
mentioning that
it would be possible to incorporate a \emph{shock-capturing
  capability} in the \DG{} method in a very natural and straightforward
way by using limiters in the calculation of numerical fluxes.


\medskip
The outline of the paper is as follows.
In Section~\ref{sec:Vlasov-Maxwell:eqs}, we introduce the mathematical
model considered in this paper that describes the transport phenomena
of different charged particle species in a collisionless plasma under
the action of the self-consistent electromagnetic field.
The behavior of each particle species is modeled through its
distribution function that satisfies a time-dependent Vlasov equation
in the six-dimensional phase space, i.e. three dimensions in space and
velocity, respectively.
The self-consistent electromagnetic field generated by the charged
plasma particles satisfies Maxwell's equations.
In Section~\ref{sec:Transform:method}, we apply the spectral
and \DG{} discretizations to the Vlasov equations.
The expansion in Hermite basis functions removes the velocity dependence by
transforming the Vlasov equations of each particle species in a
nonlinear hyperbolic system of partial differential equations for the coefficients of the expansion that are still dependent
on time and space.
Then, this system of coefficients is discretized in space by applying
the DG method.
In Section~\ref{sec:maxwell}, we apply the \DG{} method to the
discretization of the Maxwell equations reformulated in divergence
form.
Here, we introduce the central and upwind numerical flux as possible
alternatives in the scheme.
In Section~\ref{sec:conservation}, we present the semi-discrete conservation properties of the method. 
In Section~\ref{sec:time}, the final system of time-dependent ordinary
differential equations for the various expansion coefficients is
advanced in time by applying a standard Runge-Kutta (RK) method.
In Section~\ref{sec:implementation}, we discuss some important aspects
of the implementation that are crucial to obtain a computationally
efficient solver.
We also investigate the parallel scalability of the current
implementation, and show that the algorithm is scalable on a particular high-performance-computing platform.
In Section~\ref{sec:numerical:results}, we assess the performance of
the method in terms of accuracy
and prove its reliability and robustness on a set of benchmark
problems that are representative of plasma physics modeling
situations.
In Section~\ref{sec:conclusions}, we present our final remarks and conclusions.

\medskip
\emph{Notation and Normalization.} 
We normalize the model equations as follows.
Time $t$ is normalized to the electron plasma frequency
$\ope=\sqrt{e^2 n^{e}_0\slash{}\varepsilon_0m^e}$, where $e$ is the
elementary charge, $m^e$ is the electron mass, $\varepsilon_0$ is the
permittivity of vacuum, and $n^e_0$ is a reference electron density.
The velocity coordinate $\vs$ is normalized to the speed of light
$\cs$; the spatial coordinate $\xs$ is normalized to the electron
inertial length $d_e=\cs\slash{\ope}$; the magnetic field $\vecB$ is
normalized to a reference magnetic field $\Bs_0$, and the electric
field is normalized to $\cs\Bs_0$.
We denote the quantities regarding a given plasma species by the
superscript $s$, which may take the specific values $s=e$ (electrons)
and $s=i$ (ions).
Accordingly, we denote the mass of the particles of species $s$ by
$\ms$ and their charge by $\qs$.
We normalize charge $\qs$ and mass $\ms$ to elementary charge $e$ and mass $m^e$,
respectively.
Finally, we define the cyclotron frequency of species $s$ as
$\ocs=e\Bs_0/\ms$.

\section{Vlasov-Maxwell equations}
\label{sec:Vlasov-Maxwell:eqs}
The behavior of the particles of species $s$ in a collisionless
magnetized plasma is described at any time instant $t>0$ by the
nonnegative distribution function $\fs(\xv,\vv,t)$, where $\xv$
denotes the position in the physical space $\Ox$ and $\vv$ the
position in the velocity space $\Ov$.
Under the action of the self-consistent electric and magnetic fields
$\Eve(\xv,t)$ and $\Bve(\xv,t)$ generated by the particles' motion,
the distribution function of species $s$ satisfies the (normalized)
Vlasov equation:
\begin{align}
  \frac{\partial\fse}{\partial t} + \vv\cdot\GRADX\fse + 
  \csVM\big(\Eve+\vv\times\Bve\big)\cdot\GRADV\fse = 0.
  \label{eq:Vlasov:fs}
\end{align}
The electric and magnetic fields $\Eve=(\Ex,\Ey,\Ez)^T$ and
$\Bve=(\Bx,\By,\Bz)^T$, with $T$ denoting the transpose, satisfy the time-dependent wave propagation
equations
\begin{align}
  &\frac{\partial\Eve}{\partial t} - \GRADX\times\Bve =  - \frac{\ope}{\oce}\vecJ,\label{eq:dEdt}\\[0.5em]
  &\frac{\partial\Bve}{\partial t} + \GRADX\times\Eve =  0,    \label{eq:dBdt}
\end{align}
and the divergence equations
\begin{align}
  &\GRADX\cdot\Eve = \frac{\ope}{\oce}\rho, \label{eq:divE}\\[0.5em]
  &\GRADX\cdot\Bve = 0.    \label{eq:divB}
\end{align}
In Equations~\eqref{eq:dEdt} and~\eqref{eq:divE}, $\vecJ$ and $\rho$
are the self-consistent electric current and charge density, respectively,
\begin{align}
  \rho (\xv,t) &= \sum_{s}\qs\int_{\Ov}   \fse(\xv,\vv,t)\dvv,\label{eq:rho}\\[1em]
  \vecJ(\xv,t) &= \sum_{s}\qs\int_{\Ov} \vv\fse(\xv,\vv,t)\dvv,\label{eq:vecJ}
\end{align}
where the summation is over all the plasma species denoted by $s$.
We consider the unbounded velocity space $\Ov=\REAL^{3}$ and we assume
that each distribution function $\fse$ is rapidly decaying for
$\ABS{\vv}\to\infty$, i.e., it decays proportionally to
$\exp(-\ABS{\vv}^2)$~\cite{Glassey:1996}.
This assumption is physically consistent with the Maxwellian velocity
distribution of a plasma in thermodynamic
equilibrium~\cite{Grad:1949}.
Similarly, we consider the closed bounded subset $\Ox\subset\REAL^3$
with boundary $\Gamma_x$, and we assume that suitable
problem-dependent boundary conditions for $\fse$, $\Eve$, and $\Bve$
are provided at $\Gamma_x$ for any time $t$ and any value of $\vv$ in
$\Ov$.
Moreover, physically meaningful initial conditions must be provided for
the unknown fields $\fse$, $\Eve$, $\Bve$ at the initial time $t=0$.
Finally, when periodic boundary conditions are used, 
\begin{align*}
  \int_{\Ox}\rho(\xv,t)\dxv=0
  \qquad t \geq 0,
\end{align*}
and 
\begin{align*}
  \int_{\Ox}\vecJ(\xv,t)\dxv=0
  \qquad t \geq 0,
\end{align*}
must be satisfied so that the charge density $\rho$ satisfies the global charge
neutrality condition and the total current is zero.

\section{Hermite-\DG{} discretization of the Vlasov equation}
\label{sec:Transform:method}

\subsection{Spectral discretization in velocity space using Hermite functions}
We expand the distribution function $\fs$ on the multidimensional
Hermite dual basis functions
\begin{align*}
  \Psi_{n,m,p}(\xiv^s)=\vpsi_{n}(\xi^{s}_{x})\vpsi_{m}(\xi^{s}_{y})\vpsi_{p}(\xi^{s}_{z})
  \quad\textrm{and}\quad
  \Psi^{n,m,p}(\xiv^s)=\vpsi^{n}(\xi^{s}_{x})\vpsi^{m}(\xi^{s}_{y})\vpsi^{p}(\xi^{s}_{z}),
\end{align*}
for $n=0,\ldots,\Nn$, $m=0,\ldots,\Nm$, $p=0,\ldots,\Np$, where
\begin{align}
  \xiv^{s} = \big(\xi^{s}_{x},\xi^{s}_{y},\xi^{s}_{z}\big)^T,\quad
  \xi^{s}_{x}=\frac{\vsx-\usx}{\asx},
  \quad\xi^{s}_{y}=\frac{\vsy-\usy}{\asy},
  \quad\xi^{s}_{z}=\frac{\vsz-\usz}{\asz}.
\end{align}
In this paper, the quantities $\usx,\usy,\usz$ and $\asx,\asy,\asz$ are
constant factors that depend on the plasma species and that are provided by the user for a specific problem. (Note that, in general, it can be beneficial to allow these quantities to vary in both space and time, see  for instance Refs.~\cite{Vencels_thesis,Di2019}, but this is left for future work). 
The Hermite functions $\Psi_{n,m,p}$ and $\Psi^{n,m,p}$ are given by
the tensor product of the univariate asymmetrically weighted Hermite
functions
\begin{align}
  \psi_{\zeta}(\xi^s_{\beta})
  &= \big(\pi\,2^{\zeta}\,\zeta!\big)^{-\frac{1}{2}}\Hs_{\zeta}(\xi^s_{\beta})\exp\big(-(\xi^s_{\beta})^2\big),\label{eq:Psi-dw}\\[0.5em]
  \psi^{\zeta}(\xi^s_{\beta}) &=
  \big(2^{\zeta}\,\zeta!\big)^{-\frac{1}{2}}\Hs_{\zeta}(\xi^s_{\beta}),\label{eq:Psi-up}
\end{align}
where $\Hs_{\zeta}(\xi^s_{\beta})$ is the $\zeta$-th univariate
Hermite polynomial for $\zeta\in\{n,m,p\}$ and
$\beta(\zeta)\in\{\xs,\ys,\zs\}$.
The orthogonality of the Hermite polynomials
$\Hs_{\zeta}(\xi^s_{\beta})$ and $\Hs_{\zeta'}(\xi^s_{\beta})$ with
respect to the weighted $\LTWO$-inner product with weight
$\exp(-(\xi^s_{\beta})^2)$ induces the duality relation between
$\Psi_{n,m,p}$ and $\Psi^{n',m',p'}$
\begin{align}
  \left<\Psi_{n,m,p},\Psi^{n',m',p'}\right>
  = \int_{\Ov}\Psi_{n,m,p}(\xiv^s)\Psi^{n',m',p'}(\xiv^s)d\xiv^s
  = \delta_{n,n'}\delta_{m,m'}\delta_{p,p'}.
  \label{eq:Psi:duality}
\end{align}
The normalization factors in~\eqref{eq:Psi-dw}-\eqref{eq:Psi-up} are
chosen to insure orthonormality in~\eqref{eq:Psi:duality}.
The recursive property and the derivative formula of the Hermite
polynomials imply the following relations for the Hermite functions:
\begin{align}
  \vs_{\beta}\psi_{\zeta}(\xi^s_{\beta})
  &= 
  \as_{\beta}\sqrt{\frac{\zeta+1}{2}}\psi_{\zeta+1}(\xi^s_{\beta}) +
  \as_{\beta}\sqrt{\frac{\zeta}{2}}\psi_{\zeta-1}(\xi^s_{\beta}) +
  \us_{\beta}\psi_{\zeta}(\xi^s_{\beta}),\\[0.5em]
  \frac{d\psi_{\zeta}}{d\vs_{\beta}}(\xi^s_{\beta})
  &= 
  -\frac{\sqrt{2(\zeta+1)}}{\as_{\beta}}\psi_{\zeta+1}(\xi^s_{\beta}).
\end{align}

The numerical approximation at $t\ge0$ of the distribution function
$\fs$ is given by the finite expansion
\begin{align}
  \fs(\xv,\vv,t) \approx \fsH(\xv,\vv,t)
  = \sum_{n=0}^{\Nn}\sum_{m=0}^{\Nm}\sum_{p=0}^{\Np}\CS_{n,m,p}(\xv,t)\vpsi_{n}(\xi^{s}_{x})\vpsi_{m}(\xi^{s}_{y})\vpsi_{p}(\xi^{s}_{z}),
  \label{eq:fs:expansion}
\end{align}
where the summation on $n$, $m$, and $p$ is truncated at $\Nn$, $\Nm$,
and $\Np$, respectively.
The expansion coefficients $\CS_{n,m,p}(\xv,0)$ at the initial time
$t=0$ are
\begin{align}
  \CS_{n,m,p}(\xv,0) 
  = \int_{\Ov}\fs(\xv,\vv,0)\Psi^{n,m,p}(\xiv^s)d\xiv^s.
\end{align}

To derive the time-dependent nonlinear system for the Hermite
expansion coefficients $\CS=\{\CS_{n,m,p}\}$, we
multiply~\eqref{eq:Vlasov:fs} by $\Psi^{n,m,p}(\xiv^{s})$ and
integrate over $\Ov$,
\begin{align}
  \int_{\Ov}\left( \frac{\partial\fsH}{\partial t} + \vv\cdot\GRADX\fsH + 
    \csVM\big(\vecE+\vv\times\vecB\big)\cdot\GRADV\fsH \right)\Psi^{n,m,p}(\xiv^{s})d\xiv^{s} = 0.
  \label{eq:weak:DG}
\end{align}
Then, we substitute the finite expansion~\eqref{eq:fs:expansion} of
$\fsH$, integrate by parts the derivative term in $\vv$, and obtain:
\begin{align}
  \frac{\partial\CS_{n,m,p}(\xv,t)}{\partial t} 
  + \cL_{n,m,p}(\CS) + \cN_{n,m,p}(\CS) = 0,
  \label{eq:Vlasov:compact}
\end{align}
where the ``linear'' term $\cL_{n,m,p}$ and the ``nonlinear'' term
$\cN_{n,m,p}$ are given by
\begin{align}
  \cL_{n,m,p}(\CS) 
  &= \int_{\Ov}\vv\cdot\GRADX\fsH(\xv,\vv,t)\,\Psi^{n,m,p}(\xiv^s)\,d\xiv^s,
  \label{eq:cL:def}
  \intertext{and}
  \cN_{n,m,p}(\CS) 
  &= 
  \csVM
  \int_{\Ov}\Big(\big(\vecE(\xv,t)+\vv\times\vecB(\xv,t)\big)\cdot\GRADV\fsH(\xv,\vv,t)\Big)\,\Psi^{n,m,p}(\xiv^s)\,d\xiv^{s}.
  \label{eq:cN:def}
\end{align}
A straightforward calculation using formulas~\eqref{eq:Psi-dw} and
\eqref{eq:Psi-up}, together with the orthogonality property \eqref{eq:Psi:duality} of the
Hermite basis functions
yields the set
of evolution equations for the Hermite coefficient
$\CS_{n,m,p}(\xv,t)$, for any triplet $(n,m,p)$, which reads
\begin{align}
  \frac{\partial\CS_{n,m,p}}{\partial t} 
  &+ \asx\left(
    \sqrt{\frac{n+1}{2}}\frac{\partial\CS_{n+1,m,p}}{\partial x} +
    \sqrt{\frac{n}{2}  }\frac{\partial\CS_{n-1,m,p}}{\partial x} +
    \frac{\usx}{\asx}\frac{\partial\CS_{n,m,p}}{\partial x}
  \right)\nonumber\\[0.5em]
  &+ \asy\left(
    \sqrt{\frac{m+1}{2}}\frac{\partial\CS_{n,m+1,p}}{\partial y} +
    \sqrt{\frac{m}{2}  }\frac{\partial\CS_{n,m-1,p}}{\partial y} +
    \frac{\usy}{\asy}\frac{\partial\CS_{n,m,p}}{\partial y}
  \right)\nonumber\\[0.5em]
  &+ \asz\left(
    \sqrt{\frac{p+1}{2}}\frac{\partial\CS_{n,m,p+1}}{\partial z} +
    \sqrt{\frac{p  }{2}}\frac{\partial\CS_{n,m,p-1}}{\partial z} +
    \frac{\usz}{\asz}\frac{\partial\CS_{n,m,p}}{\partial z}
  \right)\nonumber\\[0.5em]
  &-\csVM 
  \left(
    \frac{\sqrt{2n}}{\asx}\Ex\CS_{n-1,m,p}+
    \frac{\sqrt{2m}}{\asy}\Ey\CS_{n,m-1,p}+
    \frac{\sqrt{2p}}{\asz}\Ez\CS_{n,m,p-1}
  \right)\nonumber\\[0.5em]
  & -\csVM 
  \Bx\left[
    \sqrt{mp}   \left(\frac{\asz}{\asy}-\frac{\asy}{\asz}\right)\CS_{n,m-1,p-1}
    + \sqrt{m(p+1)}\frac{\asz}{\asy}\CS_{n,m-1,p+1}
  \right.\nonumber\\[0.5em]&\phantom{ +\frac{\qs}{e}\,\frac{\ocs}{\ope}\Bx\hspace{5mm} }\left.
    - \sqrt{(m+1)p}\frac{\asy}{\asz}\CS_{n,m+1,p-1} 
    + \sqrt{2m}\frac{\usz}{\asy}\CS_{n,m-1,p} 
    - \sqrt{2p}\frac{\usy}{\asz}\CS_{n,m,p-1}
  \right]\nonumber\\[0.5em]
  & -\csVM
  \By\left[
    \sqrt{pn}   \left(\frac{\asx}{\asz}-\frac{\asz}{\asx}\right)\CS_{n-1,m,p-1}
    + \sqrt{p(n+1)}\frac{\asx}{\asz}\CS_{n+1,m,p-1}
  \right.\nonumber\\[0.5em]&\phantom{ +\frac{\qs}{e}\,\frac{\ocs}{\ope}\Bx\hspace{5mm} }\left.
     - \sqrt{(p+1)n}\frac{\asz}{\asx}\CS_{n-1,m,p+1} 
     + \sqrt{2p}\frac{\usx}{\asz}\CS_{n,m,p-1} 
     - \sqrt{2n}\frac{\usz}{\asx}\CS_{n-1,m,p}
  \right]\nonumber\\[0.5em]
  & -\csVM
  \Bz\left[
    \sqrt{nm}\left(\frac{\asy}{\asx}-\frac{\asx}{\asy}\right)\CS_{n-1,m-1,p}
    + \sqrt{n(m+1)}\frac{\asy}{\asx}\CS_{n-1,m+1,p}
  \right.\nonumber\\[0.5em]&\phantom{ +\frac{\qs}{e}\,\frac{\ocs}{\ope}\Bx\hspace{5mm} }\left.
    - \sqrt{(n+1)m}\frac{\asx}{\asy}\CS_{n+1,m-1,p} 
    + \sqrt{2n}\frac{\usy}{\asx}\CS_{n-1,m,p} 
    - \sqrt{2m}\frac{\usx}{\asy}\CS_{n,m-1,p}
  \right] = 0.
  \label{eq:Vlasov-system:long}
\end{align}
By comparison with~\eqref{eq:Vlasov:compact}, \eqref{eq:cL:def},
\eqref{eq:cN:def}, one can observe that the first three terms in the
parenthesis after the time derivative, which contain the spatial
derivatives of the Hermite coefficients, derive from $\cL(\CS)$ and
the subsequent ones ensue from $\cN(\CS)$.
Details about the derivation of Eq.~\eqref{eq:Vlasov-system:long} are
reported in the final appendix for completeness.


\subsection{Discontinuous Galerkin approximation in configuration space}

We adopt the usual notation from finite difference schemes on
Cartesian grids.
We partition the space domain $\Ox$ into $N_c=N_xN_yN_z$ cubic or
regular hexahedral cells, so that we have $N_x$ mesh elements in the
$x$-direction, $N_y$ elements in the $y$-direction, and $N_z$
elements in the $z$-direction.
These partitions are labeled by the latin indices $i,j,k$
running from $1$ to $N_x$, $N_y$, and $N_z$, respectively.
For convenience of exposition, we label the generic mesh cell by the letter
$I$ and express the summation over all mesh cells by $\sum_{I}$
(without specifying the summation bounds).
With some abuse of notation, we may subindex $\Is$ as $\Is_{i,j,k}$,
so that, for example, two consecutive cells in the $x$-direction are
denoted by $\Is_{i,j,k}$ and $\Is_{i+1,j,k}$ and are separated by the
cell interface $\Fip$.
We denote the position of the cell center $\Is_{i,j,k}$ by
$\xv_{i,j,k}=(\xs_{i,j,k},\ys_{i,j,k},\zs_{i,j,k})$, and the size of such cell along
the $x$-, $y$-, and $z$-directions by $\Delta\xs_{i,j,k}$,
$\Delta\ys_{i,j,k}$, and $\Delta\zs_{i,j,k}$, so that
\begin{align*}
  \Is\equiv
  \Is_{i,j,k} =
  &\left[-\frac{\Delta\xs_{i,j,k}}{2}+\xs_{i,j,k},\frac{\Delta\xs_{i,j,k}}{2}+\xs_{i,j,k}\right]\times
  \left[-\frac{\Delta\ys_{i,j,k}}{2}+\ys_{i,j,k},\frac{\Delta\ys_{i,j,k}}{2}+\ys_{i,j,k}\right]\times\nonumber\\[0.5em]
  &\left[-\frac{\Delta\zs_{i,j,k}}{2}+\zs_{i,j,k},\frac{\Delta\zs_{i,j,k}}{2}+\zs_{i,j,k}\right].
\end{align*}
For the exposition's sake, we may assume that the cells are all equispaced, 
and consider $\Delta\xs$,
$\Delta\ys$, and $\Delta\zs$ as the mesh size steps in the three spatial directions.
Accordingly, triplets with an half-integer index, e.g.
$(i\pm\frac{1}{2},j,k)$, $(i,j\pm\frac{1}{2},k)$ and
$(i,j,k\pm\frac{1}{2})$, denote the cell interfaces that are
orthogonal to the $x$-, $y$-, and $z$-direction, respectively, and
delimiting cell $\Is_{i,j,k}$.  
The faces are oriented such that the normal vector to each face always
points outwards.

Next, we consider the space of polynomials of degree up to $\NDG$
defined on $I$, which we denote by $\PS{\NDG}(I)$.
We do not assume any continuity or weaker regularity condition at the
interface shared by two consecutive cells.
Therefore, any function defined on $\Ox$ whose restriction to any mesh
cell is the product of univariate polynomials of degree (at most) $\NDG$, 
may be discontinuous at any
cell interface.
We denote the basis for the local polynomial space on cell $I$ by
$\{\varphi^{\Il}\}$ for $l=1,\ldots,\Nl$, where $\Nl$ is the
cardinality of $\PS{\NDG}(I)$, so that
$\PS{\NDG}(I)=\SPAN\big\{\varphi^{\Il}\}_{l}$.
We recall that $\Nl=\NDG+1$ for univariate polynomials,
$\Nl=(\NDG+1)(\NDG+2)\slash{2}$ for bivariate polynomials, and
$\Nl=(\NDG+1)(\NDG+2)(\NDG+3)\slash{6}$ for trivariate
polynomials.

The discontinuous Galerkin approximation of the spectral coefficients
$\CS_{n,m,p}(\xv,t)$, for a fixed triple $(n,m,p)$ and in every cell $I$, is given by
expanding the Hermite coefficient in the local polynomial basis of the space
$\PS{\NDG}(I)$:
\begin{align*}
  \Cs^{s}_{n,m,p}(\xv,t)\approx
  \Cs^{s,\DG{}}_{n,m,p}(\xv,t)
  :=\sum_{l=1}^{\Nl}C_{n,m,p}^{s,\Il}(t)\varphi^{\Il}(\xv),
  \quad\forall\,\xv\in I.
\end{align*}
Collecting the local expansions, we obtain the (possibly discontinuous)
global approximation of the spectral coefficients
\begin{align}
  \Cs^{s,\DG{}}_{n,m,p}(\xv,t) = \sum_{I}\sum_{l=1}^{N_l}\C_{n,m,p}^{s,\Il}(t)\varphi^{\Il}(\xv),
  \qquad\forall\,\xv\in\Ox,
  \label{eq:DG:local:Cs}
\end{align}
and, accordingly, the Hermite-\DG{} approximation of the distribution
function $\fs$ is given by
\begin{align}
  \fsN(\xv,\vv,t) :=
  \sum_{n,m,p}\C_{n,m,p}^{s,\DG{}}(\xv,t)\Psi_{n,m,p}(\xiv^s) =
  \sum_{n,m,p}\sum_{\Il}\C_{n,m,p}^{s,\Il}(t)\Psi_{n,m,p}(\xiv^s)\varphi^{\Il}(\xv),
  \label{eq:fsN:def}
\end{align}
where, to ease the notation, we did not specify the summation bounds.

\medskip
\begin{remark}
  To avoid using a cumbersome notation, hereafter, we will remove the
  superscript $\DG{}$ from the symbols denoting the approximate ``$\CS$''
  coefficients.
  Therefore, we keep using $\CS_{n,m,p}$ instead of
  $\C_{n,m,p}^{s,\DG{}}$, but intending that these quantities refer to
  the \DG{} expansion with a finite number of spatial modes.
\end{remark}

\medskip
\begin{remark}
  It is worth noting that $\fsH$ and $\fsN$ are
  different functions.
  Indeed, $\fsH$ for each species $s$ is the solution of the
  variational problem~\eqref{eq:weak:DG} and is an approximation of
  the corresponding distribution function $\fs$
  solving~\eqref{eq:Vlasov:fs} after the truncation of the expansion
  on the Hermite velocity basis.
  Instead, $\fsN$ is the discontinuous Galerkin approximation of
  $\fsH$ and is the solution of the discrete variational problem that
  is constructed in the rest of this section.
  Since the numerical method is formulated in terms of the finite set
  of ``$\CS$'' coefficients, we do not need to know the specific
  discrete equation satisfied by $\fsN$ for the implementation.
 Such equation is used in the analysis of the
  conservation properties of the semi-discrete method in Appendix \ref{sec:variational:formulation}.
\end{remark}

\medskip
Let $\Ls_\zeta$ denote the univariate Legendre polynomial of degree
$\zeta$ in the interval $[-1,1]$.  
The set of polynomials $\{\Ls_\zeta\}_{\zeta=0}^{\NDG}$ forms an
orthogonal basis for $\PS{\NDG}([-1,1])$
\cite{Funaro:2008}.
We construct the multidimensional functions $\varphi^{\Il}(\xv)$ as
the tensor product of rescaled and translated univariate Legendre
polynomials as follows,
\begin{equation}
  \varphi^{\Il}(\xv) = 
  \Ls_{\lx}\left(2\frac{\xs - \xsI}{\Delta\xs}\right) 
  \Ls_{\ly}\left(2\frac{\ys - \ysI}{\Delta\ys}\right) 
  \Ls_{\lz}\left(2\frac{\zs - \zsI}{\Delta\zs}\right),
  \label{eq:vphi:Legendre}
\end{equation}
where $\lx,\ly,\lz=0,\ldots,\NDG$ and $\lx+\ly+\lz\leq\NDG$.
The index $l$ is a convenient mapping onto integer numbers that enumerates
the triplets $(\lx,\ly,\lz)$ from $1$ to $\Nl$.
The orthogonality of the Legendre polynomials allows us to simplify
the calculation
in the discontinuous Galerkin formulation.
Indeed, we perform the integration analitically whenever we can reduce
the multidimensional integration involving the basis functions
$\varphiIl$ to the exact one-dimensional integration formulas
\begin{subequations}
  \begin{align}
    \int_{-1}^{1} \Ls_i(\sigma)\Ls_j(\sigma)d\sigma &= \frac{2}{2i+1} \delta_{i,j},
    \label{eq:DG:Ls:Ls}
    \\[1em]
    \int_{-1}^{1} \Ls_i(\sigma) \frac{d\Ls_j(\sigma)}{d\sigma}d\sigma &= 
    \begin{cases}
      2,& \mbox{if~} i<j \mbox{~and~} j-i \mbox{~is odd,~}\\ 
      0,& \mbox{otherwise,}
    \end{cases}\label{eq:DG:Ls:Ls2}
    \\[1em]
    \int_{-1}^{1} \Ls_i(\sigma) \Ls_j(\sigma) \Ls_k(\sigma) d\sigma &= 2  
    \begin{pmatrix}
      i & j & k \\ 
      0 & 0 & 0  
    \end{pmatrix}^2,
    \label{eq:DG:Ls:dLsdx}
  \end{align}
\end{subequations}
where the special form of Wigner 3-j symbol in the right-hand side
of~\eqref{eq:DG:Ls:dLsdx} is defined as~\cite{Wigner:1951,Abramowitz-Stegun:1972}:
\begin{align}
  \begin{pmatrix}
    i & j & k \\ 
    0 & 0 & 0  
  \end{pmatrix} =
  \begin{cases}
    \displaystyle
    (-1)^\ell\frac{\ell!}{(\ell-i)!(\ell-j)!(\ell-k)!}
    \sqrt{\frac{(2\ell-2i)!(2\ell-2j)!(2\ell-2k)!}{(2\ell+1)!}},
    & \mbox{~for~} 2\ell = i + j + k \mbox{~even,}\\[0.5em] 
    0,
    & \mbox{ otherwise.}
  \end{cases}
  \label{eq:DG:Ls:dLsdx:aux}
\end{align}

To derive the discontinuous Galerkin approximation of the Vlasov
equations, we multiply~\eqref{eq:Vlasov:compact} by the generic basis
function $\varphi^{\Il}$ and integrate over the space domain $\Ox$
\begin{align}\label{eq:semidisc}
  \int_{I}\left(
    \frac{\partial\Cs^{s}_{n,m,p}}{\partial t} + \cL_{n,m,p}(\CS) + \cN_{n,m,p}(\CS)
  \right)\varphiIl(\xv)\dxv = 0.
\end{align}
Note that the integral above has been restricted to the mesh element
$I$ since $\varphi^{\Il}$ is zero outside $I$.
We now consider each term separately.
In the first term a direct substitution of \eqref{eq:DG:local:Cs}
together with the orthogonality of the Legendre functions yields
\begin{equation}\label{eq:dCdt}
  \int_{I}\frac{\partial\Cs^{s}_{n,m,p}(t)}{\partial t}\varphiIl(\xv)\dxv =
      \sum_{l'=1}^{N_l}\frac{d\Cs^{s,\Ilp}_{n,m,p}(t)}{\dt}\int_{I}\varphiIlp(\xv)\varphiIl(\xv)\dxv
    = \mu_{I,l}
  \frac{d\Cs^{s,\Il}_{n,m,p}(t)}{\dt},
\end{equation}
where the multiplicative factor
$\mu_{I,l}:=\int_{\Is}\left(\varphiIl\right)^2\dxv$ is the $(l,l)$
entry of the \DG{} mass matrix (the mass matrix is diagonal in this case
in view of the orthogonality properties of the Legendre polynomials).

\subsection{Linear terms}
In order to reformulate~\eqref{eq:Vlasov-system:long} in a more
compact form we introduce the
vector-valued function $\Cvs(\xv,t)\in\mathbb{R}^N$, with $N:=(\Nn+1)(\Nm+1)(\Np+1)$,
that collects the
Hermite spectral coefficients for all $(\xv,t)$.
For a given vector $\Vv\in\mathbb{R}^N$, we denote
by $\big[\Vv\big]_{n,m,p}$ the
entry corresponding to the $(n,m,p)$-th triple, so that, for
example, $\big[\Cvs\big]_{n,m,p}:=\CS_{n,m,p}$.
With this notation, we reformulate the linear term
in~\eqref{eq:cL:def} as,
\begin{align}
  \cL(\Cvs)
  = \left(
    \matAx\frac{\partial}{\partial x} 
    + \matAy\frac{\partial}{\partial y} 
    + \matAz\frac{\partial}{\partial z} 
  \right)\Cvs,
  \label{eq:cL:def:b}
\end{align}
where 
$\big[\cL(\Cvs)\big]_{n,m,p}=\cL_{n,m,p}(\CS)$,
and the matrices $\matAx, \matAy, \matAz\in\mathbb{R}^{N\times N}$
are defined according to \eqref{eq:Vlasov-system:long} as,
\begin{subequations}
  \begin{align}
    \left[\matAx\frac{\partial\Cvs}{\partial x}\right]_{n,m,p}
    &= \asx\left(
      \sqrt{\frac{n+1}{2}}\frac{\partial\CS_{n+1,m,p}}{\partial x}
      + \sqrt{\frac{n}{2}}\frac{\partial\CS_{n-1,m,p}}{\partial x}
      + \frac{\usx}{\asx}\frac{\partial\CS_{n,m,p}}{\partial x}
    \right),
    \\[0.5em]
    \left[\matAy\frac{\partial\Cvs}{\partial y}\right]_{n,m,p}
    &= \asy\left(
      \sqrt{\frac{m+1}{2}}\frac{\partial\CS_{n,m+1,p}}{\partial y}
      + \sqrt{\frac{m}{2}}\frac{\partial\CS_{n,m-1,p}}{\partial y}
      + \frac{\usy}{\asy}\frac{\partial\CS_{n,m,p}}{\partial y}
    \right),
    \\[0.5em]
    \left[\matAz\frac{\partial\Cvs}{\partial z}\right]_{n,m,p}
    &= \asz\left(
      \sqrt{\frac{p+1}{2}}\frac{\partial\CS_{n,m,p+1}}{\partial z}
      + \sqrt{\frac{p}{2}}\frac{\partial\CS_{n,m,p-1}}{\partial z}
      + \frac{\usz}{\asz}\frac{\partial\CS_{n,m,p}}{\partial z}
    \right).
  \end{align}
\end{subequations}
Since $\matAx$, $\matAy$, and $\matAz$ are constant-valued matrices, 
we can directly consider their action on the vector of unknowns $\Cvs$ 
and write
\begin{align}
  \cL(\Cvs) 
  = \frac{\partial}{\partial\xs}\big(\matAx\Cvs\big) 
  + \frac{\partial}{\partial\ys}\big(\matAy\Cvs\big)
  + \frac{\partial}{\partial\zs}\big(\matAz\Cvs\big),
  \label{eq:nabla:cL:Cvs}
\end{align}
with the component-wise definitions
\begin{subequations}
  \begin{align}
    \left[\matAx\Cvs\right]_{n,m,p}
    &= \asx\left(
      \sqrt{\frac{n+1}{2}}\CS_{n+1,m,p}
      + \sqrt{\frac{n}{2}}\CS_{n-1,m,p}
      + \frac{\usx}{\asx}\CS_{n,m,p}
    \right),
    \label{eq:Ax:Cvs}\\[0.5em]
    \left[\matAy\Cvs\right]_{n,m,p}
    &= \asy\left(
      \sqrt{\frac{m+1}{2}}\CS_{n,m+1,p}
      + \sqrt{\frac{m}{2}}\CS_{n,m-1,p}
      + \frac{\usy}{\asy}\CS_{n,m,p}
    \right),
    \label{eq:Ay:Cvs}\\[0.5em]
    \left[\matAz\Cvs\right]_{n,m,p}
    &= \asz\left(
      \sqrt{\frac{p+1}{2}}\CS_{n,m,p+1}
      + \sqrt{\frac{p}{2}}\CS_{n,m,p-1}
      + \frac{\usz}{\asz}\CS_{n,m,p}
    \right).
    \label{eq:Az:Cvs}
  \end{align}
\end{subequations}
Relations~\eqref{eq:Ax:Cvs}-\eqref{eq:Az:Cvs} are used in the next
section to introduce the upwind flux discretization of the Vlasov
equations in the discontinuous Galerkin framework.

We split the cell boundary as
$\partial\Is=\partial\Is_x\cup\partial\Is_y\cup\partial\Is_z$, where
$\partial\Is_x=\fip\cup\fim$,
$\partial\Is_y=\fjp\cup\fjm$, and
$\partial\Is_z=\fkp\cup\fkm$.
Integrating by parts the linear term of \eqref{eq:semidisc}
and using~\eqref{eq:nabla:cL:Cvs} yields
\begin{align}
  \int_{\Is}\cL_{n,m,p}(\CS)\varphi^{\Il}(\xv)\,\dxv
  &=\int_{\Is}\bigg[\sum_{\beta\in\{x,y,z\}} \matAb\dfrac{\partial \Cvs(\xv,t)}{\partial\beta}
  \bigg]_{n,m,p} \varphi^{\Il}(\xv)\,\dxv
  \nonumber\\[0.5em]
  &= \int_{\Is}\bigg[\sum_{\beta\in\{x,y,z\}} \dfrac{\partial}{\partial\beta}(\matAb\Cvs(\xv,t))\bigg]_{n,m,p} \varphi^{\Il}(\xv)\,\dxv
  \nonumber\\[0.5em]
    &= \sum_{\beta\in\{x,y,z\}}\int_{\Is}
    \dfrac{\partial}{\partial\beta}[\matAb\Cvs(\xv,t)]_{n,m,p} \varphi^{\Il}(\xv)\,\dxv
  \nonumber\\[0.5em]
  &= -\sum_{\beta\in\{x,y,z\}}\int_{I}
  [\matAb\Cvs(\xv,t)]_{n,m,p}
  \dfrac{\partial \varphi^{\Il}(\xv)}{\partial \beta}\,\dxv\nonumber\\[0.5em]
  &\qquad + \sum_{\beta\in\{x,y,z\}}\Diffv{\beta}{[\matAb\Cvs(\xv,t)]_{n,m,p}
  \,\varphi^{\Il}(\xv)}_{\partial I_{\beta}},
  \label{eq:intg-by-parts}
\end{align}
where
\begin{align}
  \Diffv{\beta}{[\matAb\Cvs(\xv,t)]_{n,m,p}\,\varphi^{\Il}(\xv)}_{\partial I_{\beta}}
  := \int_{\partial I_{\beta}}[\ns_{\beta}\matAb\Cvs(\xv,t)]_{n,m,p}\,\varphi^{\Il}(\xv)\dS,
\end{align}
are the terms at $\partial I$, the boundary of cell $I$,
originating from the integration by parts, and $\ns_{\beta}$, $\beta\in\{x,y,z\}$, is the $\beta$ component of
$\nv=(n_x,n_y,n_z)^T$, the unit vector orthogonal to $\partial I$.
Therefore, the right-hand side of~\eqref{eq:intg-by-parts} contains
two integral terms: a volume integration term on $I$ and a surface
integration term on $\partial I$.
The volume integral is the sum of three independent volume integrals
associated with the derivatives of $\varphiIl$ in the three directions
$x$, $y$, and $z$.
To exploit the orthogonality properties of the Legendre polynomials \eqref{eq:DG:Ls:Ls} and \eqref{eq:DG:Ls:Ls2},
we substitute expansion~\eqref{eq:DG:local:Cs} and use
formulas~\eqref{eq:Ax:Cvs}, \eqref{eq:Ay:Cvs}, and~\eqref{eq:Az:Cvs}.
Let us
introduce the matrices $\Phiv_{\beta}\in\mathbb{R}^{N_l\times N_l}$,
for $\beta\in\{x,y,z\}$, defined as
  \begin{align}\label{eq:term:2:05:x}
    \big(\Phiv_{\beta}\big)_{l,l'} = \int_{I}\frac{\partial\varphi^{\Il}(\xv)}{\partial\beta}
    \,\,\varphi^{\Ilp}(\xv)\dxv.
  \end{align}
Then,
\begin{subequations}
  \begin{align}
    \big[\calI^{x,s}\big]_{n,m,p}
    &:=\int_{I}\big[\matAx\Cvs(\xv,t)\big]_{n,m,p}
    \frac{\partial\varphi^{\Il}(\xv)}{\partial x}\dxv
    \nonumber\\[0.5em]
    &=\sum_{l'=1}^{N_l}\big(\Phiv_{x}\big)_{l,l'}
    \bigg(
    \asx\sqrt{\frac{n+1}{2}}C_{n+1,m,p}^{s,\Ilp}(t) + \asx\sqrt{\frac{n}{2}}C_{n-1,m,p}^{s,\Ilp}(t) + \usx C_{n,m,p}^{s,\Ilp}(t)
    \bigg),
    \label{eq:Ix}\\[0.5em]
    \big[\calI^{y,s}\big]_{n,m,p}
    &:=\int_{I}\big[\matAy\Cvs(\xv,t)\big]_{n,m,p}
    \frac{\partial\varphi^{\Il}(\xv)}{\partial y}\dxv
    \nonumber\\[0.5em]
    &=\sum_{l'=1}^{N_l}\big(\Phiv_{y}\big)_{l,l'}
    \bigg(
    \asy\sqrt{\frac{m+1}{2}}C_{n,m+1,p}^{s,\Ilp}(t) + \asy\sqrt{\frac{m}{2}}C_{n,m-1,p}^{s,\Ilp}(t) + \usy C_{n,m,p}^{s,\Ilp}(t)
    \bigg),
    \label{eq:Iy}\\[0.5em]
    \big[\calI^{z,s}\big]_{n,m,p}
    &:=\int_{I}\big[\matAz\Cvs(\xv,t)\big]_{n,m,p}
    \frac{\partial\varphi^{\Il}(\xv)}{\partial z}\dxv 
    \nonumber\\[0.5em]
    &=\sum_{l'=1}^{N_l}\big(\Phiv_{z}\big)_{l,l'}
    \bigg(
    \asz\sqrt{\frac{p+1}{2}}C_{n,m,p+1}^{s,\Ilp}(t) + \asz\sqrt{\frac{p}{2}}C_{n,m,p-1}^{s,\Ilp}(t) + \usz C_{n,m,p}^{s,\Ilp}(t)
    \bigg),\label{eq:Iz}
  \end{align}
\end{subequations}
where, for future reference, we introduced the symbols $\calI^{\beta,s}$.
After splitting the integral in \eqref{eq:term:2:05:x} in the one-dimensional integrals
for the Legendre polynomials, formulas~\eqref{eq:DG:Ls:Ls} and
\eqref{eq:DG:Ls:Ls2} can easily be applied.

\begin{figure}
  \begin{center}
    \begin{overpic}[width=10cm]{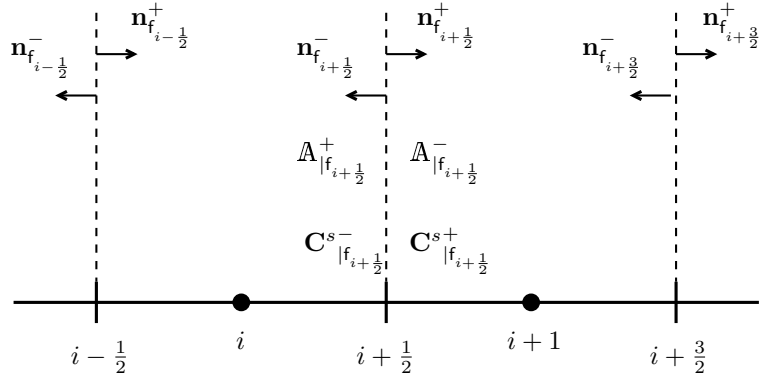}
      \put(85,-5){$i+\frac32$}
      \put(46,-5){$i+\frac12$}
      \put( 8, -5){$i-\frac12$}
      \put(66,-3){$i+1$}
      \put(30,-3){$i$}
      \put(76,36){$\mathbf{n}^{-}_{\F_{i+\frac32}}$}
      \put(92,40) {$\mathbf{n}^{+}_{\F_{i+\frac32}}$}
      \put(38,36){$\mathbf{n}^{-}_{\F_{i+\frac12}}$}
      \put(54,40){$\mathbf{n}^{+}_{\F_{i+\frac12}}$}
      \put(  0,36){$\mathbf{n}^{-}_{\F_{i-\frac12}}$}
      \put(16,40){$\mathbf{n}^{+}_{\F_{i-\frac12}}$}
      \put(38,22){$\matA^{+}_{|\F_{i+\frac12}}$}
      \put(53,22){$\matA^{-}_{|\F_{i+\frac12}}$}
      \put(39,10){${\Cvs}^{-}_{|\F_{i+\frac12}}$}
      \put(53,10){${\Cvs}^{+}_{|\F_{i+\frac12}}$}
    \end{overpic}
  \end{center}
  \vspace{0.75cm}
  \caption{Notation of the numerical flux functions along direction $x$.
    To ease the notation, we do not show the indices $j$ and $k$.
    We denote the
    unit normal vector to face $\F_{\ell}$ by $\nv_{\F_{\ell}}$,
    $\ell=i\pm\frac12,i+\frac32$.
    Vectors $\nv^{+}_{\F_{\ell}}$ are oriented along the positive real
    axis (from left to right); vectors $\nv^{-}_{\F_{\ell}}$ are in
    the opposite sense.
    Similarly, at interface $\F_{i+\frac12}$ matrix
    $\matAx=\matAx^{+}+\matAx^{-}$ is decomposed into characteristics
    waves traveling from left to right, $\matAx^{+}$, which transport
    solution ${\Cvs}^{-}$ from inside cell $i$ toward the cell
    interface, and from right to left, $\matAx^{-}$, which transport
    the solution ${\Cvs}^{+}$ from inside cell $i+1$ towards the cell
    interface.
    Let ${\Cvs}^{\pm}_{|\F_{i+\frac12}}$ be a shortcut for
    $\Cvs(x^{\pm}_{i+\frac{1}{2}},y,z,t)$.
    According to this notation, the upwind flux is given by
    $\widehat{\matAx\Cvs}{\big\rvert_{\F_{i+\frac12}}}=\matAx^{+}{\Cvs}^{-}_{|\F_{i+\frac12}}+\matAx^{-}{\Cvs}^{+}_{|\F_{i+\frac12}}$.}
  \label{fig:numflux:notation}
  \vspace{0.5cm}
\end{figure}

\medskip
On the other hand, the calculation of the boundary integral term
of~\eqref{eq:intg-by-parts} deserves special attention since we need
to introduce upwind numerical flux functions.
The choice of upwinding the fluxes is dictated by considerations on
the numerical stability of the scheme, \emph{cf.}
\cite{Cockburn-Shu:1989}, since upwinding  make the proper
information propagation possible in the computational domain.
Figure \ref{fig:numflux:notation} illustrates the meaning of the main
symbols that we adopt in rest of the section and throughout the paper.
In view of the face orientation and noting that $\ns_{\beta}$ can only be $\pm1$ or zero, the total fluxes along the directions
$x$, $y$, and $z$ are given by
\begin{subequations}
  \begin{align}
    \Diffx{\big[\matAx\Cvs(\xv,t)\big]_{n,m,p}\varphi^{\Il}(\xv)}_{\pxI}
    &= \int_{\Fip}\big[\matAx\Cvs(x_{i+\frac{1}{2}},y,z,t)\big]_{n,m,p}\varphi^{\Il}(x_{i+\frac{1}{2}},y,z)\dy\dz
    \nonumber\\[0.5em]
    &- \int_{\Fim}\big[\matAx\Cvs(x_{i-\frac{1}{2}},y,z,t)\big]_{n,m,p}\varphi^{\Il}(x_{i-\frac{1}{2}},y,z)\dy\dz,
    \label{eq:diff:x}\\[0.5em]
    \Diffy{\big[\matAy\Cvs(\xv,t)\big]_{n,m,p}\varphi^{\Il}(\xv)}_{\pyI}
    &= \int_{\Fjp}\big[\matAy\Cvs(x,y_{j+\frac{1}{2}},z,t)\big]_{n,m,p}\varphi^{\Il}(x,y_{j+\frac{1}{2}},z)\dx\dz
    \nonumber\\[0.5em]
    &- \int_{\Fjm}\big[\matAy\Cvs(x,y_{j-\frac{1}{2}},z,t)\big]_{n,m,p}\varphi^{\Il}(x,y_{j-\frac{1}{2}},z)\dx\dz,
    \label{eq:diff:y}\\[0.5em]
    \Diffz{\big[\matAz\Cvs(\xv,t)\big]_{n,m,p}\varphi^{\Il}(\xv)}_{\pzI}
    &= \int_{\Fkp}\big[\matAz\Cvs(x,y,z_{j+\frac{1}{2}},t)\big]_{n,m,p}\varphi^{\Il}(x,y,z_{j+\frac{1}{2}})\dx\dy
    \nonumber\\[0.5em]
    &- \int_{\Fkm}\big[\matAz\Cvs(x,y,z_{j-\frac{1}{2}},t)\big]_{n,m,p}\varphi^{\Il}(x,y,z_{j-\frac{1}{2}})\dx\dy.
    \label{eq:diff:z}
  \end{align}
\end{subequations}

\medskip
Let $\Cvs(\xpp,y,z,t)$ denote $\lim_{\epsilon\rightarrow
  0^+}\Cvs(x_{i+\frac{1}{2}}+\epsilon,y,z,t)$, and similarly for the
trace at the other faces. 
The integral on face $\Fip$ is approximated by,
\begin{align}
  \delta_x^{s,+} := \int_{\Fip}
  \left[
    \NFLX{\matAx\Cvs}(x_{i+\frac{1}{2}},y,z,t)
  \right]_{n,m,p}
  \varphi^{\Il}(x_{i+\frac{1}{2}},y,z)\dy\dz,
  \label{eq:flux-x:10}
\end{align}
and the upwind numerical flux in the integral argument is defined as
\begin{align}
  \NFLX{\matAx\Cvs}(x_{i+\frac{1}{2}},y,z,t) = 
  \underbrace{\matAx^{+}\Cvs(x^{-}_{i+\frac{1}{2}},y,z,t)}_{\textrm{from cell $I=I_{i,j,k}$}} + 
  \underbrace{\matAx^{-}\Cvs(x^{+}_{i+\frac{1}{2}},y,z,t)}_{\textrm{from cell $I^{+}=I_{i+1,j,k}$}}.
  \label{eq:flux-x:15} 
\end{align}
In this formula, we outlined the contributions from the two cells sharing the face
$\Fip$, namely $I=I_{i,j,k}$ and $I^{+}:=I_{i+1,j,k}$, and denoted 
the upwind matrices given by the characteristic decomposition of matrix $\matAx$ by 
$\matAx^{+}$ and $\matAx^{-}$. 
We recall that matrix $\matA^{+}$ is built from the 
positive eigenvalues of matrix $\matA$, and matrix $\matA^{-}$ from the negative ones, 
so that $\matA=\matA^{+}+\matA^{-}$.
We refer to~\cite{LeVeque:1992} for a presentation of this topic, while we discuss
the major details of how to implement this algorithm efficiently
in Remark~\ref{remark:upwind:flux:implementation} at the end of the section.
Similarly, the
numerical fluxes at the six faces $\F_{i\pm\frac{1}{2},j,k}$,
$\F_{i,j\pm\frac{1}{2},k}$, and $\F_{i,j,k\pm\frac{1}{2}}$ are given by
\begin{subequations}
  \begin{align}
    \NFLX{\matAx\Cvs}{\big\rvert_{\Fipm}} & = \matAx^{\pm}\Cvs(x^{\mp}_{i\pm\frac{1}{2}},y,z,t) + \matAx^{\mp}\Cvs(x^{\pm}_{i\pm\frac{1}{2}},y,z,t),\label{eq:upw:x}\\[0.5em]
    \NFLX{\matAy\Cvs}{\big\rvert_{\Fjpm}} & = \matAy^{\pm}\Cvs(x,y^{\mp}_{j\pm\frac{1}{2}},z,t) + \matAy^{\mp}\Cvs(x,y^{\pm}_{j\pm\frac{1}{2}},z,t),\label{eq:upw:y}\\[0.5em]
    \NFLX{\matAz\Cvs}{\big\rvert_{\Fkpm}} & = \matAz^{\pm}\Cvs(x,y,z^{\mp}_{k\pm\frac{1}{2}},t) + \matAz^{\mp}\Cvs(x,y,z^{\pm}_{k\pm\frac{1}{2}},t),\label{eq:upw:z}
  \end{align}
\end{subequations}
where $\matAx^{\pm}$, $\matAy^{\pm}$, and $\matAz^{\pm}$ are the \emph{upwind matrices}
given by the characteristic decomposition of matrices $\matAx$, $\matAy$, $\matAz$,
respectively.
In the following, we continue detailing the procedure for the numerical flux
across face $\Fip$, the extension to the other five faces being
deemed straightforward.
Using the expansion of the Hermite coefficients in the local
polynomial basis functions, we can rewrite the
contribution from cell $I=I_{i,j,k}$ in \eqref{eq:flux-x:15} as follows,
\begin{equation*}
  \begin{aligned}
    \big[\matAxp\Cvs(\xmp,y,z,t)\big]_{n,m,p}
    &= \sum_{n'=0}^{\Nn}(\Av_x^+)_{n+1,n'+1}\,\CS_{n',m,p}(\xmp,y,z,t)
    &\big[\mbox{\textrm{since~$\xmp\in I$}}\big]\\[1em]
    &= \sum_{n'=0}^{\Nn}(\Av_x^+)_{n+1,n'+1}\,\sum_{l'=1}^{N_l}C^{s,\Il'}_{n',m,p}(t)\,\varphi^{\Il'}(\xmp,y,z)
    &\big[\mbox{\textrm{rearrange~the~summation}}\big]\\[0.5em]
    &= \sum_{l'=1}^{N_l} \bigg(\sum_{n'=0}^{\Nn}(\Av_x^+)_{n+1,n'+1}\,
    C^{s,\Il'}_{n',m,p}(t)\bigg) \varphi^{\Il'}(\xmp,y,z).
  \end{aligned}
\end{equation*}
Analogously, using the local polynomials
$\varphi^{\Icp,l'}$ defined on cell $\Icp$,
\begin{equation*}
  \begin{aligned}
    \big[\matAxm\Cvs(\xpp,y,z,t)\big]_{n,m,p}
    &= \sum_{n'=0}^{\Nn}(\Av_x^-)_{n+1,n'+1}\,\CS_{n',m,p}(\xpp,y,z,t)
    &\big[\mbox{\textrm{since~$\xpp\in \Icp$}}\big]\\[0.5em]
    &= \sum_{n'=0}^{\Nn}(\Av_x^-)_{n+1,n'+1}
    \sum_{l'=1}^{N_l}C^{s,\Icp,l'}_{n',m,p}(t)\,\varphi^{\Icp,l'}(\xpp,y,z)
    &\big[\mbox{\textrm{rearrange~the~summation}}\big]\\[0.5em]
    &= \sum_{l'=1}^{N_l} \bigg(\sum_{n'=0}^{\Nn}(\Av_x^-)_{n+1,n'+1}\,
    C^{s,\Icp,l'}_{n',m,p}(t)\bigg) \varphi^{\Icp,l'}(\xpp,y,z).
  \end{aligned}
\end{equation*}
The matrices $\Av_x^{\pm}$ corresponding to the linear operators $\matAx^{\pm}$
are implemented as explained in Remark~\ref{remark:upwind:flux:implementation}
and are indexed from $1$ to $\Nn+1$.
We compute $\delta_x^{s,+}$ using ~\eqref{eq:flux-x:10} and splitting
\eqref{eq:flux-x:15}, namely
\begin{align}
  \delta_x^{s,+}
  &= 
  \sum_{l'=1}^{N_l} \left(\sum_{n'=0}^{\Nn}(\Av_x^+)_{n+1,n'+1}C^{s,\Il'}_{n',m,p}(t)\right)
  \underbrace{\left(\int_{\Fip}\varphi^{\Il'}(\xmp,y,z)\,\varphi^{\Il}(\xmp,y,z)\,\dy\dz\right)}_{\textrm{only cell $I$}}
  \nonumber\\[0.5em]&\phantom{=}
  +
  \sum_{l'=1}^{N_l} \left(\sum_{n'=0}^{\Nn}(\Av_x^-)_{n+1,n'+1}C^{s,\Icp,l'}_{n',m,p}(t)\right)
  \underbrace{\left(\int_{\Fip}\varphi^{\Icp,l'}(\xpp,y,z)\,\varphi^{\Il}(\xmp,y,z)\,\dy\dz\right)}_{\textrm{two distinct cells $I$ and $\Icp$}}.
  \label{eq:upwind:flux}
\end{align}
We compute the face integrals in~\eqref{eq:upwind:flux} by using
orthogonality properties of the Legendre polynomials
\eqref{eq:DG:Ls:Ls}.
To summarize, the local spectral Hermite-\DG{} discretization of the
linear term of the Vlasov equation reads
\begin{align}
  \int_{\Is}\cL_{n,m,p}(\CS)\varphi^{\Il}(\xv)\,\dxv
  = \sum_{\beta\in\{x,y,z\}}\bigg(- \big[\calI^\beta\big]_{n,m,p} + (\delta_\beta^{s,+} - \delta_\beta^{s,-})\bigg),
  \label{eq:Vlasov:cL:def}
\end{align}
where the volume terms $\big[\calI^x\big]_{n,m,p}$,
$\big[\calI^y\big]_{n,m,p}$ and $\big[\calI^z\big]_{n,m,p}$ are
defined in \eqref{eq:Ix}, \eqref{eq:Iy} and \eqref{eq:Iz},
respectively.
The boundary term $\delta_x^{s,+}$ defined in \eqref{eq:flux-x:10} is
computed as in \eqref{eq:upwind:flux}, while similar derivations can
be carried out for the terms $\delta_x^{s,-}$, $\delta_y^{s,\pm}$,
$\delta_z^{s,\pm}$.

\medskip
\begin{remark}
  \label{rem:upwind:Riemann:pblm}
  The upwind integrated fluxes given by combining
  \eqref{eq:diff:x}-\eqref{eq:diff:z} and
  \eqref{eq:upw:x}-\eqref{eq:upw:z} are equivalent to solving the
  Riemann problems defined by the discontinuous traces of the Hermite
  coefficients at the cell interfaces for the linear hyperbolic
  system~\eqref{eq:weak:DG}, see, e.g.,~\cite{LeVeque:1992}.
\end{remark}

\medskip
\begin{remark}
  \label{remark:upwind:flux:implementation}
  For an efficient implementation of \eqref{eq:upw:x},
  \eqref{eq:upw:y}, \eqref{eq:upw:z}, the vector $\Cvs$ of the Hermite
  coefficients is stored as a 3rd order tensor.
  The implemented matrices
  $\Av_{\beta}\in\mathbb{R}^{(N_{v_\beta}+1)\times (N_{v_\beta}+1)}$
  corresponding to the linear operators $\matA_{\beta}$ for
  $\beta\in\{x,y,z\}$, are real, symmetric, and tridiagonal with
  entries,
  \begin{align*}
    (\Av_{\beta})_{i,i}   &= u^s_\beta,
    & \mbox{for}\quad i=1,\ldots,N_{v_\beta}+1,\\
    (\Av_{\beta})_{i,i-1} &=  \alpha^s_{\beta}\sqrt{(i-1)/2},
    & \mbox{for}\quad i=2,\ldots,N_{v_\beta}+1,\\
    (\Av_{\beta})_{i,i+1} &=  \alpha^s_{\beta}\sqrt{i/2},
    &\mbox{for}\quad i=1,\ldots,N_{v_\beta}.
  \end{align*}
  The matrices $\{\Av_{\beta}\}_{\beta\in\{x,y,z\}}$ can be
  diagonalized, namely there exists an orthogonal matrix $\matRb$
  (i.e., $\matRb^T=\matRb^{-1}$), whose columns are the eigenvectors
  of matrix $\Av_{\beta}$ and a diagonal matrix $\matDb$, whose
  diagonal elements $\{\lambda_{\beta,i}\}_{i=1}^{N_\beta}$ are the
  eigenvalues of $\Av_{\beta}$.
  Therefore, it holds that $\Av_{\beta}=\matRb^T\matDb\matRb$, and we
  can define the ``upwind'' matrices
  $\Av_{\beta}^{\pm}=\matRb^T\matDb^{\pm}\matRb$, where
  $\matDb^{\pm}=\frac{1}{2}\mathrm{diag}(\lambda_{\beta,i}\pm\ABS{\lambda_{\beta,i}})$.
  Note that the computation of the eigenvalues and eigenvectors of the
  matrices $\{\Av_{\beta}\}$ is not computationally expensive: the
  matrices are tridiagonal, do not change at each mesh interface, and
  are independent of time.
  Consequently, they only need to be computed once for a single cell
  and stored in memory at the beginning of each numerical simulation.
\end{remark}

\subsection{Nonlinear terms}
Consider the local contribution of the nonlinear terms in the semi-discretization \eqref{eq:Vlasov-system:long}
of Vlasov equation, namely
\begin{align*}
  \int_{\Is}\cN_{n,m,p}(\CS)\varphi^{\Il}(\xv)\,\dxv.
\end{align*}
The electric and magnetic field $\vecE$ and $\vecB$,
which are involved in these terms,
are approximated using piecewise polynomials.
In particular, in the
mesh element $I$ we consider the multivariate Legendre polynomial basis
$\{\varphi^{\Il}(\xv)\}_{l=1}^{N_l}$.
To ease the exposition, we assume that the degree of the local
polynomial spaces are the same, so that,
\begin{align}
  \vecEN(\xv,t)
  &= \begin{pmatrix}
    \ExN(\xv,t)\\
    \EyN(\xv,t)\\
    \EzN(\xv,t)
    \end{pmatrix}
  = \begin{pmatrix}
      \sum_{\Il}\ExIl(t)\varphiIl(\xv)\\[0.5em]
      \sum_{\Il}\EyIl(t)\varphiIl(\xv)\\[0.5em]
      \sum_{\Il}\EzIl(t)\varphiIl(\xv)
    \end{pmatrix}
  = \sum_{\Il}
    \begin{pmatrix}
        \ExIl(t)\\[0.5em]
        \EyIl(t)\\[0.5em]
        \EzIl(t)
    \end{pmatrix}
    \varphiIl(\xv)
  =: \sum_{\Il}\vecEIl(t)\,\varphiIl(\xv),
  \label{eq:Ev:expansion}
  \\[0.5em]
  \vecBN(\xv,t)
  &= \begin{pmatrix}
  \BxN(\xv,t)\\
  \ByN(\xv,t)\\
  \BzN(\xv,t)
  \end{pmatrix}
  =\begin{pmatrix}
      \sum_{\Il}\BxIl(t)\varphiIl(\xv)\\[0.5em]
      \sum_{\Il}\ByIl(t)\varphiIl(\xv)\\[0.5em]
      \sum_{\Il}\BzIl(t)\varphiIl(\xv)
    \end{pmatrix}
  = \sum_{\Il} \begin{pmatrix}
  \BxIl(t)\\[0.5em]
  \ByIl(t)\\[0.5em]
  \BzIl(t)
  \end{pmatrix}
  \varphiIl(\xv)
  =: \sum_{\Il}\vecBIl(t)\,\varphiIl(\xv).
  \label{eq:Bv:expansion}
\end{align}
To keep the presentation focused we only present the spatial discretization of the terms
involving the electric and magnetic fields in the $x$-direction. In the other
directions the derivation follows straightforwardly.

From \eqref{eq:Vlasov-system:long}, we define
\begin{align*}
  \big[\calI^{\Ex,s}\big]_{n,m,p} :=
  \frac{\qs}{\ms}\,\frac{\oce}{\ope}
  \int_{I}\frac{\sqrt{2n}}{\asx}\ExN(\xv,t)\CS_{n-1,m,p}(\xv,t)\varphi^{\Il}(\xv)\,d\xv.
\end{align*}
Using the spatial \DG{} discretization of the Hermite coefficients and
the electric field yields,
\begin{equation}\label{eq:IEx}
\begin{aligned}
  \big[\calI^{\Ex,s}\big]_{n,m,p} 
  & = \frac{\qs}{\ms}\,\frac{\oce}{\ope}\frac{\sqrt{2n}}{\asx}\sum_{l'=1}^{N_l}C^{s,\Il'}_{n-1,m,p}(t)\int_{I}\ExN(\xv,t)\,\varphi^{\Il'}(\xv)\,\varphi^{\Il}(\xv)\,d\xv\\[0.5em]
  & = \frac{\qs}{\ms}\,\frac{\oce}{\ope}\frac{\sqrt{2n}}{\asx}\sum_{l'=1}^{N_l}C^{s,\Il'}_{n-1,m,p}(t)\sum_{l''=1}^{N_l}\Ex^{\Il''}(t)\int_I\varphi^{I,l''}(\xv)\varphi^{\Il'}(\xv)\,\varphi^{\Il}(\xv)\,d\xv,
\end{aligned}
\end{equation}
where the integral appearing in the last term can be computed using
formula \eqref{eq:DG:Ls:dLsdx}.
We define and compute the other nonlinear terms
$\big[\calI^{\Ey,s}\big]_{n,m,p}$ and
$\big[\calI^{\Ez,s}\big]_{n,m,p}$ involving $\Ey$ and
$\Ez$, respectively, in the same manner.

To deal with the terms involving the magnetic field $\vecB$, we define
\begin{align*}
  &\big[\calI^{\Bx,s}\big]_{n,m,p}:=\frac{\qs}{\ms}\,\frac{\oce}{\ope}\int_{I}
  \BxN(\xv,t)\left(
    \sqrt{mp}\left(\frac{\asz}{\asy}-\frac{\asy}{\asz}\right)\CS_{n,m-1,p-1}
    + \sqrt{m(p+1)}\,\frac{\asz}{\asy}\CS_{n,m-1,p+1}
     \right.\nonumber\\[0.5em]
     &\qquad\left.
    - \sqrt{(m+1)p}\,\frac{\asy}{\asz}\,\CS_{n,m+1,p-1} 
    + \sqrt{2m}\,\frac{\usz}{\asy}\,\CS_{n,m-1,p} 
    - \sqrt{2p}\,\frac{\usy}{\asz}\,\CS_{n,m,p-1}
  \right)\varphi^{\Il}(\xv)\,d\xv,
\end{align*}
and we introduce the quantity
\begin{align*}
  \zeta^{s,\Il'}_{x,n,m,p}(t) := 
    &\, \sqrt{mp}\left(\frac{\asz}{\asy}-\frac{\asy}{\asz}\right)C^{s,\Il'}_{n,m-1,p-1}(t)
    + \sqrt{m(p+1)}\,\frac{\asz}{\asy}\,C^{s,\Il'}_{n,m-1,p+1}(t)
    \nonumber\\[0.5em]
  &
    - \sqrt{(m+1)p}\,\frac{\asy}{\asz}\,C^{s,\Il'}_{n,m+1,p-1} (t)
    + \sqrt{2m}\,\frac{\usz}{\asy}\,C^{s,\Il'}_{n,m-1,p}(t)
    - \sqrt{2p}\,\frac{\usy}{\asz}\,C^{s,\Il'}_{n,m,p-1}(t).
\end{align*}
Expanding the magnetic field $\Bx$ in the \DG{} basis, one
gets
\begin{equation}\label{eq:IBx}
\begin{aligned}
  \big[\calI^{\Bx,s}\big]_{n,m,p}
  &= \frac{\qs}{\ms}\,\frac{\oce}{\ope}\int_{I}\BxN(\xv,t)
  \sum_{l'=1}^{N_l}\zeta^{s,\Il'}_{x,n,m,p}(t)\varphi^{\Il'}(\xv)\varphi^{\Il}(\xv)\,d\xv\\[0.5em]
  &= \frac{\qs}{\ms}\,\frac{\oce}{\ope}\sum_{l'=1}^{N_l}\zeta^{s,\Il'}_{x,n,m,p}(t)
  \sum_{l''=1}^{N_l}\Bx^{\Il''}(t)
  \int_I \varphi^{I,l''}(\xv)\varphi^{\Il'}(\xv)\varphi^{\Il}(\xv)\, d\xv.
\end{aligned}
\end{equation}
An analogous approach yields the approximation of the terms
$\big[\calI^{\By,s}\big]_{n,m,p}$ and
$\big[\calI^{\Bz,s}\big]_{n,m,p}$ involving $\By$ and $\Bz$,
respectively.

To summarize, the local spectral Hermite-\DG{} discretization of the
nonlinear term of the Vlasov equation reads
\begin{align*}
  \int_{\Is}\cN_{n,m,p}(\CS)\varphi^{\Il}(\xv)\,\dxv
  = -\sum_{\beta\in\{x,y,z\}}\bigg(
  \big[\calI^{E_{\beta},s}\big]_{n,m,p} +
  \big[\calI^{B_{\beta},s}\big]_{n,m,p}
  \bigg),
\end{align*}
where the terms involving the electric field are defined
in~\eqref{eq:IEx} and those involving the magnetic field are
defined in~\eqref{eq:IBx}.

\section{\DG{} method for Maxwell's equations}
\label{sec:maxwell}

In this section, we discuss the discontinuous Galerkin approximation
of the Maxwell equations~\eqref{eq:dEdt}-\eqref{eq:dBdt}.
In the continuum framework, the constraint equations~\eqref{eq:divE}
and~\eqref{eq:divB} are satisfied at any time $t>0$ if they are
satisfied at the initial time $t=0$.
Since we do not approximate such equations directly, we expect that a
violation occurs, i.e., the numerical approximations to the
electromagnetic fields $\vecE$, $\vecB$, $\vecJ$ and $\rho$ provided
by the discontinuous Galerkin method satisfy them only in an
approximate manner.
However, this error is expected to be small and to decrease to zero at
least at the same convergence rate of the numerical approximations
used in the other two equations.

To perform the \DG{} approximation of the Maxwell equations,
we first reformulate them in the divergence form.
It is worth noting that our approach differs from that proposed
in~\cite{Cheng-Gamba-Li-Morrison:2014,Cheng-Christlieb-Zhong:2014},
where a local integration by parts is carried out on the \emph{curl}
terms.
Starting from the divergence form, we, then, integrate by part on every element 
and apply the Gauss-Green divergence theorem according to the original 
\DG{} method proposed in~\cite{Cockburn-Shu:1989,Cockburn-Shu:1991,Cockburn-Lin-Shu:1991}.
The local integration by part and the discontinuous nature of the
approximate electromagnetic fields at the cell interface allows us to
introduce the numerical flux functions, which can be central or
upwind.
The upwinding of the numerical flux is performed numerically, even
though the associated Riemann problem can be easily solved exactly, 
cf.~\cite{Cheng-Christlieb-Zhong:2014}.
Such choice does not reduce the computational efficiency because the
mesh is Cartesian and the few terms that are needed for the flux
estimation can be precomputed and stored at negligible cost.

\medskip
Eqs.~\eqref{eq:dEdt}-\eqref{eq:dBdt} in
conservative form read as
\begin{align}
  \frac{\partial\UvC}{\partial t} + \nabla_{\xv}\cdot\FF(\UvC) = \Sv(\UvC),
  \label{eq:Maxwell:conservative:00}
\end{align}
where the components of the electric and magnetic fields form the
vector of conservative unknowns $\UvC$, $\Sv$ represents the source
terms, $\FF(\UvC)$ is the flux function, and the operator
$\nabla_{\xv}\,\cdot\,$ denotes the row-wise divergence.
Using
\eqref{eq:dEdt} and \eqref{eq:dBdt} we define
\begin{align}
  \UvC := 
  \left(
    \begin{array}{c}
      \vecE \\[0.5em]
      \vecB
    \end{array}
  \right)\in\mathbb{R}^6,
  \qquad
  \Sv = \left(
    \begin{array}{c}
      \Sv_{\vecE} \\[0.5em]
      \Sv_{\vecB}
    \end{array}
  \right):= -\frac{\ope}{\oce} \left(
    \begin{array}{c}
      \vecJ \\[0.5em]
      \mathbf{0}
    \end{array}
  \right)\in\mathbb{R}^6,
    \label{eq:Maxwell:conservative:05}
\end{align}
and
\begin{align}
  \qquad
  \FF(\UvC) := 
  \left(
    \begin{array}{c}
      \FFE(\UvC)\\[0.5em]
      \FFB(\UvC)
    \end{array}
  \right)\in\mathbb{R}^{6\times 3},
  \qquad
  \nabla_{\xv}\cdot\FF(\UvC) =
  \left(
    \begin{array}{c}
      \nabla_{\xv}\cdot\FFE(\UvC)\\[0.5em]
      \nabla_{\xv}\cdot\FFB(\UvC)
    \end{array}
  \right) =
  \left(
    \begin{array}{c}
      -\nabla_{\xv}\times\vecB\\[0.5em]
      \nabla_{\xv}\times\vecE
    \end{array}
  \right).
  \label{eq:divg:FF}
\end{align}
We partition the vector flux $\FF(\UvC)$ in a columnwise form so that
\begin{align}
  \nabla_{\xv}\cdot\FF(\UvC) 
  = \dfrac{\partial}{\partial x}\Fvx(\UvC)
  +\dfrac{\partial}{\partial y}\Fvy(\UvC)
  +\dfrac{\partial}{\partial z}\Fvz(\UvC).
  \label{eq:FF:def}
\end{align}
The terms $\Fvx(\UvC)$, $\Fvy(\UvC)$, and $\Fvz(\UvC)$ can be written via
flux matrices $\matF_{x}, \matF_{y}, \matF_{z}\in\mathbb{R}^{6\times
  6}$ as follows,
\begin{align}\label{eq:Fx}
  \Fvx(\UvC) = 
  \left(\begin{array}{r}
      -\ev_x\times\vecB\\[0.5em] 
      \ev_x\times\vecE
    \end{array}\right) =
  \left(\begin{array}{c}
      0\\[0.5em] \Bz\\[0.5em] -\By\\[0.5em] \hline 0\\[0.5em] -\Ez\\[0.5em] \Ey
    \end{array}\right) =
  \left(\begin{array}{rrr|rrr}
      &   &    &  0\,\, &  0\,\, &\phantom{-} 0 \\[0.5em]
      &   &    &  0\,\, &  0\,\, &\phantom{-} 1 \\[0.5em]
      &   &    &  0\,\, & -1\,\, &\phantom{-} 0 \\[0.5em]
      \hline
      0 &\phantom{-} 0\,\, &\phantom{-}  0 &   &    &  \\[0.5em]
      0 &\phantom{-} 0\,\, &            -1 &   &    &  \\[0.5em]
      0 &\phantom{-} 1\,\, &\phantom{-}  0 &   &    &
    \end{array}\right)
  \left(\begin{array}{c}
      \Ex\\[0.5em] \Ey\\[0.5em] \Ez\\[0.5em] \hline \Bx\\[0.5em] \By\\[0.5em] \Bz
    \end{array}\right) =
  \matF_{x}\UvC,
\end{align}
\begin{align}\label{eq:Fy}
  \Fvy(\UvC) = 
  \left(\begin{array}{r}
      -\ev_y\times\vecB\\[0.5em] 
      \ev_y\times\vecE
    \end{array}\right) = 
  \left(\begin{array}{c}
      -\Bz\\[0.5em] 0\\[0.5em] \Bx\\[0.5em] \hline \Ez\\[0.5em] 0\\[0.5em] -\Ex
    \end{array}\right) =
  \left(\begin{array}{rrr|rrr}
      &   &    &  0\,\, &  0\,\, & -1 \\[0.5em]
      &   &    &  0\,\, &  0\,\, &  0 \\[0.5em]
      &   &    &  1\,\, &  0\,\, &  0 \\[0.5em]
      \hline
      0\phantom{-} &  0\,\, &\phantom{-}  1 &   &    &  \\[0.5em]
      0\phantom{-} &  0\,\, &\phantom{-}  0 &   &    &  \\[0.5em]
     -1\phantom{-} &  0\,\, &             0 &   &    &
    \end{array}\right)
  \left(\begin{array}{c}
      \Ex\\[0.5em] \Ey\\[0.5em] \Ez\\[0.5em] \hline \Bx\\[0.5em] \By\\[0.5em] \Bz
    \end{array}\right) =
  \matF_{y}\UvC,
\end{align}
\begin{align}\label{eq:Fz}
  \Fvz(\UvC) = 
  \left(\begin{array}{r}
      -\ev_z\times\vecB\\[0.5em] 
      \ev_z\times\vecE
    \end{array}\right) =
  \left(\begin{array}{c}
      \By\\[0.5em] -\Bx\\[0.5em] 0\\[0.5em] \hline -\Ey\\[0.5em] \Ex\\[0.5em] 0
    \end{array}\right) =
  \left(\begin{array}{rrr|rrr}
      &   &    &  0\,\, &\phantom{-}  1\,\, &\phantom{-}  0 \\[0.5em]
      &   &    & -1\,\, &\phantom{-}  0\,\, &\phantom{-}  0 \\[0.5em]
      &   &    &  0\,\, &\phantom{-}  0\,\, &\phantom{-}  0 \\[0.5em]
      \hline
      0 &            -1\,\, &\phantom{-}  0 &   &    &  \\[0.5em]
      1 &\phantom{-}  0\,\, &\phantom{-}  0 &   &    &  \\[0.5em]
      0 &\phantom{-}  0\,\, &\phantom{-}  0 &   &    &
    \end{array}\right)
  \left(\begin{array}{c}
      \Ex\\[0.5em] \Ey\\[0.5em] \Ez\\[0.5em] \hline \Bx\\[0.5em] \By\\[0.5em] \Bz
    \end{array}\right) =
  \matF_{z}\UvC,
\end{align}
where vectors $\ev_x=(1,0,0)^T$, $\ev_y=(0,1,0)^T$, and $\ev_z=(0,0,1)^T$
form the canonical basis of $\mathbb{R}^3$.
Let
$\nv=(\ns_x,\ns_y,\ns_z)^T$ 
be a generic vector in $\mathbb{R}^3$.
We denote
$\FF(\UvC)\nv=\Fvx(\UvC)\ns_x+\Fvy(\UvC)\ns_y+\Fvz(\UvC)\ns_z=\sum_{\beta\in\{x,y,z\}}\Fv_{\beta}(\UvC)\ns_{\beta}$,
and use the same definition for $\FFE(\UvC)\nv$ and $\FFB(\UvC)\nv$.
A straightforward calculation shows that
\begin{align}
  \FF(\UvC)\nv =
  \left(
    \begin{array}{c}
      \FFE(\UvC)\nv\\[0.5em]
      \FFB(\UvC)\nv
    \end{array}
  \right) =
  \left(
    \begin{array}{r}
      -\nv\times\vecB\\[0.5em]
      \nv\times\vecE
    \end{array}
  \right),
  \label{eq:FFE:FFB:def}
\end{align}
so that
\begin{align}\label{eq:vecidentity}
  \FFE(\UvC)\nv\cdot\vecE
  & = \FFB(\UvC)\nv\cdot\vecB = \nv\cdot\big(\vecE\times\vecB\big)\in\mathbb{R}.
\end{align}
To derive the discontinuous Galerkin approximation, we multiply
Eq.~\eqref{eq:Maxwell:conservative:00} by the local basis function
$\varphiIl$ associated with cell $I$ and integrate over the spatial
domain $\Ox$.
Let $\Uv=\big(\Uv_{\vecE}^T,\Uv_{\vecB}^T\big)^T$ denote the \DG{}
approximation of the vector-valued function $\UvC$ solution of the
Maxwell equations.
Since the support of $\varphiIl$ is given by cell $I$, we can restrict
the integration to such cell.
For $g\in\{\vecE,\vecB\}$,
\begin{align}
  \int_{I}\frac{\partial\Uv_{g}}{\partial t}(\xv,t)\varphiIl(\xv)\dxv + 
  \int_{I}\nabla_{\xv}\cdot\FF_{g}(\Uv)\,\varphiIl(\xv)\dxv = 
  \int_{I}\Sv^N_{g}\,\varphiIl(\xv)\dxv,
\end{align}
where $\Sv^N_{\vecE}:=-\ope \vecJN /\oce$ and $\Sv^N_{\vecB}=\mathbf{0}$.
Then, we integrate by parts the divergence term to get,
\begin{align}
  \int_{I}\frac{\partial\Uv_{g}}{\partial t}(\xv,t)\varphiIl(\xv)\dxv 
  -\int_{I}\FF_{g}(\Uv)\nabla_{\xv}\varphiIl(\xv)\,\dxv 
  +\int_{\partial I}\FF_{g}(\Uv)\nv\,\varphiIl(\xv)\dS 
  = \int_{I}\Sv^N_{g}\,\varphiIl(\xv)\dxv.
  \label{eq:Maxwell:DG}
\end{align}

\medskip
We substitute the expansions \eqref{eq:Ev:expansion},
\eqref{eq:Bv:expansion} of the electric field $\vecE$ and the magnetic
induction $\vecB$ in~\eqref{eq:Maxwell:DG}.
The first term in~\eqref{eq:Maxwell:DG} contains the time-derivatives
of $\vecE^{\Il}(t)$ and $\vecB^{\Il}(t)$.
A direct substitution and the orthogonality of Legendre functions
$\varphiIl$ yields
\begin{multline}
  \int_{I}\frac{\partial\Uv(t)}{\partial t}\varphiIl\dxv =
  \left(
    \begin{array}{c}
      \displaystyle\sum_{l'=1}^{N_l}\frac{d\vecE^{\Ilp}(t)}{\dt}\int_{I}\varphiIlp\varphiIl\dxv\\[0.5em]
      \displaystyle\sum_{l'=1}^{N_l}\frac{d\vecB^{\Ilp}(t)}{\dt}\int_{I}\varphiIlp\varphiIl\dxv
    \end{array}
  \right) = \mu_{I,l}
  \frac{d\Uv^{\Il}(t)}{\dt}
  \quad\textrm{with}\quad
  \Uv^{\Il}(t) := \left(
    \begin{array}{c}
      \vecE^{\Il}(t)\\[0.5em]
      \vecB^{\Il}(t)
    \end{array}
  \right),
  \label{eq:term:1}
\end{multline}
where the multiplicative factor $\mu_{I,l}:=\int_I\left(\varphiIl\right)^2\dxv$
is as in \eqref{eq:dCdt}.
The second term in~\eqref{eq:Maxwell:DG} is the volume integral of the
generalized vector flux function $\FF(\Uv)$, and entails computing the
term
\begin{align}
  \calI_F:=\int_{I}\left(\Fvx(\Uv)\frac{\partial\varphiIl(\xv)}{\partial\xs}
    +\Fvy(\Uv)\frac{\partial\varphiIl(\xv)}{\partial\ys}
    +\Fvz(\Uv)\frac{\partial\varphiIl(\xv)}{\partial\zs}\right)\dxv.
  \label{eq:term:2:00}
\end{align}
To reformulate each addendum in~\eqref{eq:term:2:00} as the product of
an integral involving only the \DG{} basis functions and a term with
the \DG{} coefficients $\vecE^{\Il}(t)$ and $\vecB^{\Il}(t)$, we split
the volume integral into the three distinct contributions, for each
$\beta\in\{x,y,z\}$.
It holds
\begin{align}
  \calI_F=\sum_{\beta\in\{x,y,z\}}\sum_{l'=1}^{N_l}
  \left[\int_{I}\frac{\partial\varphiIl(\xv)}{\partial \beta}\varphiIlp(\xv)\dxv\right]
  \!\!
  \left(\begin{array}{r}
      -\ev_{\beta}\times\vecB^{\Ilp} \\[0.5em] 
      \ev_{\beta}\times\vecE^{\Ilp}
    \end{array}\right)
  \!=\!
  \sum_{\beta\in\{x,y,z\}}
  \sum\limits_{l'=1}^{N_l}(\Phiv_{\beta})_{l,l'}
  \left(\begin{array}{r}
      \displaystyle-\ev_{\beta}\times\vecB^{\Ilp}\\[0.5em]
      \displaystyle\ev_{\beta}\times\vecE^{\Ilp}
    \end{array}\right),
  \label{eq:term:2:10:x}
\end{align}
where the matrix $\Phiv_{\beta}$ is defined in \eqref{eq:term:2:05:x}.
The third term in~\eqref{eq:Maxwell:DG} is the integral at the cell
boundary $\partial I$.
At the six faces of $\partial I$, we approximate the flux with the
numerical flux as follows
\begin{align}
  \Fvx(\Uv)\big\rvert_{\Fipm}\approx\NFLX{\matF_x\Uv}(\xv_{i\pm\frac{1}{2},j,k},t),\\
  \Fvy(\Uv)\big\rvert_{\Fjpm}\approx\NFLX{\matF_y\Uv}(\xv_{i,j\pm\frac{1}{2},k},t),\\
  \Fvx(\Uv)\big\rvert_{\Fkpm}\approx\NFLX{\matF_z\Uv}(\xv_{i,j,k\pm\frac{1}{2}},t).
  \label{eq:upwind:FF}
\end{align}
The size of the matrices $\matF_{x}$, $\matF_{y}$, and $\matF_{z}$ is
$6\times6$, and again they are the same for all the mesh interfaces,
\emph{cf.} Figure \ref{fig:numflux:notation}, so they can be computed
and stored once for a single cell at a negligible cost.
We consider two different kind of numerical fluxes, the \emph{central
  numerical flux} and the \emph{upwind numerical flux}.
We provide a detailed description of the numerical treatment of the
boundary integral in the $x$ direction, and, in particular, at the
face $\Fip$.
The extension to the other directions and faces is deemed
straightforward.
Let $\nv=(n_x,n_y,n_z)^T$ be the unit vector that is orthogonal to the
boundary of $I$.
Noting that, by construction, $n_x=\pm 1$, $n_y=n_z=0$ on the two
faces $\F_{i\pm\frac{1}{2},j,k}$, we obtain,
\begin{align}\label{eq:Flux:Maxwell}
  \int_{\partial_x I}n_x\Fvx(\Uv)\,\varphiIl(\xv)\dS = 
  \int_{\F_{i+\frac{1}{2},j,k}}\Fvx(\Uv)\,\varphiIl(\xs_{i+\frac12},\ys,\zs)\dy\dz -
  \int_{\F_{i-\frac{1}{2},j,k}}\Fvx(\Uv)\,\varphiIl(\xs_{i-\frac12},\ys,\zs)\dy\dz.
\end{align}
\medskip
\noindent
$\bullet$ The \textit{central numerical flux} is given by the formula:
\begin{align}\label{eq:fluxC}
  \Fvx(\Uv)\big\rvert_{\Fip}
  \approx\NFLX{\matF_x\Uv}(\xv_{i+\frac{1}{2},j,k},t) 
  =\matF_x\,\frac{1}{2}\big(\Uv(\xv^{-}_{i+\frac{1}{2},j,k},t)+\Uv(\xv^{+}_{i+\frac{1}{2},j,k},t)\big).
\end{align}
The implementation is straightforward.
  
\medskip
\noindent
$\bullet$ The \textit{upwind numerical flux} is based on the
characteristic decomposition of the flux matrices:
$\matF_{\beta}=\matF_{\beta}^{+}+\matF_{\beta}^{-}$, for each
$\beta\in\{x,y,z\}$.
We perform such decomposition numerically using the same procedure
described in Remark \ref{remark:upwind:flux:implementation} for the matrices
$\mathbbm{A}_{\beta}$.
On face $\F_{i+\frac{1}{2},j,k}$, we approximate the flux integral
using the upwind flux decomposition of the matrix $\matF_x$ as follows
\begin{align}
  \Fvx(\Uv)\big\rvert_{\Fip}\approx\NFLX{\matF_x\Uv}(\xv_{i+\frac{1}{2},j,k},t) 
  = \underbrace{\matF_x^{+}\Uv(x^{-}_{i+\frac12},y,z,t)}_{\textrm{from cell $I=I_{i,j,k}$}}
  + \underbrace{\matF_x^{-}\Uv(x^{+}_{i+\frac12},y,z,t)}_{\textrm{from cell $I^{+}=I_{i+1,j,k}$}}.
  \label{eq:upwind:FF:short}
\end{align}
Thus, the third integral in~\eqref{eq:Maxwell:DG} can be approximated by
\begin{align}
  &\int_{\F_{i+\frac{1}{2},j,k}}\NFLX{\matF_{x}\Uv}(x_{i+\frac12},y,z,t)\,
  \varphiIl(x_{i+\frac12},y,z)\dy\dz\nonumber\\[0.5em]
  &\qquad\qquad
  =
  \matF_x^{+}\sum_{l'=1}^{N_l}\Uv^{\Ilp}(t)
  \underbrace{\int_{\F_{i+\frac{1}{2},j,k}}\varphiIlp(x_{i+\frac12},y,z)\varphiIl(x_{i+\frac12},y,z)\dy\dz}_{\textrm{only cell $I$}}
  \nonumber\\[1.0em]
  &\qquad\qquad\quad
  +
  \matF_x^{-}\sum_{l'=1}^{N_l}\Uv^{\Icp,l'}(t)\underbrace{\int_{\F_{i+\frac{1}{2},j,k}}\varphi^{\Icp,l'}(x_{i+\frac12},y,z)\varphiIl(x_{i+\frac12},y,z)\dy\dz}_{\textrm{two distinct cells $I$ and $\Icp$}}.
  \nonumber\\[1.0em]
\end{align}

\medskip
The semi-discrete \DG{} scheme for the Maxwell equations reads
\begin{align}
  \mu_{\Il}\frac{d\Uv^{\Il}(t)}{\dt}
  - \sum_{\beta\in\{x,y,z\}}
  \sum\limits_{l'=1}^{N_l}(\Phiv_{\beta})_{l,l'}
      \left(\begin{array}{r}
        \displaystyle-\ev_{\beta}\times\vecB^{\Ilp}(t)\\[0.5em]
        \displaystyle\ev_{\beta}\times\vecE^{\Ilp}(t)
      \end{array}\right)
  +\int_{\partial I}\NFLX{\FF(\Uv)\nv}\,\varphiIl\dS 
  = 
  \int_{I}\Sv^N(\Uv)\,\varphiIl\dxv.
  \label{eq:time:Maxwell}
\end{align}
The contribution at the mesh faces is given in \eqref{eq:fluxC} or
\eqref{eq:upwind:FF:short} depending on the choice of the numerical
flux, while the last term in \eqref{eq:time:Maxwell} can be easily
computed by applying a quadrature rule.
    
Alternatively, using the \DG{} approximation of the electric current
$\vecJ$ given by
\begin{equation}\label{eq:vecJN:def}
    \vecJ^N(\xv,t) =
    \sum_{s} \qs \int_{\Ov} \vv \fsN(\xv,\vv,t)\dvv = 
    \sum_s \qs
    \,\alpha_x^s \alpha_y^s \alpha_z^s
    \sum_{\Il}\varphiIl(\xv)
    \left ( 
    C^{s,\Il}_{0,0,0} 
    \begin{pmatrix}
        u_x^s \\[0.5em]
        u_y^s \\[0.5em]
        u_z^s
    \end{pmatrix}
    + \frac{1}{\sqrt{2}} 
    \begin{pmatrix} 
        \alpha_x^s\, C^{s,\Il}_{1,0,0} \\[0.5em]
        \alpha_y^s\, C^{s,\Il}_{0,1,0} \\[0.5em]
        \alpha_z^s\, C^{s,\Il}_{0,0,1}
    \end{pmatrix}
    \right ), 
\end{equation}
the source term in \eqref{eq:time:Maxwell} can be easily reduced to
the volume integral of the polynomial basis functions, and still be
computed by applying formula~\eqref{eq:DG:Ls:Ls} independently in each
spatial direction.
In an analogous way, the \DG{} approximation of the charge density
\eqref{eq:rho} reads
\begin{equation}
    \label{eq:rho_her}
    \rho^N(\xv,t)
    = \sum_s \qs\int_{\Ov}\fsN(\xv,\vv,t)\dvv
    = \sum_s \qs \, \alpha_x^s \alpha_y^s \alpha_z^s\, \sum_{\Il} C^{s,\Il}_{0,0,0}\varphiIl(\xv).
\end{equation}
We use the \DG{} approximation $\rho^N$ in the statement of
Theorem~\ref{theorem:semi-discrete:momentum:conservation} to
characterize the behavior of the (discrete) total momentum.

\section{Conservation properties of the semi-discrete Hermite-\DG{}
  method for periodic boundary conditions}
\label{sec:conservation}

We list here three theorems that fully characterize
the behavior of number of particles, total momentum and total energy
in the semi-discrete formulation, where, for simplicity, we focus on periodic boundary conditions.
Their proofs are reported in the final Appendix
\ref{sec:variational:formulation}.

\subsection{Conservation of the number of particles}
\label{subsec:conservation:number-of-particles}

The total number of particles $N^{tot}(t)$ at any time $t$ in the
semi-discrete formulation of the Vlasov-Maxwell
equations is given by summing
the number of particles $N^{s}(t)$ of species $s$ over all the
species.
Since the Vlasov-Maxwell system does not have any physical process
that may transform particles of a species into particles of another
species, the number of particles of each species is also conserved
separately.
We show the conservation of the total number of particles
in the following theorem.

\medskip
\begin{theorem}[Conservation of the number of particles]
  \label{theo:number-of-particles}
  Let $\fsN(\xv,\vv,t)$ be the numerical solution of the semi-discrete
  Vlasov-Maxwell problem.
  Then, the total number of particles in the plasma, $\cN^{tot}(t)$,
  and the number of particles of each plasma species $s$,
  $\cN^{s}(t)$, which are respectively defined as
  \begin{align}
    \cN^{tot}(t) = \sum_{s}\cN^{s}(t),
    \quad\textrm{and}\quad
    \cN^{s}(t) = \sum_{I}\int_{I}\int_{\Ov}\fsN(\xv,\vv,t)\dxv\dvv
    \quad\forall\ss,
    \label{eq:number-of-particles}
  \end{align}
  are constant in time, i.e., $d\cN^{tot}\slash{\dt}=0$ and
  $d\cN^{s}\slash{\dt}=0$\ for each species $s$.
\end{theorem}

\subsection{Conservation properties of the total momentum}
\label{subsec:conservation:total-momentum}


We define the discrete total momentum as the sum of two terms which
represent the momentum from the plasma kinetics and the momentum from
the electromagnetic fields, i.e.,
$\Pv^{tot,N}=\Pv^{N}_{\Vlasov}(t)+\Pv^{N}_{\Maxwell}(t)$, where
\begin{align}
  \Pv^{N}_{\Vlasov}(t)  &:= \sum_{s}\ms\sum_I\int_{I}\left(\int_{\Ov}\fsN(\xv,\vv,t)\vv\dvv\right)\,\dxv,\label{eq:PvN:Vlasov}\\[0.5em]
  \Pv^{N}_{\Maxwell}(t) &:= \left(\frac{\oce}{\ope}\right)^2\sum_I\int_{I}\vecEN(\xv,t)\times\vecBN(\xv,t)\dxv.\label{eq:PvN:Maxwell}
\end{align}
The total momentum in the continuum framework is given by the same
definitions above, but using the fields $\fs$, $\Ev$ and
$\Bv$ instead of the discrete fields $\fsN$, $\vecEN$ and $\vecBN$,
and is conserved in the Vlasov-Maxwell system for periodic boundary conditions.
We characterize the behavior in time of the discrete total momentum by
the following theorem, which shows that a violation of the momentum
conservation occurs.
However, if the solution fields are sufficiently regular (at least
continuous), such a violation is expected to be small according to the
adopted space and time resolution and the order of the method.

\medskip
\begin{theorem}[Conservation of momentum]
  \label{theorem:semi-discrete:momentum:conservation}
  Let $\fsN$, $\vecEN$ and $\vecBN$ be the numerical solution of the
  semi-discrete variational formulation of the Vlasov-Maxwell
  problem. 
  Let 
  $\Pv^{tot,N}(t) := \Pv^{N}_{\Vlasov}(t)+\Pv^{N}_{\Maxwell}(t)$ be
  the discrete total momentum, where $\Pv^{N}_{\Vlasov}$ and
  $\Pv^{N}_{\Maxwell}$ are defined in
  \eqref{eq:PvN:Vlasov} and \eqref{eq:PvN:Maxwell}, respectively.  Then,
  $\Pv^{tot,N}(t)$ satisfies the ordinary differential equation
  \begin{align}
    \dfrac{d}{\dt}\big(\Pv^{N}_{\Vlasov}(t)+\Pv^{N}_{\Maxwell}(t)\big)
    = \Rv(t),
  \end{align}
  where the residual term on the right-hand side is given by
  \begin{align}
    \Rv(t) = 
    -\dfrac{\oce}{\ope}\sum_I\int_{I}\mathcal{G}_I(\xv,t)\dxv 
    -\sum_I\int_{\partial\Is}\big(\NFLX{\nv\cdot\gv^{N}_{\vv}(\xv,t)}+\mathcal{B}_I(\xv,t)\big)\dS.
  \end{align}
  Here, $\NFLX{\nv\cdot\gv^{N}_{\vv}}$ is the numerical flux associated with
  \begin{align}
    \nv\cdot\gv^{N}_{\vv} (\xv,t) &= 
    \sum_s\ms\int_{\Ov}(\nv\cdot\vv)\vv\,\fsN(\xv,\vv,t)\dvv,
    \\[0.5em]
    \intertext{and}
    \mathcal{G}_I(\xv,t) &= \frac{\oce}{\ope}\bigg(\vecBN\times\big(\nabla_{\xv}\times\vecBN\big)+
    \vecEN\times\big(\nabla_{\xv}\times\vecEN\big)\bigg) - \rho^{N}\vecEN,\\[0.5em]
    \mathcal{B}_I(\xv,t) &= 
    \left(\dfrac{\oce}{\ope}\right)^2\bigg(\NFLX{\FFE(\Uv)\nv}\times\vecBN -
    \NFLX{\FFB(\Uv)\nv}\times\vecEN\bigg),
  \end{align}
    are a bulk and an elemental boundary term that depends on the
  approximate electromagnetic fields $\vecEN$, $\vecBN$, and $\rhoN$,
  and $\NFLX{\FFE(\Uv)\nv}$ and $\NFLX{\FFB(\Uv)\nv}$ are the
  numerical fluxes of $\FFE(\Uv)\nv$ and $\FFB(\Uv)\nv$ defined
  in~\eqref{eq:FFE:FFB:def}.
\end{theorem}

\medskip
We refer to Appendix \ref{ssec:Mom} for the proof of the theorem.
Here,
we further elaborate the result to investigate the
meaning of the residual term.
We consider the following formula from differential calculus: for a
sufficiently smooth vector field $\vz$, it holds
$\vz\times(\nabla\times\vz)=\dfrac12\nabla\ABS{\vz}^2-\nabla\cdot(\vz\otimes\vz)+\vz\nabla\cdot\vz$.
Hence,
\begin{align}
  \dfrac{\ope}{\oce}\mathcal{G}_I 
  &=
  \dfrac12\nabla\big(\ABS{\vecEN}^2+\ABS{\vecBN}^2\big)
  -\nabla\cdot\big(\vecEN\otimes\vecEN+\vecBN\otimes\vecBN\big)
  +\vecEN\nabla\cdot\vecEN
  +\vecBN\nabla\cdot\vecBN
  -\dfrac{\ope}{\oce}\rhoN\vecEN
  \nonumber\\[0.5em]
  &\qquad=
  \mathcal{H}_I  
  +\vecEN\left(\nabla\cdot\vecEN-\dfrac{\ope}{\oce}\rhoN\right)
  +\vecBN\nabla\cdot\vecBN,
  \label{eq:semi-discrete:GI:00}
\end{align}
with
\begin{align*}
  \mathcal{H}_I 
  = \dfrac12\nabla\big(\ABS{\vecEN}^2+\ABS{\vecBN}^2\big)
  -\nabla\cdot\big(\vecEN\otimes\vecEN+\vecBN\otimes\vecBN\big).
\end{align*}
We sum term $\mathcal{H}_I$ over all the cells $I$, use the
Gauss-Green formula, and rearrange the summation over the faces using
the periodicity in space, to get
\begin{align}
  \sum_I\int_{\Is}\mathcal{H}_I\dxv 
  &= \sum_I\int_{\partial\Is}
  \Big(\dfrac12
  \nv\big(\ABS{\vecEN}^2+\ABS{\vecBN}^2\big) -
  \nv\cdot\big(\vecEN\otimes\vecEN+\vecBN\otimes\vecBN\big)
  \Big)\dxv 
  \nonumber\\[0.5em]
  &= 
  \dfrac12\sum_{\F}\int_{\F}\Big(\JUMP{\ABS{\vecEN}^2}_{\F}+\JUMP{\ABS{\vecBN}^2}_{\F}\Big)\dS -
  \sum_{\F}\int_{\F}\Big(\JUMP{\vecEN\otimes\vecEN}_{\F}+\JUMP{\vecBN\otimes\vecBN}_{\F}\Big)\cdot\nv_{\F}\dS,
\end{align}
where in the last equation we set $\nv_{\F}=\nv^{+}_{\F}$, and the
jumps of $\vecEN\otimes\vecEN$ and $\vecBN\otimes\vecBN$ are defined
as
\begin{align*}
  \JUMP{\vecEN\otimes\vecEN}_{\F} = \nv^{+}\cdot\big(\vecEN\otimes\vecEN\big)^{+}\cdot\nv^{+} + \nv^{-}\cdot\big(\vecEN\otimes\vecEN\big)^{-}\cdot\nv^{-},\\[0.5em]
  \JUMP{\vecBN\otimes\vecBN}_{\F} = \nv^{+}\cdot\big(\vecBN\otimes\vecBN\big)^{+}\cdot\nv^{+} + \nv^{-}\cdot\big(\vecBN\otimes\vecBN\big)^{-}\cdot\nv^{-},
\end{align*}
with the interpretation that
$\nv^{+}\cdot\big(\vecEN\otimes\vecEN\big)^{+}\cdot\nv^{+}=(\nv^{+})^T\big(\vecEN(\vecEN)^T\big)^{+}\nv^{+}$,
and similarly for the other terms.

\medskip
We can analogously transform the integral term involving
$\mathcal{B}_{I}$ by rearranging the summation of the elemental
boundary integrals as summation of elemental interface integrals
\begin{align*}
  \sum_{I}\int_{\partial\Is}\mathcal{B}_I\dS
  &= \left(\dfrac{\oce}{\ope}\right)^2\sum_{I}\int_{\partial\Is}\left(\NFLX{\FFE(\Uv)\nv}\times\vecBN + \NFLX{\FFB(\Uv)\nv}\times\vecEN\right)\dS\\[0.5em]
  &= \sum_{\F}\int_{\F}\left(\JUMP{\NFLX{\FFE(\Uv)\nv}_{\F}\times\vecBN} + \JUMP{\NFLX{\FFB(\Uv)\nv}_{\F}\times\vecEN}\right)\dS,
\end{align*}
with the jumps defined as
\begin{align*}
  \JUMP{\NFLX{\FFE(\Uv)\nv}\times\vecBN}_{\F} &:= \NFLX{\FFE(\Uv^{-})\nv^{+}}\times(\vecBN)^- + \NFLX{\FFE(\Uv^{+})\nv^{-}}\times(\vecBN)^+,\\[0.5em]
  \JUMP{\NFLX{\FFB(\Uv)\nv}\times\vecEN}_{\F} &:= \NFLX{\FFB(\Uv^{-})\nv^{+}}\times(\vecEN)^- - \NFLX{\FFB(\Uv^{+})\nv^{-}}\times(\vecEN)^+.
\end{align*}
Since the numerical fluxes $\NFLX{\FFE(\Uv)\nv}$ and
$\NFLX{\FFB(\Uv)\nv}$ depend on $\vecBN$ and $\vecEN$, respectively,
the integral term involving $\mathcal{B}_{I}$ also depends on the
jumps of the electromagnetic fields, and it is quadratic 
on $\vecBN$ and $\vecEN$.

\medskip
These developments show that if the exact solutions $\fs$, $\Ev$ and
$\Bv$ are sufficiently regular (at least continuous) the residual term
must tend to zero at least at the convergence rate determined by
the order of the approximation.
Indeed, $\mathcal{H}_I$ depends on the square of the jumps of the
electromagnetics fields, which must tend to zero if $\vecEN$ and
$\vecBN$ are approximations of continuous functions.
The
last two terms in~\eqref{eq:semi-discrete:GI:00} must tend to zero as
$\mathcal{O}(\hs^{N_{DG}})$, where
$\hs=\max(\Delta\xs,\Delta\ys,\Delta\zs)$.
Similar considerations hold for the term $\mathcal{B}_{I}$.

\subsection{Conservation properties of the total energy}
\label{subsec:conservation:total-energy}

The total energy of the plasma is defined as $\calE(t)=
\calE_{\KIN}(t) + \calE_{\ELE}(t)$, where
\begin{align}
  \calE_{\KIN}(t) & := 
  \frac{1}{2} \sum_{s}\ms\sum_{\Is}\int_{\Is}
  \bigg(\int_{\Ov}\ABS{\vv}^2\fs(\xv,\vv,t)\,d\vv\bigg)\dxv,\\[0.5em]
  \calE_{\ELE}(t) & :=
  \frac{1}{2} \left(\frac{\oce}{\ope}\right)^2\int_{\Ox}\Big(\ABS{\vecE(\xv,t)}^2+\ABS{\vecB(\xv,t)}^2\Big)\dxv.
  \label{eq:energy:def}
\end{align}

Let $\ABS{\matF_{\beta}}=\matF^+_{\beta}-\matF^-_{\beta}$,
$\beta\in\{x,y,z\}$, and
$\ABS{\FF}=\ABS{\matF_x}+\ABS{\matF_y}+\ABS{\matF_z}$.  
For a generic mesh face $\F$ we introduce the quantity
\begin{align}\label{eq:JF}
  \JMPF(t) :=
  \begin{cases}
    0 
    & \qquad\textrm{for central numerical flux},\\[0.5em]
    \displaystyle\frac{1}{2}\int_{\F}\JUMP{\Uv(t)}_{\F}\cdot\ABS{\FF}\cdot\JUMP{\Uv(t)}_{\F}\,\dS
    & \qquad\textrm{for upwind numerical flux},
  \end{cases}
\end{align}
where
\begin{align*}
  \JUMP{\Uv(t)}_{\F}
  = \nv^+\cdot\Uv(\xv^+_{\F},t)+\nv^-\cdot\Uv(\xv^-_{\F},t)
  =: \nv^+\cdot\Uv^+_{\F}(t)+\nv^-\cdot\Uv^-_{\F}(t),
\end{align*}
is the jump of $\Uv$ across the mesh face $\F$, and
$\Uv^{\pm}_{\F}(t)=\Uv(\xv^{\pm},t)$ are shortcuts to denote the trace
of $\Uv$ on the opposite sides of $\F$.

The following result shows that the total discrete energy is exactly conserved
when central numerical fluxes \eqref{eq:fluxC}
are employed in the approximation of Maxwell's equations,
while a numerical dissipation proportional to the jump of the approximate electromagnetic
fields occurs when upwind numerical fluxes
\eqref{eq:upwind:FF:short} are used.
We refer to Appendix \ref{ssec:En} for the complete proof of the statement.
\medskip
\begin{theorem}[Conservation of the total energy]
  \label{theo:energy}
  Let ($\fsN(\xv,\vv,t)$, $\vecEN(\xv,t)$, $\vecBN(\xv,t)$) be the
  numerical solution of the semi-discrete Vlasov-Maxwell problem
  with periodic boundary conditions in space.
  Let  the total discrete energy of the plasma
  be defined as
  \begin{align}
    \calEN(t) & =
    \dfrac12 \sum_{s}\ms\sum_{\Is}\int_{\Is}
    \bigg(\int_{\Ov}\ABS{\vv}^2\fsN(\xv,\vv,t)\,d\vv\bigg)\dxv +
    \dfrac12\left(\frac{\oce}{\ope}\right)^2 \int_{\Ox}\Big(\ABS{\vecEN(\xv,t)}^2+\ABS{\vecBN(\xv,t)}^2\Big)\dxv
    \nonumber\\[0.5em]
    &=: \calEN_{\KIN}(t) + \calEN_{\ELE}(t).
    \label{eq:energyN:def}
  \end{align}
 Then, the variation in time of the total discrete energy satisfies
  \begin{align}
    \frac{d\calEN(t)}{\dt} = -\left(\frac{\oce}{\ope}\right)^2\sum_{\F}\JMPF(t)\leq 0.
    \label{eq:energy}
  \end{align}
\end{theorem} 

\section{Time discretization}
\label{sec:time}

The semi-discrete Hermite-\DG{} scheme of the Vlasov equation on a local
element $I$ for a fixed Hermite triplet $(n,m,p)$, and $l$-th Legendre
polynomial reads
\begin{align}
  \mu_{I,l}\frac{d\Cs_{n,m,p}^{s,\Il}(t)}{\dt}
  &
  +\sum_{\beta\in\{x,y,z\}}\bigg(-\big[\calI^{\beta,s}\big]_{n,m,p} + (\delta_\beta^{s,+} - \delta_\beta^{s,-})\bigg)
  -\sum_{\beta\in\{x,y,z\}}\bigg( \big[\calI^{E_\beta}\big]_{n,m,p} + \big[\calI^{B_\beta}\big]_{n,m,p}\bigg) = 0,
  \label{eq:time:Vlasov}
\end{align}
where the $\{\calI^{\beta,s}\}_{\beta}$ are defined in \eqref{eq:Ix},
\eqref{eq:Iy}, \eqref{eq:Iz}, $\{\calI^{E_\beta}\}_{\beta}$ and
$\{\calI^{B_\beta}\}_{\beta}$ are defined in \eqref{eq:IEx},
\eqref{eq:IBx}, and $\{\delta_\beta^{s,\pm}\}_{\beta}$ have been
introduced in \eqref{eq:flux-x:10}.
The semi-discrete \DG{} scheme for the Maxwell wave propagation equation
is as in \eqref{eq:time:Maxwell}, namely
\begin{align}
  \mu_{I,l}\frac{d\Uv^{\Il}(t)}{\dt}
  -\calI_F
  +\int_{\partial I}\NFLX{\FF(\Uv)\nv}\,\varphiIl\dS 
  = \int_{I}\Sv^N(\Uv)\,\varphiIl d\xv.
  \label{eq:time:Maxwell2}
\end{align}
%

Each term is defined in Section \ref{sec:maxwell}.

\bigskip 
The time discretization we propose is based on explicit adaptive Runge-Kutta (RK)
methods in the framework of the method of lines.
In particular, for the numerical experiments reported in
Section~\ref{sec:numerical:results}, we considered three Runge-Kutta
algorithms:
(i) the non-adaptive third-order accurate RK scheme of
Bogacki-Shampine, cf. \cite{Bogacki-Shampine:1989};
(ii) the adaptive third-order accurate RK scheme of Bogacki-Shampine
with second order embedded method, cf. \cite{Bogacki-Shampine:1989};
(iii) the non-adaptive fifth-order accurate Fehlberg RK scheme,
cf. \cite{Fehlberg:1970}.

Finally, we note that all explicit (and implicit) RK methods conserve
linear invariants, see, for example, \cite[Theorem 1.5]{HLW06}; hence,
we can extend the result of Theorem~\ref{theo:number-of-particles} to
the fully discrete case and conclude that the total number of particles is conserved.
Instead, \cite[Theorem 2.2]{HLW06} shows that quadratic invariants can
be conserved by explicit RK methods only under an algebraic condition
on the coefficients of the scheme.
Consequently, we do not expect that a result corresponding to
Theorem~\ref{theo:energy} may exist in the fully discrete setting for
our RK methods, implying that the total energy in the system is not
conserved.  The total momentum is also not expected to be conserved in
the fully discrete case, since it is already not an invariant
according to Theorem~\ref{theorem:semi-discrete:momentum:conservation}
in the semi-discrete case.
However, from theoretical considerations (see,
again,~\cite{Koshkarov-Manzini-Delzanno-Pagliantini-Roytershtein:2019:ConsLaws}),
we can expect that the violation in the conservation of total
momentum and energy are small and decreasing at least at the rate of
convergence of the numerical approximation.

\subsection{Artificial collisional operator}
Collisionless plasmas can develop finer and finer scales in velocity space, a phenomenon known as filamentation. This can lead to recurrence once velocity-space structures reach scales that are no longer resolved by the particular numerical algorithm considered \cite{joyce71,canosa74,cheng76}. It is therefore customary to introduce an artificial collisional operator to damp high order modes and prevent filamentation.
We adopt here the same collisional operator introduced in Ref. \cite{Delzanno2015},
\begin{equation}\label{collop}
    \begin{aligned}
        \mathcal{C}_{n,m,p}[C^s]= -\nu & \left[\frac{n(n-1)(n-2)}{N_{v_x}(N_{v_x}-1)(N_{v_x}-2)}+\frac{m(m-1)(m-2)}{N_{v_y}(N_{v_y}-1)(N_{v_y}-2)}\right.\\[0.5mm]
&\left.+\frac{p(p-1)(p-2)}{N_{v_z}(N_{v_z}-1)(N_{v_z}-2)}\right]C_{n,m,p}^s(\xv,t),
\end{aligned}
\end{equation}
where $\nu$ is the collision rate. 
The operator in Eq. \eqref{collop} is added to the right hand side of the Vlasov equation (\ref{eq:Vlasov-system:long}) (Eq. \eqref{eq:time:Vlasov} in semi-discrete form) for each Hermite mode. This type of collisional operator has the advantage that it does not act directly on the first three Hermite modes and therefore does not affect the conservation laws for total mass, momentum and energy presented in Section \ref{sec:conservation} (see also Ref.~\cite{Delzanno2015}). Obviously, the collisional operator must always be used in a convergence sense, i.e. making sure that it does not affect the collisionless physics of interest significantly.

\CP{
\subsection{Numerical Stability}
Previous investigations of the methods based on the Hermite expansion found no stability theorem for the asymmetrically-weighted Hermite representation, i.e. the $L_2$ of the distribution function is not bounded and can grow in time~\citep[e.g.][]{Schumer-Holloway:1998}. In contrast, the symmetrically-weighted expansion does conserve the $L_2$ norm and is therefore numerically stable. In practice, the absence of the stability theorem does not mean that the simulations necessarily blow up in time. In our experiments, the instability clearly manifests itself only in situations where a physical quantity significant for the dynamics is not properly resolved. For these cases, when the resolution is increased, the instability typically goes away. In particular, in the results presented in the paper, we have seen no sign of numerical instabilities. We note that while the collisional operator discussed above does provide additional stabilizing influence, its primary role is to control filamentation and recurrence in a manner which is common to all methods attempting to solve the Vlasov equation using a limited number of degrees of freedom. 
}

\section{Implementation details and parallel scalability}
\label{sec:implementation}
The implementation of the Hermite-\DG{} discretization that we designed
in the previous sections refers specifically to Eqs.
(\ref{eq:time:Vlasov})-(\ref{eq:time:Maxwell2}).
The implementation is carried out in the framework of the
\emph{Spectral~Plasma~Solver} (SPS), a computational software
currently under development at the Los Alamos National Laboratory for
the numerical modeling of multiscale phenomena in collisionless or
weakly collisional plasmas.
This software incorporates a spectral-based solver for the Vlasov
equations and the coupled electromagnetic models, which is an
implementation of the numerical methods described in
References~\cite{Delzanno2015,Camporeale-Delzanno-Bergen-Moulton:2015,Vencels-Delzanno-Manzini-Markidis-BoPeng-Roytershteyn:2015,Vencels-Delzanno-Johnson-BoPeng-Laure-Markidis:2015}.
The implementation of the algorithms discussed in this paper is a
sub-branch of SPS, and, for brevity, it will be referred to as
\emph{SPS-\DG{}}.

\par 
The SPS-\DG{} code is based on the method of lines.
The numerical discretization is expressed by the system of ordinary
differential equations (ODEs)
\begin{align}
  \frac{d\Sv}{\dt} = \Gv\big(\Sv\big),
\end{align}
where $\Sv(t)$ is the state vector collecting the time-dependent
coefficients $C_{n,m,p}^{s,I,l}(t)$ and $\Uv^{\Il}(t)$ resulting from
the Hermite-DG expansion of the species distribution functions $\fs$
and the electromagnetic fields $\Ev$ and $\Bv$; $\Gv$ is the
vector-valued function containing all information about the phase
space discretization.

\par 
In this section, we discuss three aspects of the design and
implementation of SPS-\DG{}: 
(i) the data structure storing the global state vector $\Sv$;
(ii) the action of function $\Gv$;
(iii) the temporal discretization.
Finally, we discuss the parallel scalability of the SPS-\DG{} code and
we show by a numerical experiment that the Hermite-DG approach leads
to efficient implementations on high-performance-computing architectures.

\subsection{Data structure of the global state vector and parallelization}
The data structure used to store the global state vector $\Sv$ was
chosen to achieve efficient parallelization of the code.
The parallel version of SPS-DG is designed to run on distributed
memory architectures and is currently based on the domain
decomposition of the physical space while the velocity space
is not decomposed.
This choice guarantees high locality of the data structures of the
code, and hence, high parallel efficiency as demonstrated by the
scalability study shown at the end of this section.

\par 
The parallel implementation of the domain decomposition in SPS-\DG{}
relies on the Portable, Extensible Toolkit for Scientific Computation
(PETSc)
developed at Argonne National Laboratory
for the scalable (parallel) solution of scientific applications
modeled by partial differential equations.
Parallelism in PETSc is achieved through the Message Passing Interface
(MPI) standard for all message-passing communications.

\medskip
We exploit DMDA, which is the native PETSc parallel data structure for
structured grids, to accommodate the \DG{} cells $I$, see PETSc user
manual~\cite{petsc-user-ref}.
DMDA is a PETSc object that manages an abstract grid object and its
interactions with the algebraic solvers.
The other degrees of freedom located within the \DG{} cells are
internally stored in each DMDA point.

\par 
The total number of degrees of freedom (DOFs) that we need to store to
represent the state vector $\Sv$ at each time instant is
\begin{align*}
   \#\mbox{DOFs} = N_x N_y N_z N_l \big(6 + N_s(\Nn+1)(\Nm+1)(\Np+1)\big),
\end{align*}
where $N_s$ is the number of different plasma species.
For example, in the Orszag-Tang vortex test that we consider below for
the scalability study and in the section of the numerical experiments,
we consider $N_s=2$, $N_x\sim N_y\sim N_z\sim 100$,
$N_l\sim\Nn\sim\Nm\sim\Np\sim 10$, and total number of DOFs may be in
the range from $\sim 0.02$ up to $1$ trillions.
Modern 
supercomputers can perform kinetic plasma simulations with trillions
of DOFs~\cite{Bowers:2008}.


\par 
Since the space decomposition is performed along each spatial
dimension, i.e., along $x$, $y$ and $z$, the maximum number of MPI processes
that SPS-DG can use is $N_x N_y N_z$, which is of the order of
$10^6$---$10^7$ even for the smallest 3D simulations, ensuring that a high degree of parallelism is available. 
At the same time, partitioning only the spatial directions reduces the
amount of required communications between processes, thus
improving the parallelization efficiency, since the DOFs pertained to
a selected DG cell always reside on the same processor.
Moreover, the communication is only needed to compute the numerical
flux from adjacent \DG{} cells, which requires to transfer information
only between processors with high affinity.

\par 
The DMDA data structure is also responsible for managing the boundary
conditions.
The SPS-\DG{} code can incorporate two distinct types of boundary
conditions:
(i) periodic boundary conditions, which are automatically handled by
DMDA; and
(ii) ghost cells, where the user is responsible for
providing the state vector $\Sv$ on the boundary at each time step.

\subsection{Phase space discretization}
The procedure to evaluate the function $\Gv$ at each time step is determined by the
Hermite-\DG{} discretization of the phase space that we detailed in
Sections~\ref{sec:Transform:method} and~\ref{sec:maxwell}.
The SPS-\DG{} code does not perform any numerical
integration to evaluate $\Gv$.
Indeed, all integrals in the system
\eqref{eq:time:Vlasov}-\eqref{eq:time:Maxwell2} are computed
analytically by using the orthogonality relations and exact
formulas~\eqref{eq:DG:Ls:Ls}-\eqref{eq:DG:Ls:dLsdx}.
Additionally, all linear matrices in the definition of the upwind
numerical flux in the discretization of the Vlasov and Maxwell
equations, i.e., $\mathbb{A}_{\beta}$, $\matF_{\beta}$ and their
upwind decompositions $\mathbb{A}_{\beta}^{\pm}$,
$\matF_{\beta}^{\pm}$, are precomputed and stored at the beginning of
every simulation, so they are never computed during the evaluation of
$\Gv$.

\subsection{Implementation of the time-stepping scheme}
The implementation of the SPS-\DG{} code uses the external time-stepping
library TS of the PETCs framework, cf.
\cite{petsc-web-page,petsc-user-ref,petsc-efficient}, which provides
optimized and thoroughly benchmarked implementations of explicit
Runge-Kutta methods.

\subsection{Scalability}
\begin{figure}[!t]
  \centering
  \includegraphics[width=0.75\linewidth]{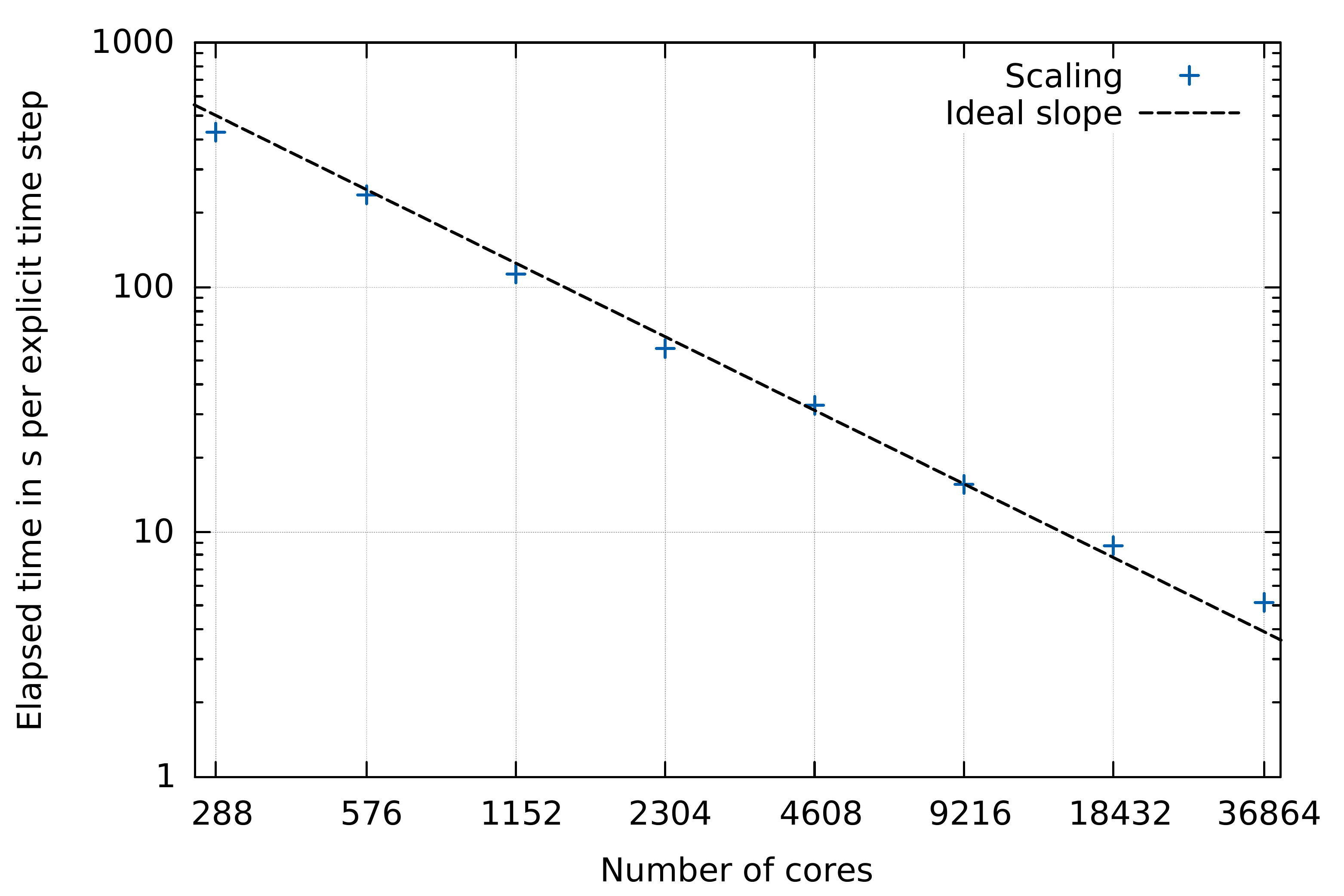}
  \caption{Strong scalability test for Orszag-Tang problem.  The test
    was performed on LANL's cluster Grizzly (CTS-1 cluster with 8-SU,
    $1490$ nodes, $53640$ cores, $191$ TB memory, $1.8$ Pflops peak
    operating speed).}
  \label{fig:sc}
\end{figure}
SPS-\DG{} achieves good parallel scalability on different high-performance-computing platforms, including the clusters available at the Los
Alamos National Laboratory (LANL).
An example of the scalability of SPS-\DG{} is shown in
Figure~\ref{fig:sc}, which reports the elapsed times versus the number
of cores in the numerical simulation of the (2D-3V) magnetized plasma
turbulence decay problem, known as Orszag-Tang vortex test.
More details about this test case and the performance of the SPS-\DG{}
code will be given in Section~\ref{sec:OT}.
We considered 
$N_x=N_y = 768$, $N_z=1$ ($L_x=L_y=200$, $L_z=1$) spatial grid cells
and a maximum local polynomial degree $\NDG=2$ in the \DG{}
approximation, and, for the velocity resolution, $\Nn = \Nm = \Np =
9$ Hermite modes.
The total number of degrees of freedom per DG cell is $12036$
corresponding to $\sim 7$ billions of DOFs in total.
Time stepping was provided by the third order RK integrator in the TS
library with constant time step $\Delta t = 0.05$.

The code was run for $20$ time steps on the LANL cluster Grizzly and
the elapsed time at the end of each run was normalized to that for one time
step.
Grizzly is a Commodity Technology System, version 1 with 8 Scalable
Units (CTS-1 cluster with 8-SU) and 18 additional compute nodes.
In total, Grizzly has $1490$ and $53640$ compute nodes and cores,
respectively, with a total of $191$TB cluster memory.
Grizzly peak operating speed is $1.8$ Pflops.
Figure~\ref{fig:sc} shows nearly ideal scalability for a broad range
of core numbers.
Perfect scalability is usually achieved when the number of degrees of
freedom is larger or comparable to $10^5$ (we recall that the absolute
minimum number of DOFs recommended by PETSc developers is
$10^4$~\cite{petsc-faq-scaling}), even if the number of DG cells per core
is small, i.e., $\sim 1$ (in the case of $36864$ cores, this test has
$1.9 \cdot 10^5$ degrees of freedom per cell and 16 DG cells per core).

\section{Numerical results}
\label{sec:numerical:results}
In this section we perform tests to benchmark and assess the accuracy
of the SPS-\DG{} framework.

\subsection{Accuracy test for spatial discretization}
We start with a manufactured solution test to assess the accuracy of
the spatial discretization in our implementation.
Instead of using a known manufactured solution that prescribes a
source term, we exploit the time reversibility of the Vlasov-Maxwell
system.
Indeed, we can integrate the Vlasov-Maxwell system forward and
backward in time and return to the initial state, possibly modified by
the numerical integration errors.
Equivalently, we integrate the Vlasov-Maxwell system forward in time
from $t=0$ up to $t=T$ for a given final time $T>0$. Then, we reverse
the velocity coordinate and the direction of the magnetic field 
(recall that $\Bv$ is a pseudovector), as follows
\begin{align*}
  \fs(\xv,\vv,t) \longrightarrow \fs(\xv,-\vv,t), \quad 
  \mathbf B \longrightarrow - \mathbf B,
\end{align*}
and continue the integration from $t=T$ to $t=2T$.
Since the Hermite expansion of the Vlasov equation does not break the
time reversibility, we can use this test to assess the numerical
errors that are introduced by the discontinuous Galerkin
discretization in space.
To this end, we need a sufficiently small time step $\Delta\ts$ and a
RK time-stepping method that is sufficiently accurate to ensure
that the time discretization error is much smaller than the spatial
discretization error.

\medskip
In this test, we discretize the velocity space by setting
$\Nn=\Nm=\Np=3$, so to have $4$ polynomials in each velocity
direction, and the physical space domain
$\Ox=[0,L_x]\times[0,L_y]\times[0,L_z]$ by setting $L_x=L_y=L_z=1$ and
$N_x=N_y=N_z=12,\,24\,,48\,,96\,,192\,,384$. The cell size is uniform in
every direction.
Periodic boundary conditions are assumed.
The degree of the local discontinuous polynomials is set to
$\NDG=0,1,2,3$.
Note that the case $\NDG=0$ corresponds to a low-order finite volume
approximation.

\medskip
We initialize the magnetic field $B_x = 1$ and the electron-proton
plasma with Maxwellian distributions
\begin{align*}
  \fs(\xv,\vv,0) = 
  \prod_{\beta\in\{x,y,z\}} \frac{1}{v_{T_\beta^s} \sqrt{2\pi}} \exp \left [ -\frac{v_\beta ^2}{2 v_{T_\beta^s}^2} \right ],
\end{align*}
where $s\in\{e,i\}$ (respectively, electrons or ions), $v_{T_\beta^s}$ is the thermal velocity
along direction $\beta\in\{x,y,z\}$ for species $s$.
The Maxwellian distribution corresponds to only one term in the
asymmetrically weighted Hermite polynomial expansion
(cf.~\eqref{eq:rho_her}), namely
\begin{align*}
  C_{0,0,0}^s (\xv) = \frac{1}{\alpha^s_x \alpha^s_y \alpha^s_z }, 
\end{align*}
where $\alpha_\beta^s = \sqrt{2} v_{T_\beta^s}$ and $u_\beta^s = 0$ for
$\beta\in\{x,y,z\}$.
We use realistic ion-to-electron mass ratio $m_i/m_e = 1836$, thermal
velocities $v_{T_\beta^e} = v_{T_\beta^i} \sqrt{m_i/m_e} = 0.125$ for
$\beta\in\{x,y,z\}$, and electron plasma frequency/gyrofrequency ratio
$\omega_{pe}/\omega_{ce} = 4$.
We also introduce a current perturbation
$J_z(\xv)=\sin(2\pi\xs+1)\sin(2\pi\ys+2)\sin(2\pi\zs+3)$ by exciting
$C_{0,0,1}^e$, cf.~\eqref{eq:vecJN:def}.
The other coefficients in the Hermite expansion are initialized to
zero.

\begin{figure}[H]
    \centering
    \includegraphics[width=0.32\linewidth]{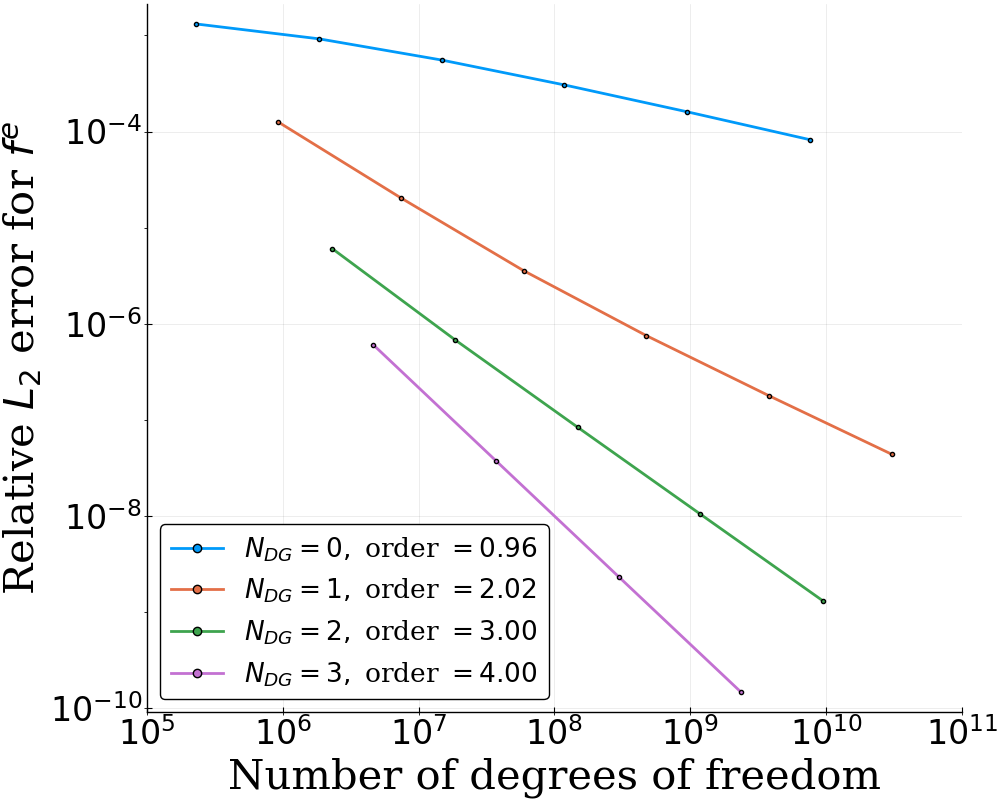}
    \includegraphics[width=0.32\linewidth]{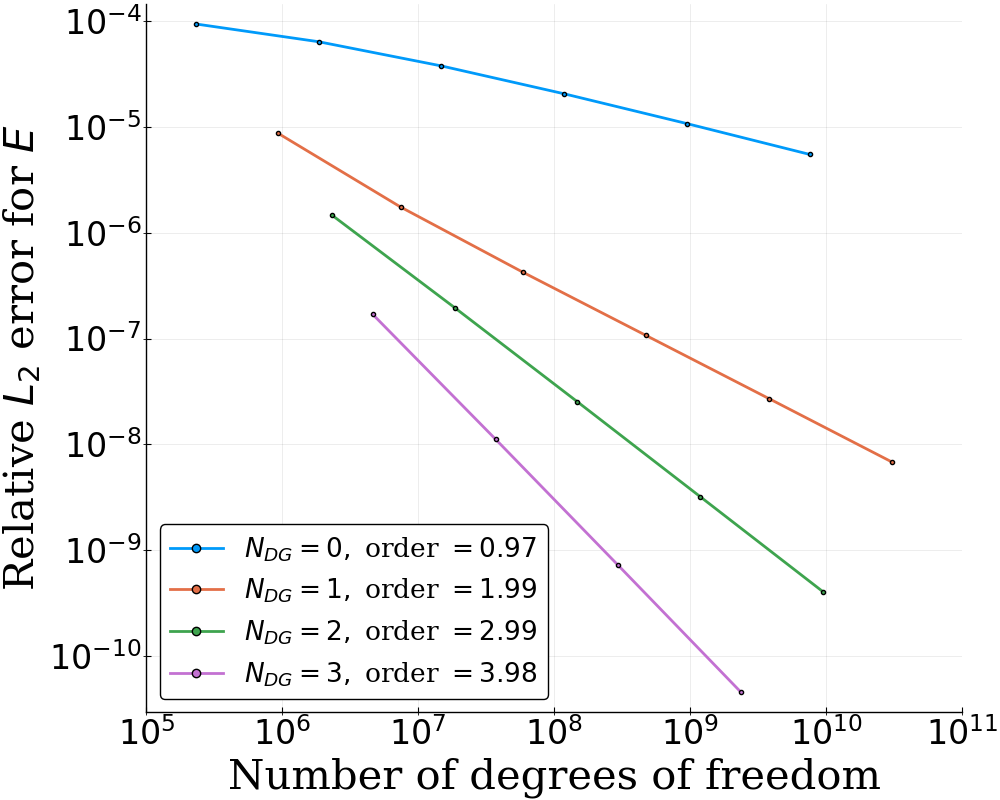}
    \includegraphics[width=0.32\linewidth]{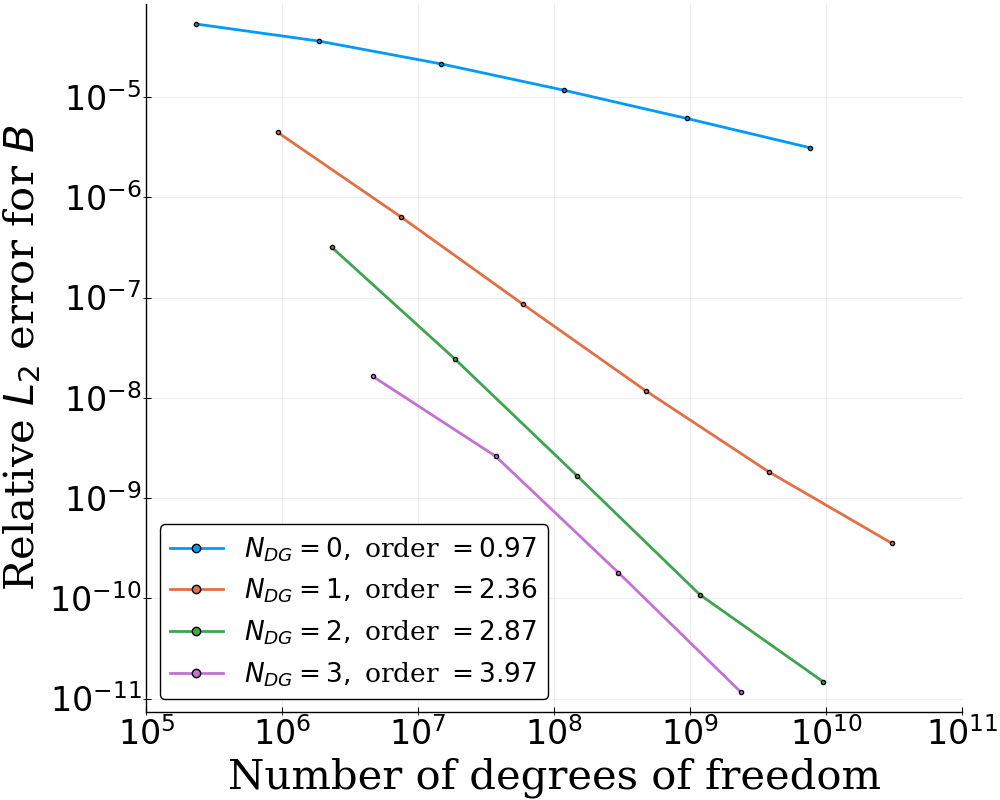}
    \caption{Manufactured solution benchmark: $L_2$ errors in electron
      particle distribution function (left panel), electric field
      (central panel), and magnetic field (right panel) using the
      central numerical flux to solve the Maxwell equations.}
    \label{fig:ms:c}
\end{figure}

\begin{figure}[H]
    \centering
    \includegraphics[width=0.32\linewidth]{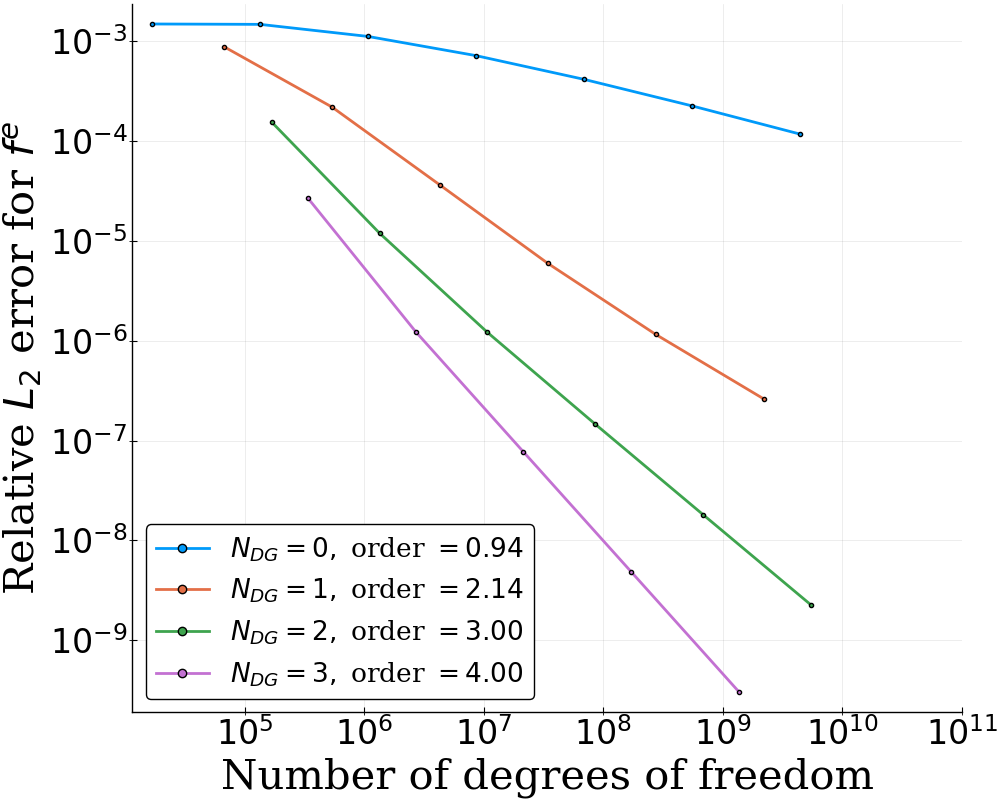}
    \includegraphics[width=0.32\linewidth]{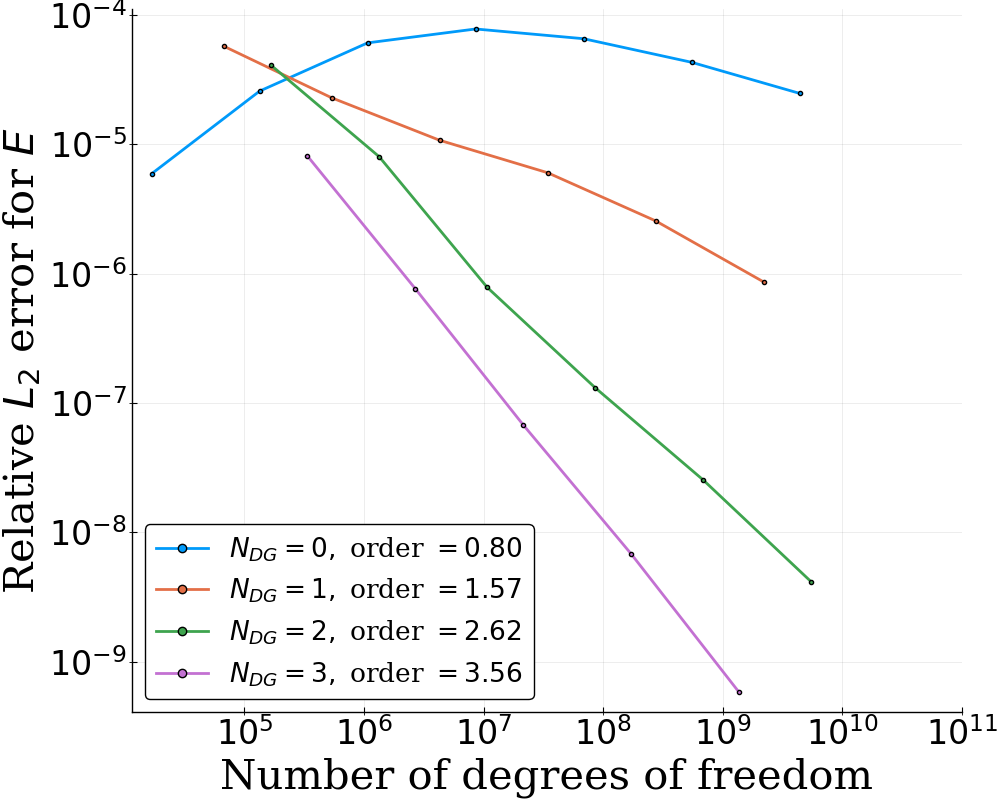}
    \includegraphics[width=0.32\linewidth]{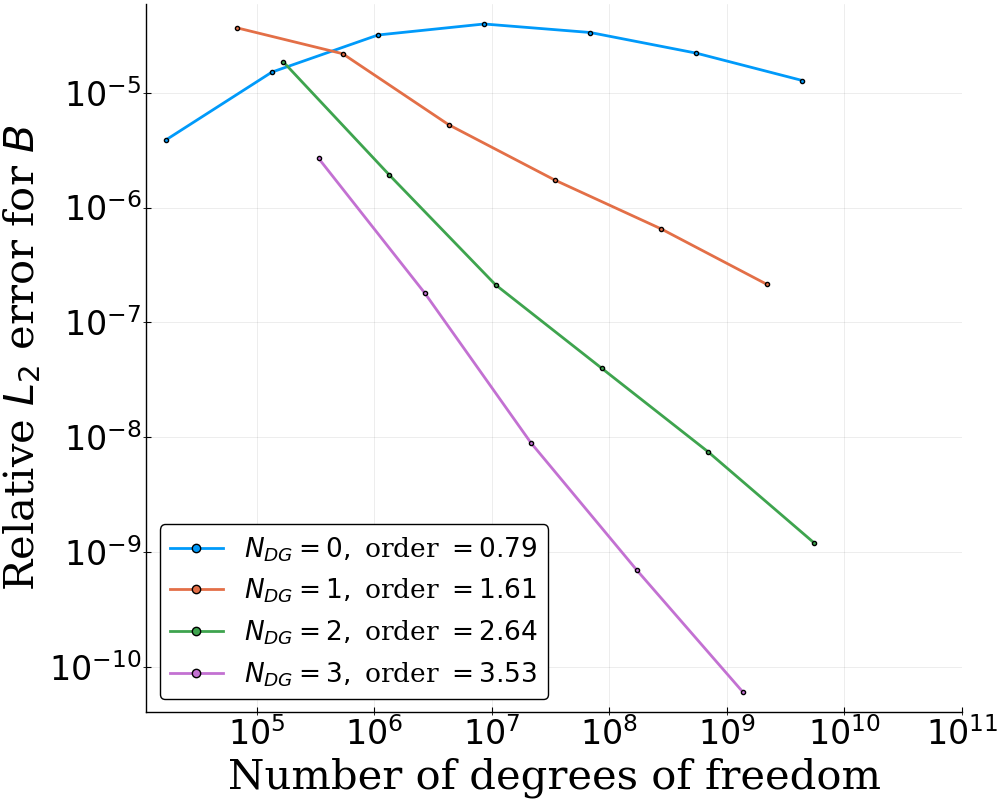}
    \caption{Manufactured solution benchmark: $L_2$ errors in electron
      particle distribution function (left panel), electric field
      (central panel), and magnetic field (right pane) using the
      upwind numerical flux to solve the Maxwell equations.}
    \label{fig:ms:u}
\end{figure}

\medskip
In order to avoid pollution of the results by time discretization
errors, we integrate the system forward up to time $T=1$ and then
backward for an equivalent time interval with highly accurate fifth-order RK method~\cite{Fehlberg:1970} with time step $\Delta t =
10^{-3}$ (two thousand explicit time steps in total). 
The resulting $L^2$ errors for the electron distribution function,
electric and magnetic fields versus number of total degrees of freedom
are shown in Figure~\ref{fig:ms:c} for the central numerical flux and
in Figure~\ref{fig:ms:u} for the upwind numerical flux in the Maxwell equations.
We do not compute the error for the ion distribution function because
ions are practically motionless on such short electron time
scales.

\medskip
In every figure in this subsection, we measure the order of convergence of the method
by fitting data from the last two (most resolved) points.
The computed orders for different $\NDG$ and numerical fluxes are
summarized in Table~\ref{tab:ms:order}.
For cases with the central numerical flux, we can see that order of convergence measured
numerically is in good agreement with the theoretical prediction,
i.e., $\NDG+1$.
In case of the upwind numerical flux, the order measured numerically for the
electron distribution function error follows the same theoretical
prediction.
However, errors in electric and magnetic fields show a consistent
reduction in the computed order, so that the electromagnetic field errors
follow approximately an $\NDG+1/2$ order of convergence.
This convergence rate is consistent with the study of
Reference~\cite{Cheng-Gamba-Li-Morrison:2014}, where the \DG{}
discretization with the upwind numerical flux was applied to the
Vlasov-Maxwell system.

\begin{table}[H]
  \centering
  \begin{tabular}{|l|c|c|c|c|}
    \hline 
    & $\NDG=0$ &  $\NDG=1$ & $\NDG=2$ & $\NDG=3$ \\ \hline
    Electron distribution function with central flux & $0.96$ & $2.02$ & $3.00$ & $4.00$  \\ \hline
    Electric field                 with central flux & $0.97$ & $1.99$ & $2.99$ & $3.98$  \\ \hline
    Magnetic field                 with central flux & $0.97$ & $2.36$ & $2.87$ & $3.97$  \\ \hline
    Electron distribution function with upwind  flux & $0.94$ & $2.14$ & $3.00$ & $4.00$  \\ \hline
    Electric field                 with upwind  flux & $0.80$ & $1.57$ & $2.62$ & $3.56$  \\ \hline
    Magnetic field                 with upwind  flux & $0.79$ & $1.61$ & $2.64$ & $3.53$  \\ \hline 
  \end{tabular}
  \caption{Manufactured solution benchmark: numerical rate of convergence of the spatial discretization for different polynomial degrees $\NDG$ using central and upwind numerical fluxes to solve the Maxwell equations.}
  \label{tab:ms:order}
\end{table}

\subsection{Whistler instability}
\label{sec:whistler}
In the second benchmark problem we investigate the whistler instability.
  \CP{Since the instability is
driven by the electron temperature
anisotropy and cyclotron resonance, cf.~\cite{Gary:2005}, this test asserts the method's ability to describe kinetic physics.
The whistler instability} is common in space plasmas and it is believed to be
behind the generation of chorus waves in the Earth's magnetosphere~\cite{thorne2010}.

We discretize the velocity space by setting $\Nn=\Nm=\Np=9$, so to
have $10$ polynomials in each velocity direction, and the physical
space domain $\Ox=[0,L_x]\times[0,L_y]\times[0,L_z]$ by setting
$L_x=2\pi$, $L_y=L_z=1$ and $N_x = 50$, $N_y = N_z = 1$, with uniform cell
size in every direction.

Further, we consider $B_x=1$ for the initial magnetic field, while the
distribution functions of electrons and ions (protons) are assumed to
be Maxwellian.
Therefore, we set
\begin{equation}
  C_{0,0,0}^s (\xv) = \frac{1}{\alpha_x^s \alpha_y^s \alpha_z^s}, 
  \label{mdistc}
\end{equation}
where $s\in\{e,i\}$ (electrons or ions),
$\alpha_\beta^s=\sqrt{2}\vs_{T_\beta^s}$, and $u_\beta^s=0$ for
$\beta\in\{x,y,z\}$. 
We use a realistic ion-to-electron mass ratio $m_i/m_e=1836$, thermal
velocities $\vs_{T_\beta^e}=\vs_{T_\beta^i}\sqrt{m_i/m_e}=0.125$,
$\beta\in\{x,y,z\}$ with the exception of the reduced electron thermal
velocity along the $x$ axis, $v_{T_x^e} = 0.056$, to create anisotropy and, thus, the source of the
instability.
The electron plasma/gyrofrequency ratio is
$\omega_{pe}/\omega_{ce}=4$, and the collisional operator has $\nu=1$.
In order to seed the whistler instability, we initialize a small
electron current perturbation along $x$,
$j^e_x(\xv)=10^{-3}\cos(\xs)$, which can be imposed through the first Hermite moment
$C^e_{1,0,0}$.
The other coefficients in the Hermite expansion are initialized to
zero.

\par 
To advance the numerical solution in time, we use the third order
non-adaptive RK scheme of Bogacki-Shampine
\cite{Bogacki-Shampine:1989} with time steps
$\Delta\ts=0.02,\,0.01,\,0.005$.
We also assess the performance of the method for the two alternative
choices of central and upwind numerical fluxes and for the two local
polynomial degrees given by setting $\NDG=1$ and $2$ in the \DG{}
approximation.

\par
\begin{figure}[H]
  \centering
  \includegraphics[width=0.45\linewidth]{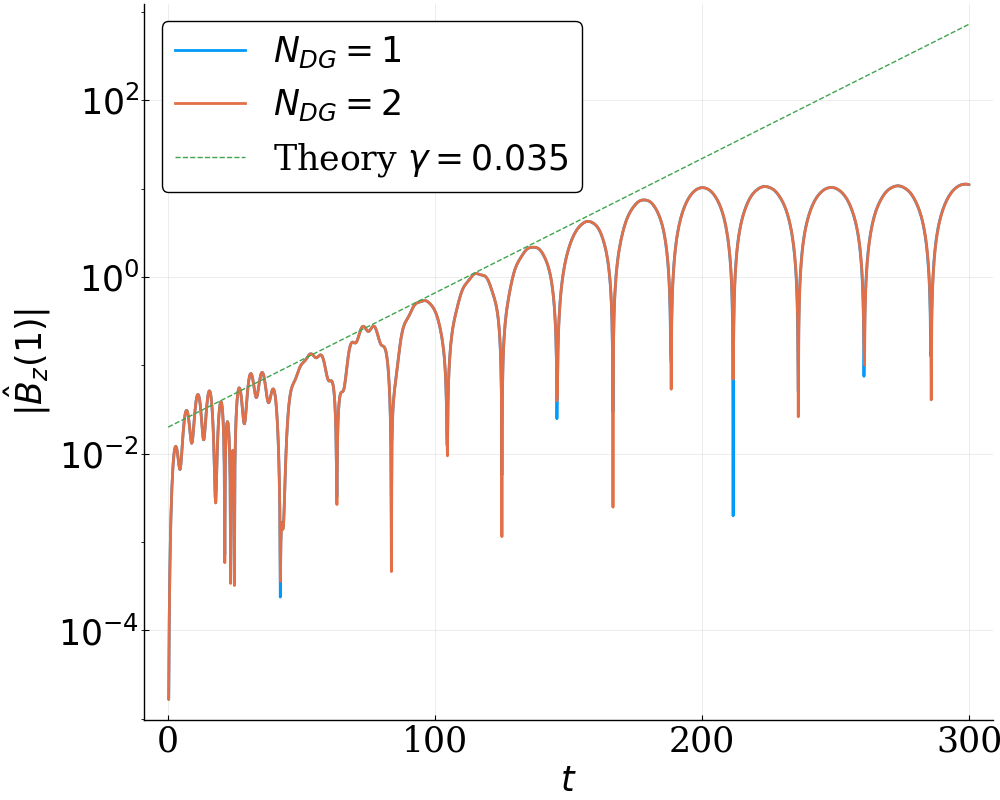}
  \includegraphics[width=0.45\linewidth]{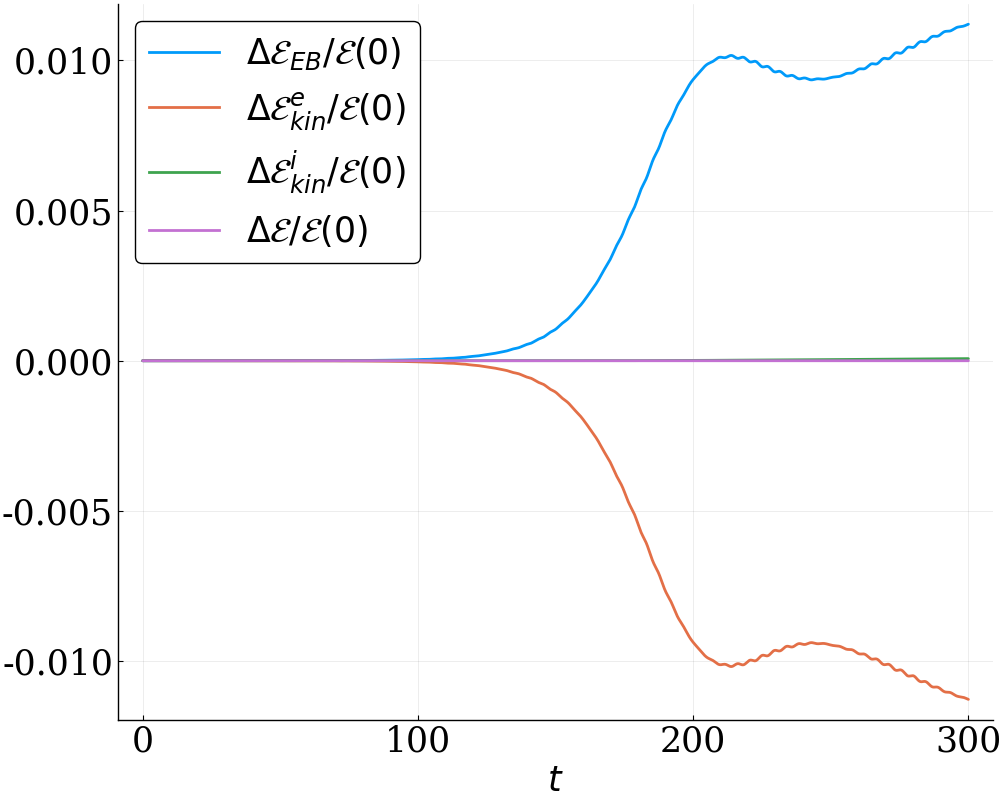}
  \caption{Whistler instability benchmark: time evolution of
    $\ABS{\hat{\Bs}_z(1)}$ (left panel); time evolution of
    electromagnetic, kinetic, and total energy (right panel). 
    Numerical results are shown when using the central numerical flux,
    $\Delta\ts=0.02$, and $\NDG=1,\,2$ (left panel) and $\NDG=2$ (right
    panel).}
  \label{fig:whistler_x}
\end{figure}

First, we verify the ability of the RK-Hermite-\DG{} method to reproduce
the whistler instability, i.e., the exponential growth of the
electromagnetic whistler wave from an initial small perturbation.
To this end, we monitor the time evolution of the first magnetic field
Fourier mode $\hat B_z (k)$ with $k=1$ and
\begin{align*}
  \hat{\Bs}_z(\ks) = \sum_I\Bs_z(\xs_c^I) e^{i\ks\xs_c^I},
\end{align*}
where $\xs_c^I$ is the center of the $I$-th \DG{} cell. 
The evolution of $\hat B_z(1)$ is shown in the left panel of
Figure~\ref{fig:whistler_x}.
The whistler wave grows exponentially with the theoretically predicted
growth rate $\gamma = 0.035$, and later saturates due to nonlinear
effects.
In the right panel of Figure~\ref{fig:whistler_x}, the evolution of
different energy contributions
\begin{equation}
  \frac{\Delta\mathcal{E}_{EB}}{\mathcal{E}(0)} = \frac{\mathcal{E}_{EB}(t) - \mathcal{E}_{EB}(0)}{\mathcal{E}(0)}, \quad
  \frac{\Delta\mathcal{E}^{s}_{kin}}{\mathcal{E}(0)} = \frac{ \mathcal{E}^{s}_{kin}(t) - \mathcal{E}^{s}_{kin}(0)}{\mathcal{E}(0)}, \quad
  \frac{\Delta\mathcal{E}}{\mathcal{E}(0)} = \frac{\mathcal{E}(t) - \mathcal{E}(0)}{\mathcal{E}(0)}, 
  \label{res:en1}
\end{equation}
is plotted, where
\begin{equation}
  \mathcal{E}_{EB}(t) = \frac{1}{2} \left ( \frac{\omega_{ce}}{\omega_{pe}} \right )^2 \int_{\Omega_x} \left (  \mathbf E^2 + \mathbf B^2  \right )   d\xv  ,\quad
  \mathcal{E}^{s}_{kin}(t)  = \frac{m^s}{2} \int_{\Omega_x} \int_{\Omega_v} v^2 f^s   d\xv d\vv ,\quad
  \mathcal{E}(t)      = \mathcal{E}_{EB}(t) + \sum_s \mathcal{E}^{s}_{kin}(t) ,
  \label{res:en2}
\end{equation}
are the electromagnetic energy, the kinetic energy of species $s$, and
the total energy, respectively.

The energy of the electromagnetic wave grows at the expense of the
kinetic energy of the electrons, while the ion kinetic energy stays
almost unchanged.
This is expected since the whistler instability is controlled by the
electron dynamics, and ions are effectively motionless on the short
electron time scales.
Our simulations produce almost indistinguishable results for two
$\NDG=1$ and $2$, central and upwind numerical fluxes in the
discretization of the Maxwell equations, and time steps
$\Delta\ts=0.02,\,0.01,\,0.005$.
For this reason, in Figure~\ref{fig:whistler_x}, we report only the
numerical results when using the central numerical flux,
$\Delta\ts=0.02$, and $\NDG=1,\,2$ (left panel) and $\NDG=2$ (right
panel).

\subsection{Tearing instability}
\label{sec:tearing}
The next benchmark problem concerns the evolution of the so-called tearing instability, which is a linear instability that produces magnetic reconnection in sheared magnetic field configurations, e.g.~\cite{Biskamp2000}. The tearing instability (and magnetic reconnection in general) exists only in the presence of finite dissipation, which in collisionless plasmas is produced by kinetic effects associated with wave-particle interactions. As such, it represents a challenging and practically important illustration of the method's ability to correctly capture kinetic effects.
Furthermore, relative to the whistler instability, which in practice involves only electron motion, this test is truly multi-scale, since both electrons and ions concur to the development of the tearing instability.

We employ a Harris sheet equilibrium~\cite{harris1962equilibrium}, adapted to satisfy periodic boundary conditions
by initializing two separate 
reconnection regions.
For this purpose,
we initialize four different plasma species
(electrons and ions for each reconnection region)
with distribution functions
\begin{align}
  f^s(\xv,\vv,0) = 
  \frac{n^s(\xv)}{(2\pi)^{3/2} v_{T_s}^3} 
  \exp \left [ -\frac{ v_x^2 + v_y^2 + (v_z - V_z^s)^2}{2 v_{T_s}^2} \right ],
\end{align}
where $s \in \{$ electrons \#1, electrons \#2, ions \#1, ions \#2 $\}$
denoted by $e1,e2,i1,i2$, respectively.
In this case, the initial plasma density is 
\begin{align}
n^{e1}(\xv) = n^{i1}(\xv) &= \mbox{sech}^2 \left ( \frac{x - 0.25 L_x}{\lambda}\right),\\
n^{e2}(\xv) = n^{i2}(\xv) &= \mbox{sech}^2 \left ( \frac{x - 0.75 L_x}{\lambda}\right),
\end{align}
and current sheets of width $\lambda$ are
created by counterstreaming electrons and ions. 
Moreover, current sheets at $x=0.25L_x$ and $x=0.75L_x$ are initialized in opposite directions, i.e., $V_z^{e1} = -V_z^{e2} =- 1/(32\sqrt{3})$ and $V_z^{i1} = -V_z^{i2}=1/(16\sqrt{3})$.
Additionally, the equilibrium magnetic field is initialized 
to satisfy the stationary Amp\`ere law
\begin{align}
    B_x = 0,\quad
    B_y =  
  \tanh \left ( \frac{x-0.25L_x}{\lambda}  \right )
-  \tanh \left ( \frac{x-0.75L_x}{\lambda}  \right )
-1, \quad
B_z = 1.
\end{align}
The equilibrium configuration has the
following dimensionless parameters 
\begin{align}
    \frac{T_i}{T_e} = 2, \quad
    \frac{\omega_{pe}}{\omega_{ce}} = 2, \quad
    \frac{m_i}{m_e} = 256, \quad
    \frac{\vs_{T_i}}{\omega_{ci} \lambda} = 1, 
\end{align}
which are sufficient to reconstruct the other equilibrium parameters.
\par 
In order to initiate reconnection, we seed an unstable perturbation in the magnetic field as
\begin{align}
    \delta B_x =  - \delta B k_y \sin  ( k_x x  ) \sin   ( k_y y ), \quad
    \delta B_y =  -\delta B k_x \cos  ( k_x x  ) \cos   ( k_y y  ), \quad  \delta B_z =0,
    \label{eq:HS:pert}
\end{align}
with $\delta B = 10^{-3}$, $k_x = 2 \pi / L_x$, and $k_y = 2 \pi / L_y$.
\par 
We discretize the velocity space by setting $\Nn=\Nm=\Np=9$, so to
have $10$ polynomials in each velocity direction. The physical
space domain $\Ox=[0,L_x]\times[0,L_y]\times[0,L_z]$ is chosen with
$L_x=200$, $L_y=4\pi \lambda = 64 \pi/\sqrt{3}$, $L_z=1$ 
and discretized with $N_x = 128$, $N_y = 36$, $N_z = 1$, $N_{DG}=2$
with uniform cell size in every direction.
Lastly, we use $\nu=1$ to avoid filamentation of the particle distribution
function and the third order
adaptive RK scheme of Bogacki-Shampine
\cite{Bogacki-Shampine:1989} 
with average $\Delta t \approx 0.14$ and final simulation time $T = 10^5$.
\par
\begin{figure}[H]
  \centering
  \includegraphics[width=0.75\linewidth]{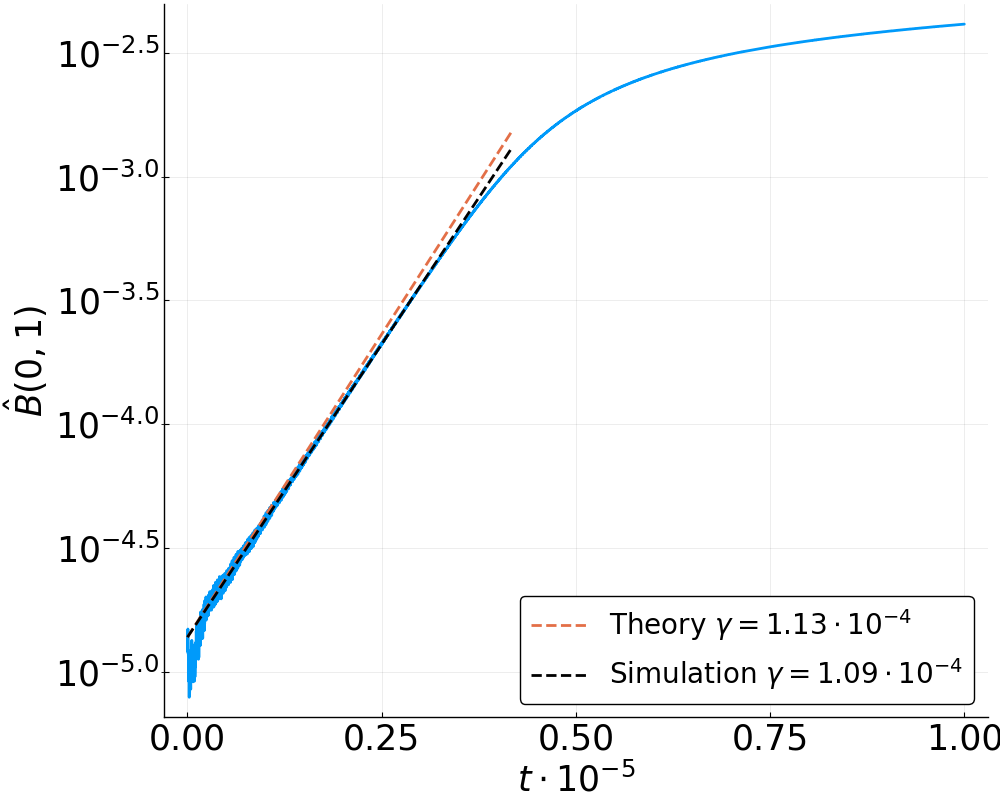}
  \caption{Tearing instability benchmark: time evolution of
  unstable mode 
    $\hat{B}(0,1)$.}
  \label{fig:HS:inst}
\end{figure}
\begin{figure}[H]
  \centering
  \includegraphics[width=0.45\linewidth]{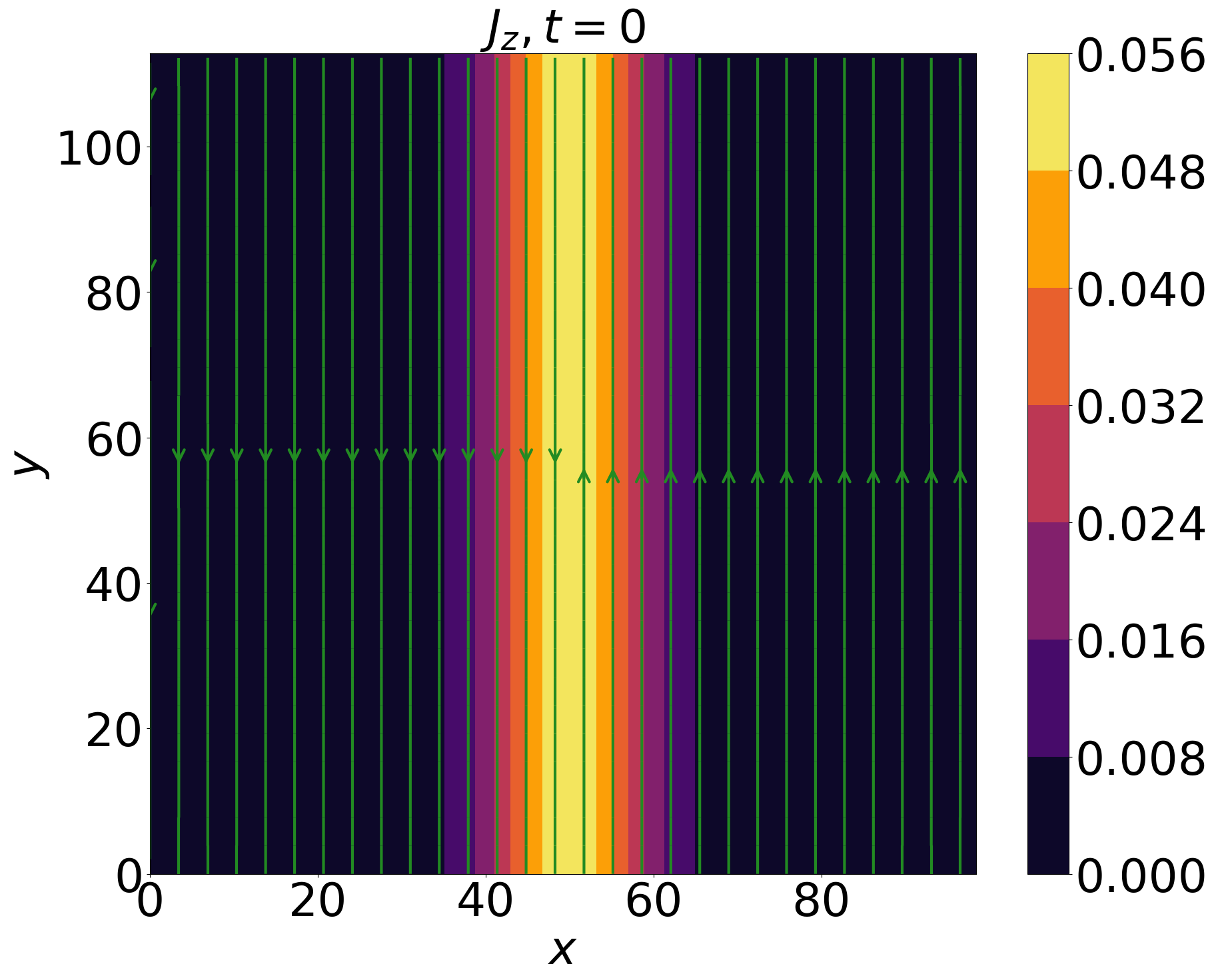}
  \includegraphics[width=0.45\linewidth]{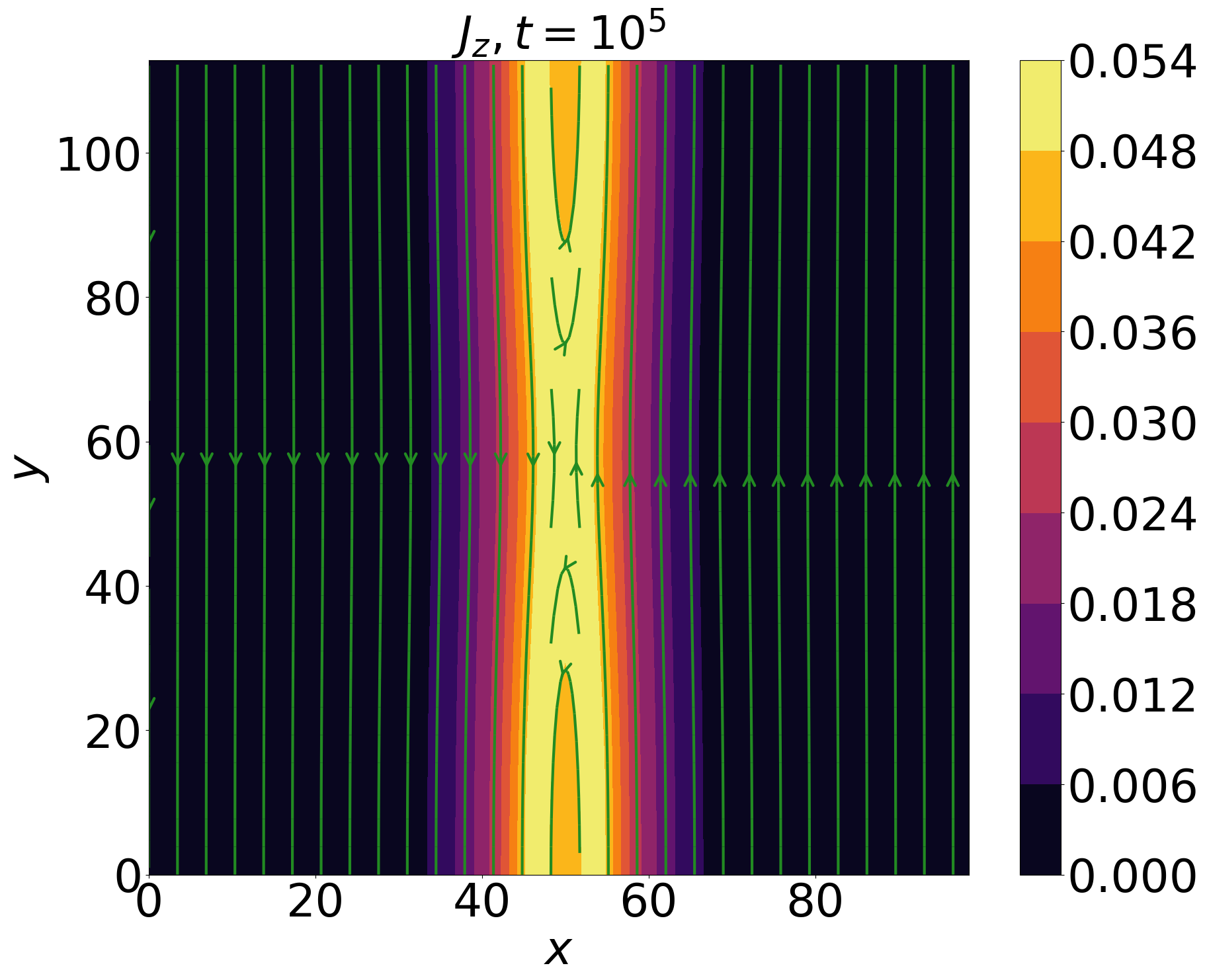}
  \caption{Tearing instability benchmark: current density $j_z$ along
    $\hat{\mathbf{z}}$ axis in the simulations at initial $t=0$
    (left panel) and final time $t=10^5$ (right panel).
    Green contour lines show the direction of magnetic field lines.}
  \label{fig:HS:curr}
\end{figure}

First, we verify the ability of SPS-DG to reproduce
the tearing instability, i.e., the exponential growth of the
excited perturbation (\ref{eq:HS:pert})
relative to the growth rate
$\gamma = 1.13 \cdot 10^{-4}$
computed by a linear Vlasov solver \cite{Gary:2005}.
(This growth rate was also verified with a different linear solver based on an Hermite expansion of the linearized Vlasov-Maxwell equations \cite{Camporeale-Delzanno-Lapenta-Daughton:2006}.)
To this end, we monitor the time evolution of the magnetic field
Fourier mode $\hat B (k_x,k_y)$ with $k_x=0$, $k_y = 1$ and
\begin{align*}
  \hat B (k_x,k_y) = \frac{1}{N_x N_y} \sqrt{ \sum_{\beta\in\{x,y,z\}} \hat B_\beta^2(k_x,k_y) },\\
  \hat{\Bs}_\beta(k_x,k_y) = \sum_{I/2}\Bs_\beta(\xs_c^I,\ys_c^I) e^{ik_x\xs_c^I}e^{ik_y\ys_c^I},
\end{align*}
where summation $\sum_{I/2}$ is performed over half of the domain, i.e.,
$[0,L_x/2]\times[0,L_y]$, so that only one current sheet is considered
(otherwise the $k_x=0$ mode cancels out due to symmetry).
The evolution of $\hat B(0,1)$ is shown in
Figure~\ref{fig:HS:inst}
where one can see that the tearing instability
grows exponentially with a growth rate that is in good agreement with the theoretically predicted
value, and later saturates due to nonlinear
effects. The growth rate from the SPS-DG simulation is $\gamma=1.09\times 10^{-4}$, with a $4\%$ relative error with respect to the value obtained by numerically solving the linearized Vlasov-Maxwell equations, $\gamma=1.13\times 10^{-4}$. We note that the latter computation is performed for a single current sheet configuration and some (minor) difference with the double-sheet configuration is expected given that the eigenfunction of the tearing instability decays weakly with distance from the center of the current sheet and in the considered configuration the current sheets are separated only by the distance of approximately $10.8 \lambda$. 
The current densities for the initial time $t=0$ and the final time $t=10^5$
are shown in Fig.~\ref{fig:HS:curr}.
The figure shows only half of the simulation domain, i.e.,
$[0,L_x/2]\times[0,L_y]$,
since the other half is symmetric.
One can see that the initial uniform current sheet 
has evolved forming the so-called X-point, which is the characteristic signature of magnetic reconnection.

\subsection{Orszag-Tang vortex}
\label{sec:OT}
The last benchmark is the Orszag-Tang vortex problem
\cite{Orszag:1979}. Here the initial condition corresponds to two large-scale vortices, which subsequently
evolve to form small-scale structures, such as current sheets. If the system size is sufficiently large, a transition to fully developed turbulence occurs via breaking of thin current sheets by magnetic reconnection. In fact, the Orszag-Tang initial conditions are often used in studies of two-dimensional plasma turbulence, see e.g.~\cite{Biskamp-Welter:1989, parashar:2009, Vencels-Delzanno-Manzini-Markidis-BoPeng-Roytershteyn:2015} and others. The problem is of particular interest for the present work, since it is an example of the interaction
between large-scale, fluid-like behavior and small-scale, dissipative processes involving kinetic physics.
We compare solutions obtained using the SPS-DG method against a reference solution obtained using a 
conventional PIC algorithm implemented in the
VPIC code~\cite{Bowers:2008,Bowers:2008b,Bowers:2009}. 
In principle, the model equations considered in the SPS-DG and VPIC
codes are different, because the VPIC code solves the relativistic version of the Vlasov-Maxwell
system, while SPS-DG does not account for relativistic effects.
However, we will consider parameters where the relativistic effects
are not important, so that a comparison is meaningful.
Since the early evolution of the system is dominated by large-scale
structures, the dynamics of the SPS-DG and VPIC solutions should be the
same.
Later in time, when small-scale structures form and kinetic
physics becomes important, we may expect the behavior of the SPS-DG
and VPIC code to differ due to the limited resolution in velocity
space of SPS-DG. Further, late-time evolution may become turbulent (stochastic), so that comparisons between the solutions are only meaningful in a statistical sense.

In this test, we discretize the velocity space by setting
$\Nn=\Nm=\Np=9$, and the physical space domain
$\Ox=[0,L_x]\times[0,L_y]\times[0,L_z]$ by setting $L_x=L_y=50$ and
$L_z=1$ and $N_x=N_y=120$, $N_z=1$, with uniform cell size in
every direction.
We impose spatially periodic boundary conditions.
The components of the initial magnetic field $\Bv(\xv,0)$ are set to
\begin{align*}
  \Bs_x(\xv,0) &=-\delta\Bs \sin(k_y y + 4.1),\\
  \Bs_y(\xv,0) &= \delta\Bs \sin(2k_x x + 2.3),\\
  \Bs_z(\xv,0) &= 1,
\end{align*}
with $\delta\Bs = 0.2$, $k_x = 2 \pi /L_x$, $k_y = 2 \pi/L_y$.  
The (randomly chosen) phases $4.1$ and $2.3$ are needed to remove any
artificial symmetry in the initial setup.  
The distribution functions for electrons and ions are initialized to
shifted Maxwellian distributions with spatially uniform density (we
omit species superscripts for clarity):
\begin{align*}
  f(\xv,\vv,0) = 
  \prod_{\beta\in\{x,y,z\}} \frac{1}{v_{T\beta} \sqrt{2\pi}} 
  \exp \left [ -\frac{ (v_\beta - V_\beta (\xv)) ^2}{2 v_{T_\beta}^2} \right ],
\end{align*}
with electron and ion velocities
\begin{align}
  V^e_x(\xv) &=-\delta B v_a \sin(k_y y + 0.5),\\
  V^e_y(\xv) &= \delta B v_a \sin(k_x x + 1.4),\\
  V^e_z(\xv) &= -\frac{ \delta B \omega_{ce}}{\omega_{pe}} \left (2 k_x \cos(2k_x x + 2.3) + k_y \cos (k_y y + 4.1 )\right ) ,\\
  V^i_x(\xv) &= U^e_x(\xv),\\
  V^i_y(\xv) &= U^e_y(\xv),\\
  V^i_z(\xv) &= 0,
\end{align}
where $v_a = 0.1$ and $\omega_{pe}/\omega_{ce} = 2$. 
The values $0.5$ and $1.4$ above are randomly chosen phases and
$V_z^e$ is set to satisfy Amp\`ere's law at time $t=0$.  
Other parameters include $m_i/m_e = 25$ and $\nu=1$.  
The shifted Maxwellian distribution is initialized according to the
formula
\begin{align}
  C_{n,m,p} =\frac{1}{\alpha} \sqrt{\frac{2^n 2^m 2^p}{n! m! p!}}
  \left(\frac{V_x}{\alpha_x}\right)^n
  \left(\frac{V_y}{\alpha_y}\right)^m
  \left(\frac{V_z}{\alpha_z}\right)^p,
\end{align}
where $\alpha_\beta^e = \sqrt{2} v_{T_\beta}^e = 0.25 $, $u_\beta^e=0$
and $\alpha_\beta^i = \sqrt{2} v_{T_\beta}^e/\sqrt{m_i/m_e} = 0.05 $,
$u_\beta^i=0$ with $\beta\in\{x,y,z\}$ for electrons and ions,
respectively.

\par 
For comparison with VPIC, we consider three different local
discontinuous polynomial degrees, $\NDG=1,2,3$, and both central and upwind
flux formulas in the \DG{} approximation of the Maxwell equations.  
We advance the numerical solution in time by using the adaptive
third-order accurate RK scheme of Bogacki-Shampine with second order
embedded method \cite{Bogacki-Shampine:1989}, where the mean time steps are
$\Delta t =0.13$, $\Delta t =0.079$, $\Delta t =0.053$, for
$N_{DG}=1$, $N_{DG}=2$, $N_{DG}=3$, respectively. 
The reference VPIC solution is computed on a grid with $N_x = N_y =
880$, $N_z = 1$, equally spaced partitions of the physical space
domain $\Ox$, time step $\Delta\ts=0.039373$ and average number of
particles per cell $N_p=4000$.
The simulations run for two Alfv\'en times, i.e., up to $T=2
L_x/v_a=1000$, where the Alfv\'en time, $L_x/v_a$, is a characteristic
dynamic time scale of the system.

\par
\begin{figure}[H]
  \centering
  \includegraphics[width=0.45\linewidth]{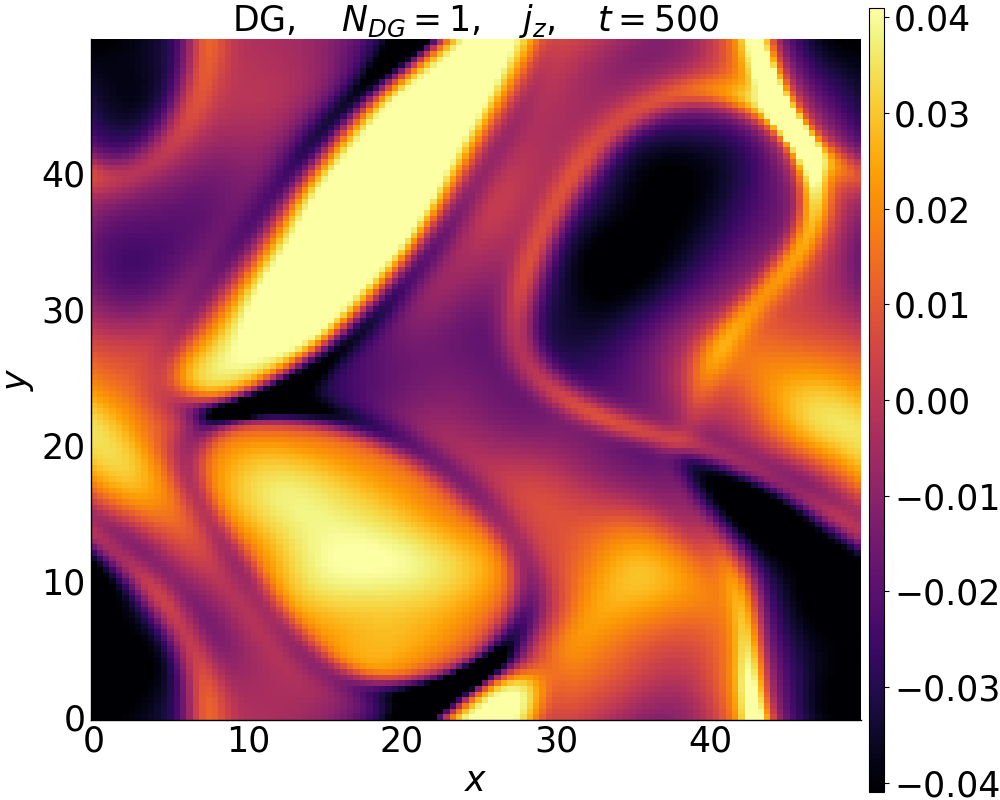}
  \includegraphics[width=0.45\linewidth]{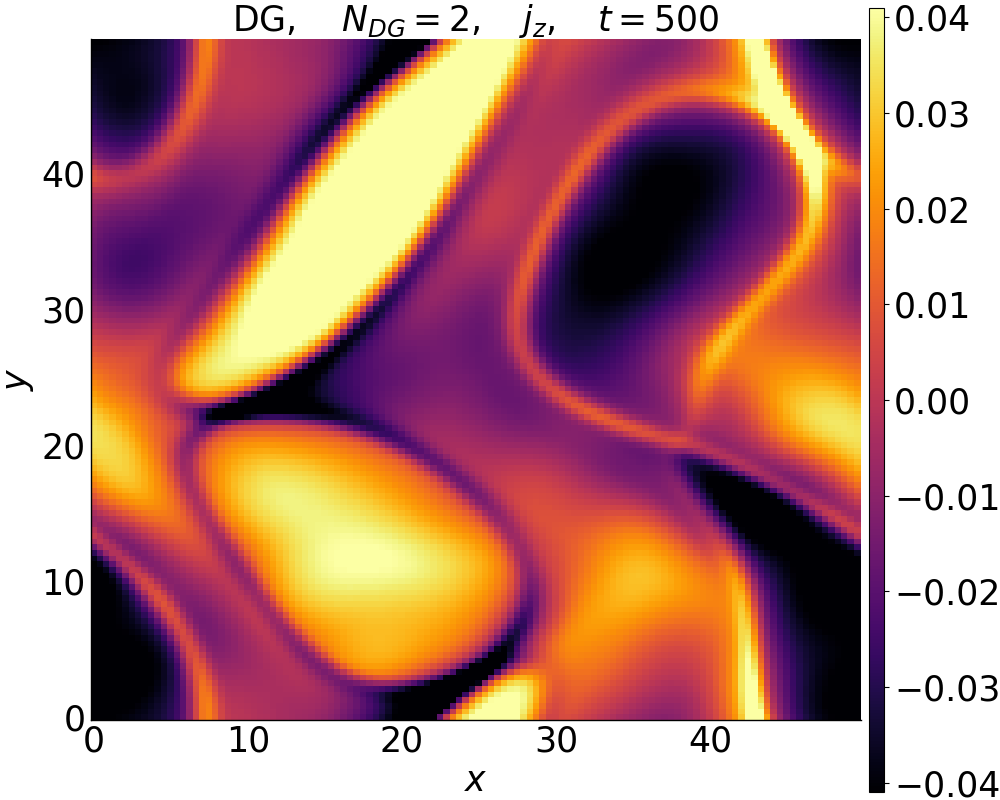} \\
  \includegraphics[width=0.45\linewidth]{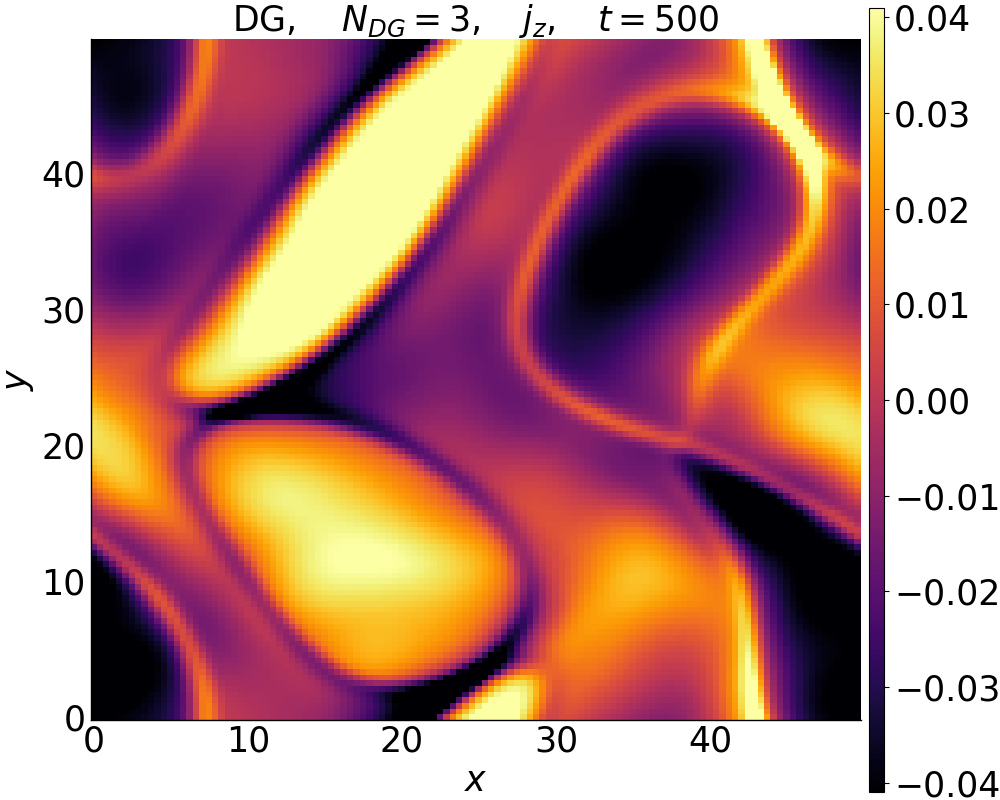}
  \includegraphics[width=0.45\linewidth]{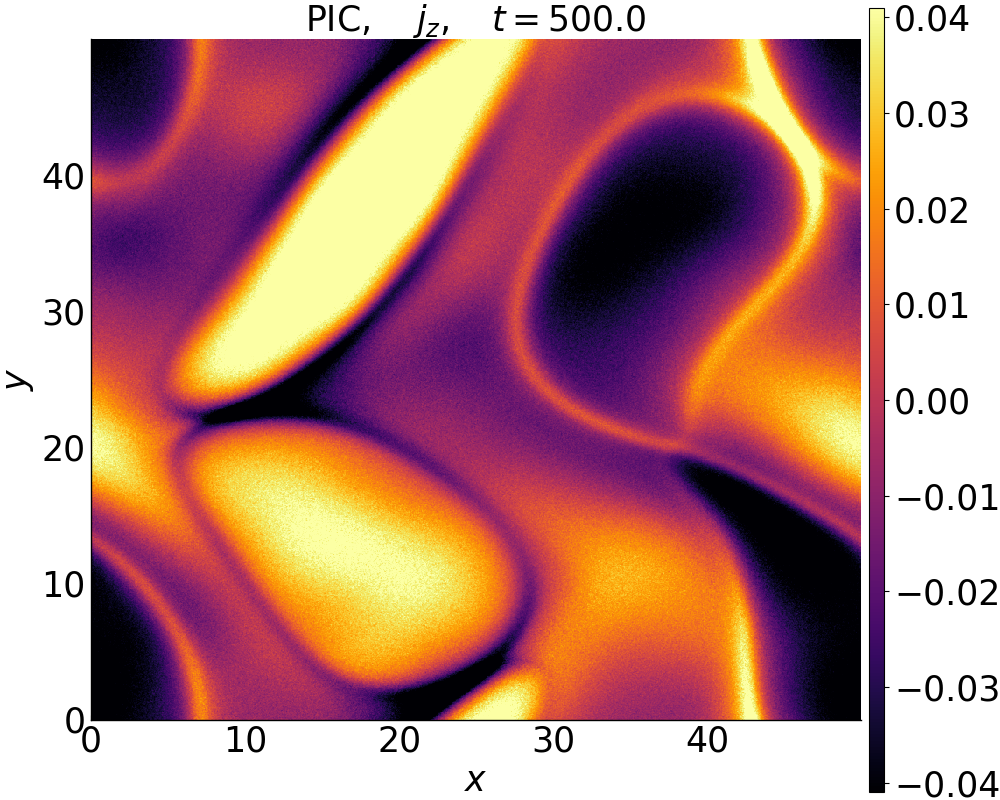}
  \caption{Orszag-Tang vortex benchmark: current density $j_z$ along
    $\hat{\mathbf{z}}$ axis in the simulations at one Alfv\'en time
    $t=L_x/v_a = 500$.}
  \label{fig:OTs:jz}
\end{figure}

The currents along $\hat{\mathbf{z}}$ formed after one Alfv\'en time (at
$t=500$) are shown in Figure~\ref{fig:OTs:jz} for SPS-DG with
$N_{DG}=1,2,3$ and for PIC.  
The figure shows a good agreement between all SPS-DG runs and the fully
kinetic PIC algorithm, indicating that even a spatial discretization
with $N_{DG}=1$ is sufficient in the early times of the simulation.

\begin{figure}[H]
  \centering
  \includegraphics[width=0.45\linewidth]{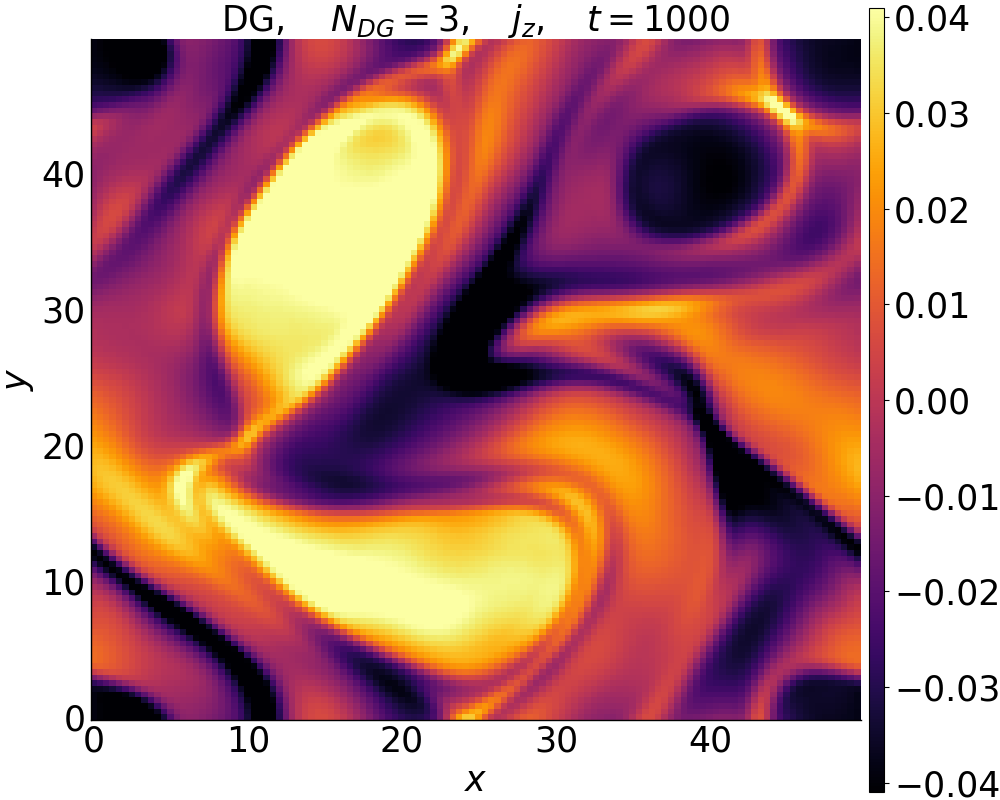}
  \includegraphics[width=0.45\linewidth]{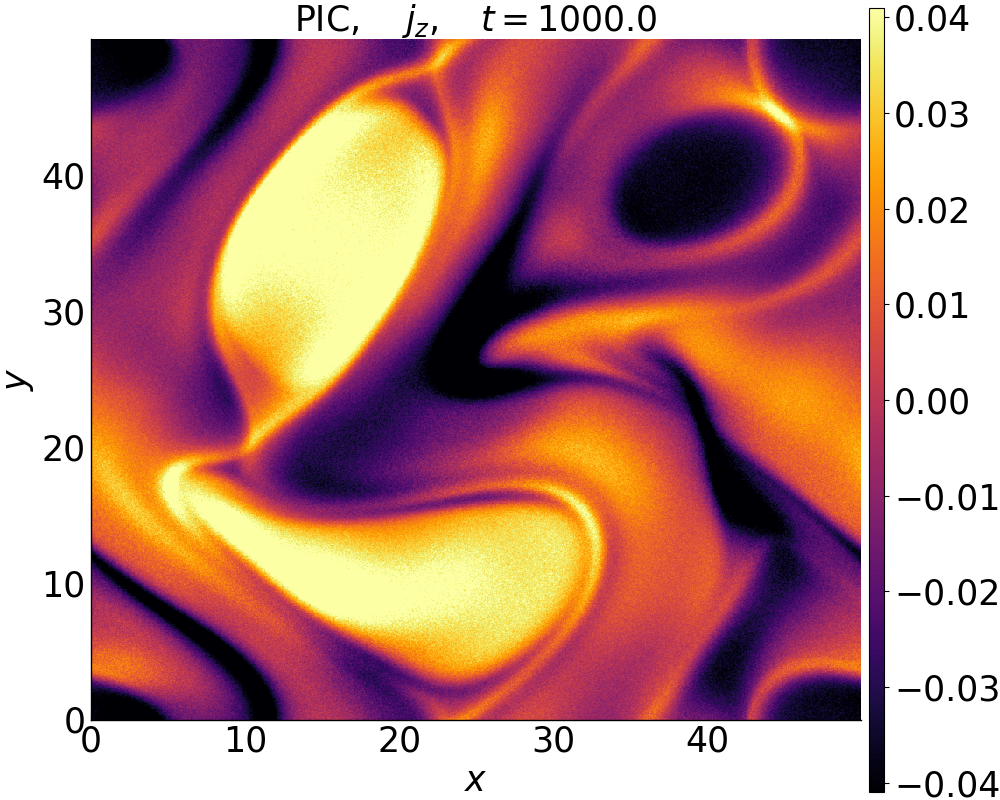} 
  \caption{Orszag-Tang vortex benchmark: current density $j_z$ along
    $\hat{\mathbf{z}}$ axis in the simulations at time $t=1000$.}
  \label{fig:OTs:jz2}
\end{figure}

Further evolution of the current density along $\hat{\mathbf{z}}$ is
shown in Figure~\ref{fig:OTs:jz2} where time snapshot $t=1000$ is shown for SPS-DG with $N_{DG}=3$ and for PIC.
We can see that the agreement between the two numerical solutions decreases with time,
with SPS-DG becoming slightly more diffusive.
We attribute this effect to the lack of velocity space resolution,
because strong flows form at later times making the plasma
distribution function strongly non-Maxwellian and requiring a larger
number of Hermite moments to capture the smaller structures.  
This interpretation is also supported by the fact that the SPS-DG
solutions for $\NDG=2$ and $\NDG=3$ are essentially identical (not
shown), indicating that the spatial and temporal resolution is
sufficient in SPS-DG (the higher order \DG{} with $\NDG=3$ has smaller
time step than that for $\NDG=2$).

\begin{figure}[H]
  \centering
  \includegraphics[width=0.45\linewidth]{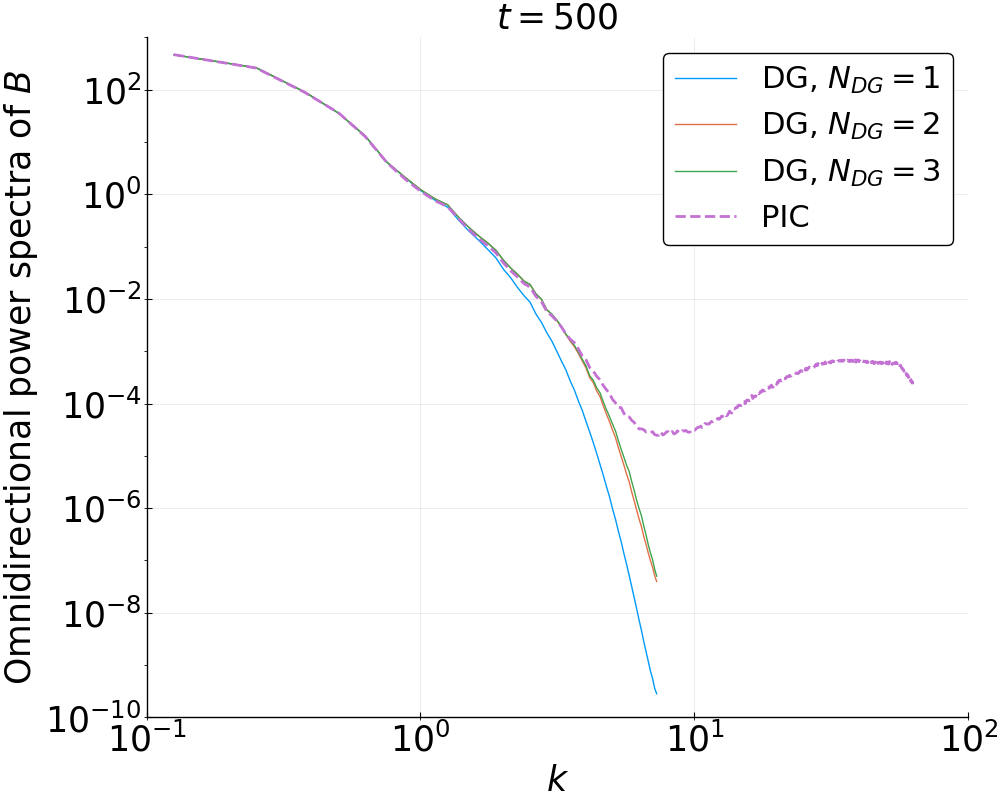}
  \includegraphics[width=0.45\linewidth]{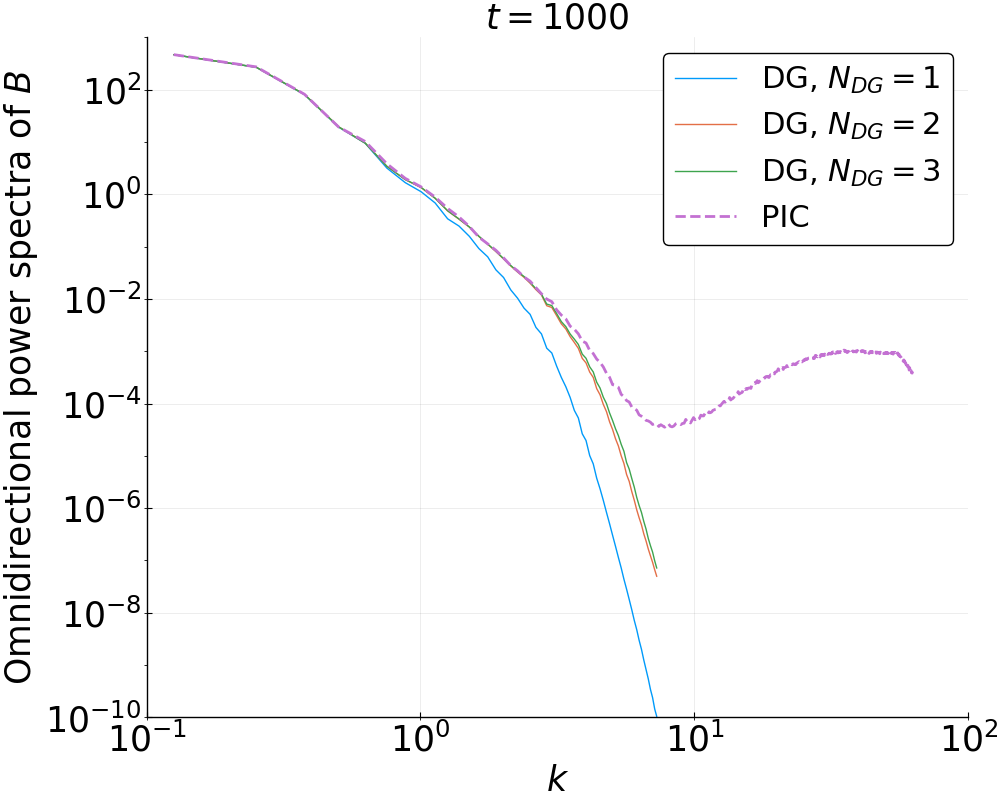} 
  \caption{Orszag-Tang vortex benchmark: Omnidirectional power
    spectrum of magnetic fluctuations in the simulations at times
    $t=500, 1000$. }
  \label{fig:OTs:spec}
\end{figure}

Next, we compare the omnidirectional spectrum of the magnetic field in
Figure~\ref{fig:OTs:spec}.
The spectra agree very well (up $k\sim 10$, for higher $k$ the PIC
spectrum is dominated by particle noise) at early times and start to
diverge slightly at later times ($t\sim1500$, not shown) for $N_{DG}\ge 2$. The case
$N_{DG}=1$ is too diffusive and the spectra start to diverge at $k\sim 1$.
The results for SPS-DG with $\NDG=2$ and $\NDG=3$ practically
coincide, confirming that the chosen spatial and temporal resolution
in SPS-DG is sufficient in these cases.

\begin{figure}[H]
    \centering
    \includegraphics[width=0.32\linewidth]{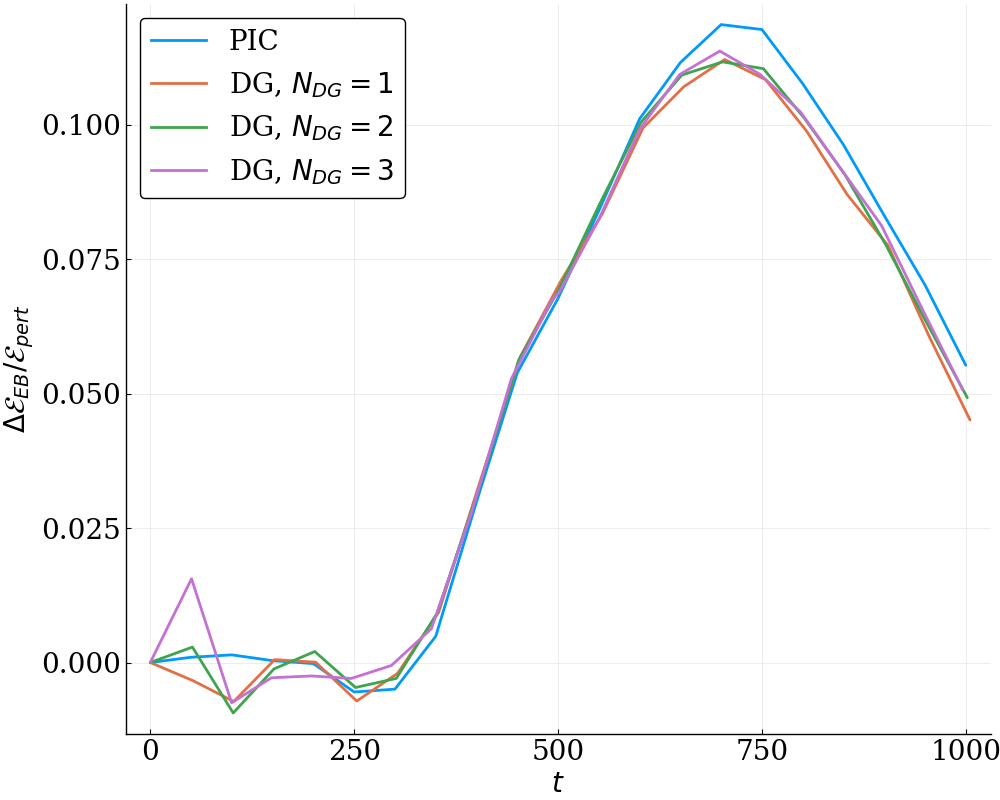} 
    \includegraphics[width=0.32\linewidth]{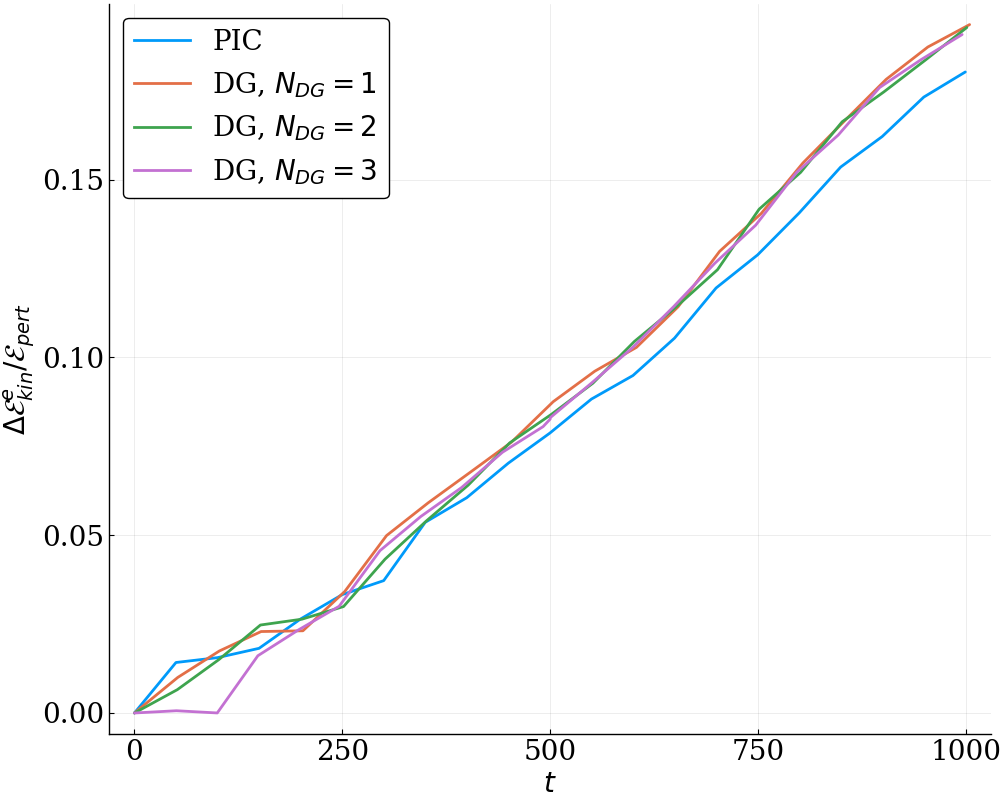}
    \includegraphics[width=0.32\linewidth]{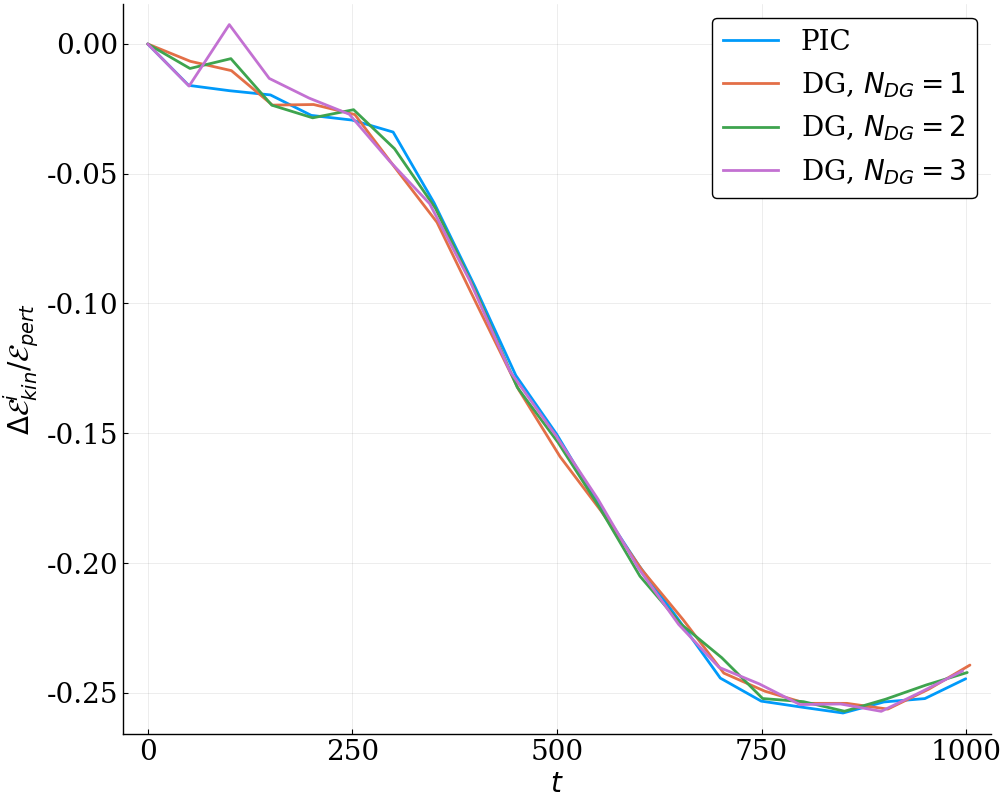} 

    \caption{Energy evolution in Orszag-Tang vortex test.}
    \label{fig:OTs:en}
\end{figure}

An important part of the comparison between SPS-DG and PIC is the
evolution of the energies defined in Eqs.~(\ref{res:en1}) and
(\ref{res:en2}).
This comparison is shown in Figure~\ref{fig:OTs:en}, which shows the various parts of
the total energy normalized to the energy of the initial perturbation
($\mathcal{E}_{pert}$), which includes the contribution of $B_x, B_y,
U^e_\beta, U^i_\beta$ for $\beta\in\{x,y,z\}$.
The evolution of the electromagnetic energy in SPS-DG closely follows PIC
with slight differences at later times.
The comparison of the kinetic energy reveals a good agreement as well,
with slight differences for electrons at later times.
Part of this disagreement can be attributed to relativistic effects in
the PIC simulations.
Indeed, particles kinetic energies in the relativistic case are bigger
than in the non-relativistic case by the Lorentz factor.
To estimate relativistic heating, we compare the measured thermal
relativistic energy in the PIC simulations with the classical
non-relativistic estimate (i.e., $3T/2$ with $T$ the temperature).
The difference for electrons is $10\%$ and for ions $0.4\%$, which is
consistent with the slight differences between electron kinetic and
electromagnetic energies between SPS-DG and PIC.
In summary, the comparison between SPS-DG and PIC on the OT test
problem reveals that, while some small-scale local differences exist
between the two at later times, 10 Hermite moments in each direction
are sufficient to capture accurately the early-time dynamics of the
system as well as the behavior of important quantities such as
magnetic-field spectra and energy partition.
\section{Conclusions}
\label{sec:conclusions}

We have presented a new spectral method for the solution of the multi-dimensional
Vlasov-Maxwell equations. The method combines an Hermite expansion in velocity space with a discontinuous Galerkin discretization for the spatial coordinates. 
In terms of DG, we analyze two schemes differing by the treatment of the flux at the interface between cells. Specifically, the Vlasov equation is discretized with upwind fluxes while Maxwell's equations can be discretized either with upwind or central fluxes. While upwind fluxes tend to improve the overall stability of the scheme, using central fluxes in Maxwell's equations can lead to the conservation of total energy in the system (under appropriate boundary conditions). 
We have further adopted
an explicit time discretization based on various Runge-Kutta methods of different orders.

The algorithms described in this paper have been implemented in the SPS-DG code. SPS-DG takes advantage of the PETSc data structure and solvers. Several numerical tests have been presented to show (a) the nearly optimal scalability of the approach up to $\sim 40,000$ cores, and (b) the accuracy of the spatial discretization, where we have recovered the appropriate order of convergence of the different DG schemes with a method of manufactured solutions. Additional tests like the whistler and tearing instabilities and the Orszag-Tang turbulence cascade demonstrate the successful application of the method to standard plasma physics problems \CP{and suggest that the method can capture kinetic behavior even with a relatively low number of modes per direction}.

Spectral methods with a suitable spectral basis, like the one considered in this paper, feature built-in fluid-kinetic coupling, i.e. they can capture the macroscopic dynamics of magnetized plasmas with the low-order moments of the expansion while the kinetic physics can be retained by adding higher-order moments only where necessary (i.e. locally in space and time). 
The DG discretization adds the ability to handle sharp, shock-like structures and extreme data locality to enable scalable implementations on high-performance-computing architectures. This removes the performance limitations of some of the earlier implementations of the Hermite spectral method, which was coupled with a Fourier spatial discretization and was hence limited in parallel scalability by the FFTs global communications.
In the future, these new algorithms might therefore enable simulations of the large-scale plasma dynamics with accurate feedback from the microscopic physics. 

\section*{Acknowledgments}
The work of GLD, OK, GM was supported by the Laboratory Directed
Research and Development - Exploratory and Research (LDRD-ER) Program
of Los Alamos National Laboratory under project number 20170207ER.
Los Alamos National Laboratory is operated by Triad National Security,
LLC, for the National Nuclear Security Administration of
U.S. Department of Energy (Contract No. 89233218CNA000001).
Computational resources for the SPS-DG simulations were provided by the Los Alamos National Laboratory Institutional Computing Program.
VR's contributions were supported by NASA grant
NNX15AR16G. Computational resources for PIC simulations were provided
by the NASA High-End Computing Program through the NASA Advanced
Supercomputing Division at Ames Research Center and by the Blue Waters
sustained-petascale computing project, which is supported by the
National Science Foundation (awards OCI-0725070 and ACI-1238993) and
the state of Illinois. Blue Waters allocation was provided by the
National Science Foundation through PRAC award 1614664.


\bibliographystyle{plain}
\bibliography{ref}


\appendix
\section{Derivation of~\eqref{eq:Vlasov-system:long}}

\subsection{Electric field terms}

We present the detailed expansion of the term containing $\Ex$
in~\eqref{eq:Vlasov-system:long}.
The final formula is obtained by using the Hermite expansion of the distribution function, substituting the
formulas for the derivatives of the Hermite basis functions, and
exploiting the orthogonality properties.
The other two terms, i.e., $\Ey$, $\Ez$, are easily obtained by a
simple permutation of indices, as shown below.
\begin{align}
  \int_{\Ov}\Ex(\xv,t)\frac{\partial\fs(\xv,\vv,t)}{\partial\vsx}\vpsi^{n}(\xi^{s}_{x})\vpsi^{m}(\xi^{s}_{y})\vpsi^{p}(\xi^{s}_{z})\,d\xi^{s}_{x}d\xi^{s}_{y}d\xi^{s}_{z}
  = -\Ex(\xv,t)\Cs_{n-1,m,p}(\xv,t)\frac{\sqrt{2n}}{\asx}.
  \label{eq:Ex:Hermite-expansion}
\end{align}
Term $\Ey$ is obtained by permuting the indices $(x,y,z)\to(y,z,x)$
and $(n,m,p)\to(m,p,n)$; term $\Ez$ is obtained by permuting the
indices $(x,y,z)\to(z,x,y)$ and $(n,m,p)\to(p,n,m)$.

\subsection{Magnetic field terms}

The magnetic field contributes to the Lorentz force by
\begin{align}
  \vv\times\Bv = \big[ 
  \vsy\Bz-\vsz\By,\,-\vsx\Bz+\vsz\Bx,\,\vsx\By-\vsy\Bx
  \big]^T.
\end{align}
We consider the terms associated with $\Bx$, that involve the following derivatives of the distribution function
\begin{align*}
\Bx\vsz\partial\fs\slash{\partial\vsy},\qquad
-\Bx\vsy\partial\fs\slash{\partial\vsz}
\end{align*}
The similar terms associated with $\By$ and $\Bz$ are derived by
permuting the indices as explained below.
The final formulas are given by substituting the
formulas for the derivatives of the Hermite basis functions, and
exploiting their orthogonality properties.
We find that
\begin{align}
  &\Bx(\xv,t)\int_{\Ov}\vsz\frac{\partial\fs(\xv,\vv,t)}{\partial\vsy}\vpsi^{n}(\xi^{s}_{x})\vpsi^{m}(\xi^{s}_{y})\vpsi^{p}(\xi^{s}_{z})\,d\xi^{s}_{x}d\xi^{s}_{y}d\xi^{s}_{z}\nonumber\\[0.5em]
  &\quad\quad
  = -\Bx(\xv,t)\left[
    \frac{\asz}{\asy}\sqrt{mp}\Cs_{n,m-1,p-1}(t) + \frac{\asz}{\asy}\sqrt{m(p+1)}\Cs_{n,m-1,p+1}(t) + \frac{\usz}{\asy}\sqrt{2m}\Cs_{n,m-1,p}(t)
  \right],
  \label{eq:Bx:1:Hermite-expansion}
  \intertext{and}
  &-\Bx(\xv,t)\int_{\Ov}\vsy\frac{\partial\fs(\xv,\vv,t)}{\partial\vsz}\vpsi^{n}(\xi^{s}_{x})\vpsi^{m}(\xi^{s}_{y})\vpsi^{p}(\xi^{s}_{z})d\xi^{s}_{x}\,d\xi^{s}_{y}d\xi^{s}_{z}
  \nonumber\\[0.5em]
  &\quad\quad
  = \Bx(\xv,t)\left[
    \frac{\asy}{\asz}\sqrt{mp}\Cs_{n,m-1,p-1} + \frac{\asy}{\asz}\sqrt{(m+1)p}\Cs_{n,m+1,p-1} + \sqrt{2p}\frac{\usy}{\asz}\Cs_{n,m,p-1}
  \right].
  \label{eq:Bx:2:Hermite-expansion}
\end{align}
We derive the two similar terms for $\By$ through the permutation
$(x,y,z)\to(y,z,x)$ and $(n,m,p)\to(m,p,n)$; and the two similar terms
for $\Bz$ through the permutation $(x,y,z)\to(z,x,y)$ and
$(n,m,p)\to(p,n,m)$.

\section{Proofs of conservation laws for the semi-discrete formulation}
\label{sec:variational:formulation}

To prove the conservation laws for the semi-discrete formulation, it
is convenient to rewrite the Hermite-DG numerical method in a
variational form.
To this end, we first introduce the following finite dimensional
spaces:
\begin{equation}\label{eq:VHN}
\begin{aligned}
  \calHN &:=\textrm{span}\Big\{\Psi_{n,m,p},\,\textrm{for~}(n,m,p)\in\big\{(0,0,0),\ldots,(N_{v_x},N_{v_y},N_{v_z})\big\}\Big\},\\[0.5em]
  \calHpN&:=\textrm{span}\Big\{\Psi^{n,m,p},\,\textrm{for~}(n,m,p)\in\big\{(0,0,0),\ldots,(N_{v_x},N_{v_y},N_{v_z})\big\}\Big\},\\[0.5em]
  \calVN &:=\textrm{span}\Big\{\vphi^{\Il},\,\textrm{for~}\Is\equiv\Is_{i,j,k},(i,j,k)\in\big\{(1,1,1),\ldots(N_x,N_y,N_z)\big\},\,\,l=1,\dots,\Nl\Big\}.
\end{aligned}
\end{equation}

\medskip
For any time $t\in[0,T]$, we assume that the numerical distribution
function $\fsN(\cdot,\cdot,t)$ defined in~\eqref{eq:fsN:def} belongs
to $\calHN\times\calVN$ for any plasma species $s$.
Similarly, we take the numerical electromagnetic fields $\vecEN$ and
$\vecBN$ that are defined in~\eqref{eq:Ev:expansion}
and~\eqref{eq:Bv:expansion} in the finite-dimensional space $\calVN$.
Finally, for convenience of exposition, we report the \DG{}
approximations of the charge and current density introduced in
\eqref{eq:rho_her} and \eqref{eq:vecJN:def}
\begin{align}
  \rhoN (\xv,t) &= \sum_{s}\rhoNs (\xv,t) = \sum_{s}\qs\int_{\Ov}   \fsN(\xv,\vv,t)\,d\vv\label{eq:rhoN:def},\\
  \vecJN(\xv,t) &= \sum_{s}\vecJNs(\xv,t) = \sum_{s}\qs\int_{\Ov}\vv\fsN(\xv,\vv,t)\,d\vv\label{eq:vecJN:def:b}.
\end{align}

\medskip 
The semi-discrete variational formulation of the Hermite-DG method
reads as:
\emph{For every species $s$, and any time $t\in[0,T]$
  find $\fsN\in\calHN\times\calVN$ and
  $\vecEN,\vecBN\in\calVN$ such that}
\begin{subequations}
  \label{eq:semi-discrete:Vlasov-Maxwell}
  \begin{align}
    \As\big( (\fsN,\vecEN,\vecBN),(\Psi,\varphi)\big) &=0
    \phantom{\Ls(\varphi)\fsNz\vecENz\vecBNz}\hspace{-1cm}
    \forall\,(\Psi,\varphi)\in\calHpN\times\calVN,
    \label{eq:semi-discrete:Vlasov}\\[0.5em]
    \Bs\big( (\vecEN,\vecBN), \varphi \big) &= \Ls(\varphi)
    \phantom{0\fsNz\vecENz\vecBNz}\hspace{-1cm}
    \forall\,\varphi\in\calVN,
    \label{eq:semi-discrete:Maxwell}
    \\[0.5em]
    \fsN(\cdot,\cdot,0)&=\fsNz   \phantom{0\Ls(\varphi)\vecENz\vecBNz}\hspace{-1cm}\textrm{in~}\Ox\times\Ov,  \\[0.5em]
    \vecEN(\cdot,0)    &=\vecENz \phantom{0\Ls(\varphi)\fsNz\vecBNz}\hspace{-1cm}  \textrm{in~}\Ox,\\[0.5em]
    \vecBN(\cdot,0)    &=\vecBNz \phantom{0\Ls(\varphi)\fsNz\vecENz}\hspace{-1cm}  \textrm{in~}\Ox,
  \end{align}
  \emph{where $\fsNz$, $\vecENz$ and $\vecBNz$ are the orthogonal
    projections of the initial conditions $\fs(\cdot,\cdot,0)$, $\Ev(\cdot,0)$ and
    $\Bv(\cdot,0)$ onto the spaces $\calHN\times\calVN$, $\calVN$ and $\calVN$,
    respectively.}
\end{subequations}

\medskip 
To define the multilinear form $A$ in \eqref{eq:semi-discrete:Vlasov},
we first introduce the auxiliary vector function
\begin{align}
  \gv^{s,N}_{\Psi}(\xv,t) := \int_{\Ov}\vv\,\fsN(\xv,\vv,t)\Psi(\xiv)d\xiv,
  \quad\forall\,\Psi\in\calHpN,
  \label{eq:aux:gv:nmp}
\end{align}
which explicitly depends on a given function in the dual Hermite space
$\calHpN$.
When $\Psi=\Psi^{n,m,p}$, we can optionally use the notation
$\gv^{s,N}_{n,m,p}$, which explicitly refers to the Hermite indices
$n,m,p$.

\medskip
Then, for any $\fsN\in\calHN\times\calVN$,
 $\vecEN\in\calVN$, $\vecBN\in\calVN$, and
$(\Psi,\varphi)\in\calHpN\times\calVN$, we define
\begin{align}
  &\As\big( \big(\fsN,\vecEN,\vecBN\big),(\Psi,\varphi)\big)
  := \sum_{I}\left(
    \int_{I}\int_{\Ov}\PT{\fsN}\,\Psi(\xiv)\,d\xiv\,\varphi(\xv)\dxv 
    -\int_{I}\gv^{s,N}_{\Psi}\cdot\GRADX\varphi(\xv)\dxv
  \right.
  \nonumber\\[0.5em] 
  &\qquad\qquad
  \left.
    +\int_{\partial\Is}\NFLX{\nv\cdot\gv^{s,N}_{\Psi}}\,\varphi(\xv)\dS
    +\csVM
    \int_{I}\int_{\Ov}\big(\vecEN+\vv\times\vecBN)\cdot\GRADV\fsN\,\Psi(\xiv)\,d\xiv\,\varphi(\xv)\dxv
  \right).
  \label{eq:bilA:def}
\end{align}
The bilinear form associated with the discretization of Maxwell's equations is
\begin{align}
  \Bs\big( (\vecEN,\vecBN), \varphi \big)
  &:= \BsE\big( (\vecEN,\vecBN), \varphi \big) + \BsB\big( (\vecEN,\vecBN), \varphi \big)
  \qquad\qquad
  \forall\,
  \vecEN,\,
  \vecBN,\,
  \varphi\in\calVN
  \label{eq:bilB:def}
\end{align}
where, for $\Uv := ((\vecEN)^T,(\vecBN)^T)^T$ denoting the DG approximation of the vector-valued function $\uv$
in \eqref{eq:Maxwell:conservative:05},
\begin{align} 
  \BsE\big( (\vecEN,\vecBN), \varphi \big) 
  &=
  \sum_{I}\left(
    \int_{I}\dfrac{\partial\vecEN}{\partial t}\varphi(\xv)\,\dxv 
    -\int_{I}\FFE(\Uv)\GRADX\varphi(\xv)\,\dxv
    +\int_{\partial\Is}\NFLX{\FFE(\Uv)\nv}\,\varphi(\xv)\,\dS
  \right),
  \label{eq:semi-discrete:Maxwell:E}\\[1.em]
  \BsB\big( (\vecEN,\vecBN), \varphi \big) 
  &=
  \sum_{I}\left(
    \int_{I}\dfrac{\partial\vecBN}{\partial t}\varphi(\xv)\,\dxv 
    -\int_{I}\FFB(\Uv)\GRADX\varphi(\xv)\,\dxv
    +\int_{\partial\Is}\NFLX{\FFB(\Uv)\nv}\,\varphi(\xv)\,\dS
  \right).
  \label{eq:semi-discrete:Maxwell:B}
\end{align}
The linear functional $L$ in \eqref{eq:semi-discrete:Maxwell} reads
\begin{align}
  \Ls(\varphi)
  :=&
  -\frac{\ope}{\oce}\sum_{I}\int_{I}\vecJN\varphi(\xv)\,\dxv,
  \qquad
  \forall
  \varphi\in\calVN.
  \label{eq:lfunL:def}
\end{align}
The quantity $\NFLX{\nv\cdot\gv^{s,N}_{\Psi}}$ in~\eqref{eq:bilA:def}
and $\NFLX{\FFE(\Uv)\nv}$ and $\NFLX{\FFB(\Uv)\nv}$
in~\eqref{eq:semi-discrete:Maxwell:E}-\eqref{eq:semi-discrete:Maxwell:B}
are the numerical fluxes at the faces of the element boundaries.
In the following analysis, we will consider the case of the
\emph{central numerical flux}, cf.~\eqref{eq:fluxC}, and the
\emph{upwind numerical flux}, cf.~\eqref{eq:upwind:FF:short}.

\medskip
\begin{remark}
  \label{rem:split:flux}
  The numerical flux function is uniquely defined at any cell
  interface up to the sign of the unit normal vector $\nv$, which is
  conventionally oriented outwards with respect to cell $\Is$.
  The integral term on the cell boundary $\partial\Is$ can be split on
  the six faces that define cell $\Is$ as follows:
  \begin{align*}
    \int_{\partial\Is}
    = \bigg(\int_{\fip} - \int_{\fim}\bigg)
    + \bigg(\int_{\fjp} - \int_{\fjm}\bigg)
    + \bigg(\int_{\fkp} - \int_{\fkm}\bigg).
  \end{align*}
  Summing over all the mesh cells provides three telescopic sums along
  the directions $x$, $y$, and $z$, respectively corresponding to the
  half-integer indices $i+\half$, $j+\half$, and $k+\half$.
  For a periodic system in three spatial directions, the summation is
  zero.
  Therefore,
  \begin{align*}
    \sum_{\Is}\int_{\partial\Is}\NFLX{\FFE(\Uv)\nv}\,\dS=
    \sum_{\Is}\int_{\partial\Is}\NFLX{\FFB(\Uv)\nv}\,\dS=0.
  \end{align*}
  This remark will be used when proving the conservation of the number
  of particles and energy.
\end{remark}

\medskip

\subsection{Proof of Theorem~\ref{theo:number-of-particles}
  (conservation of the number of particles)}
\label{ssec:N}

The invariance in time of $\cN^{tot}(t)$ and $\cN^{s}(t)$ from the
semi-discrete method can be shown by proving that the integral of
$\partial\fsN\slash{\partial\ts}$ on the phase space
$\cup_{\Is\in\Ox}\Is\times\Ov$ is zero.
Observe that $\varphi^{I,l}=1$ for $l=0$, and $\Psi^{n,m,p}=1$, for
$n=m=p=0$.
Then, we use~\eqref{eq:semi-discrete:Vlasov} and for each
particle species $s$ we note that
\begin{align*}
  \dfrac{d\cN^s(t)}{\dt} & = \sum_{I}\int_{I}\int_{\Ov}\dfrac{\partial\fsN}{\partial t}(\xv,\vv,t)d\vv\dxv 
  = \sum_{I}\int_{I}\left(\int_{\Ov}\dfrac{\partial\fsN}{\partial t}(\xv,\vv,t)
    \Psi^{0,0,0}(\xiv)d\vv\right)\varphi^{I,0}(\xv)\dxv 
  \nonumber\\[1.em] 
  & = \sum_{I}\int_{I}\bigg(\int_{\Ov}\vv\,\fsN(\xv,\vv,t)\Psi^{0,0,0}(\xiv)d\xiv\bigg)\cdot\GRADX\varphi^{I,0}(\xv)\dxv
  -\sum_{I}\int_{\partial I}\reallywidehat{\nv\cdot\gv^{s}_{0,0,0}} \,\varphi^{I,0}(\xv)\dS
  \nonumber\\[1.em] 
  &\quad\phantom{=}
  -
  \csVM
  \sum_{I}
  \int_{I}\int_{\Ov}\big(\vecEN+\vv\times\vecBN)\cdot\GRADV\fsN\,\Psi^{0,0,0}(\xiv)\,d\xiv\,\varphi^{I,0}(\xv)\dxv
  =: \textrm{T}_1 + \textrm{T}_2 + \textrm{T}_3.
\end{align*}
Next, we prove that $\textrm{T}_1=\textrm{T}_2=\textrm{T}_3=0$.
The first term is zero
because $\varphi^{I,0}\equiv 1$ on every $I$.
The second term $\textrm{T}_2$ vanishes in view of
Remark~\ref{rem:split:flux}.
To see that the third term $\textrm{T}_3$ is zero, we first transform
the integral in $d\xiv$ as follows
\begin{align}
  \int_{\Ov}\big(\vecEN+\vv\times\vecBN)\cdot\GRADV\fsN\,\Psi^{0,0,0}(\xiv)\,d\xiv
  = 
  &-\int_{\Ov}\nabla\cdot\big(\Psi^{0,0,0}(\xiv)\big(\vecEN+\vv\times\vecBN)\big)\,\fsN\,\,d\xiv
  \nonumber\\[0.5em]
  &+\mbox{\big[\textrm{zero boundary terms for $\ABS{\xiv}\to\pm\infty$}\big]},
  \label{eq:mass:conservation:T3}
  \end{align}
by an integration by parts.
We recall that the boundary terms are zero because $\fs\to0$
exponentially for $\ABS{\xiv}\to\pm\infty$.
The right-hand side of~\eqref{eq:mass:conservation:T3} is zero because
$\Psi^{0,0,0}=1$ and $\nabla\cdot\big(\vecEN+\vv\times\vecBN)=0$, thus implying that
$ \textrm{T}_3=0$.
\ENDPROOF

\subsection{Proof of
  Theorem~\ref{theorem:semi-discrete:momentum:conservation}
  (conservation of total momentum)}
\label{ssec:Mom}

To derive an evolution equation for the vector-valued momentum
$\Pv^{N}_{\Vlasov}(t)$ defined in \eqref{eq:PvN:Vlasov},
we exploit the fact that each component of the velocity field $\vv$ belongs to 
$\calHpN$.
Therefore, for every cell $\Is$ and every vector component
$\beta\in\{x,y,z\}$ we consider the semi-discrete Vlasov equation
\eqref{eq:semi-discrete:Vlasov} with
$\Psi(\xiv)\equiv\vs_{\beta}$ and $\varphi=\varphi^{\Is,0}=1$.
We change the integration variable from $\xiv$ to $\vv$ in all terms, we multiply 
by the species mass $\ms$ and we sum over the species index.
Since $\nabla_{\xv}\varphi^{I,0}(\xv)=0$, we find that
\begin{equation}\label{eq:discrete:momentum:00}
\begin{aligned}
  \dfrac{d\Pv^{N}_{\Vlasov}}{dt} & = \sum_{s}\ms\int_{I}\int_{\Ov}\PT{\fsN}\,\vv\,\dvv\dxv \\
     & = - \sum_{s}\ms\int_{\partial\Is}\NFLX{\nv\cdot\gv^{s,N}_{\vv}}\dS
  -\sum_{s} \qs\frac{\oce}{\ope}
  \int_{I}\int_{\Ov}\big(\vecEN+\vv\times\vecBN)\cdot\GRADV\fsN\,\vv\,\dvv\dxv,
\end{aligned}
\end{equation}
where $\NFLX{\nv\cdot\gv^{s,N}_{\vv}}$ is the numerical flux of
\begin{align}
  \nv\cdot\gv^{s,N}_{\vv}(\xv,t) := \int_{\Ov}\big(\nv\cdot\vv\big)\vv\,\fsN(\xv,\vv,t)\dvv.
  \label{eq:aux:nv-dot-gv}
\end{align}
Equation~\eqref{eq:discrete:momentum:00}
holds at any time
$\ts$, for every element $\Is$.
We integrate by parts the last term of
\eqref{eq:discrete:momentum:00}, and we find that
\begin{equation}
  \dfrac{d\Pv^{N}_{\Vlasov}}{dt}
  = - \sum_{s}\ms\int_{\partial\Is}\NFLX{\nv\cdot\gv^{s,N}_{\vv}}\dS
  +\sum_s\qs\frac{\oce}{\ope}
  \int_{I}\int_{\Ov}\big(\vecEN+\vv\times\vecBN\big)\fsN\,\dvv\dxv,
  \qquad\qquad
  \label{eq:discrete:momentum:10}
\end{equation}
since $\GRADV\cdot\big(\vecEN+\vv\times\vecBN)=0$ and $\GRADV\vv$ is
the identity matrix.
Using definitions~\eqref{eq:rhoN:def} and \eqref{eq:vecJN:def:b},
the last term of~\eqref{eq:discrete:momentum:10} can be 
rewritten as 
\begin{align}
  &\sum_s\qs\frac{\oce}{\ope}
  \int_{I}\int_{\Ov}\big(\vecEN+\vv\times\vecBN\big)\fsN\,\dvv\dxv
  \nonumber\\[0.5em]
  &\qquad\qquad=\frac{\oce}{\ope}\left(
  \int_{I}\left(\sum_s\qs\int_{\Ov}\fsN\,\dvv\right)\vecEN\dxv +
  \int_{I}\left(\sum_s\qs\int_{\Ov}\vv\fsN\,\dvv\right)\times\vecBN\dxv
  \right)
  \nonumber\\[0.5em]
  &\qquad\qquad=\frac{\oce}{\ope}\int_{I}\rhoN\vecEN\dxv +
  \frac{\oce}{\ope}\int_{I}\vecJN\times\vecBN\dxv.
  \label{eq:discrete:momentum:15}
\end{align}
Using~\eqref{eq:discrete:momentum:15} in~\eqref{eq:discrete:momentum:10} we find that
\begin{align}
  \sum_{s}\ms\int_{I}\int_{\Ov}\PT{\fsN}\,\vv\,\dvv\dxv
  +\sum_{s}\ms\int_{\partial\Is}\NFLX{\nv\cdot\gv^{s,N}_{\vv}}\dS
  -\frac{\oce}{\ope}\int_{I}\left(
  \rhoN\vecEN+\vecJN\times\vecBN
  \right)\dxv 
  =0.
  \label{eq:discrete:momentum:20}
\end{align}

\medskip
Then, we integrate by parts
equations~\eqref{eq:semi-discrete:Maxwell:E} and
\eqref{eq:semi-discrete:Maxwell:B} with $\varphi=\varphi^{\Il}$,
we use definition~\eqref{eq:divg:FF} to obtain
\begin{align}
  &\int_{I}\dfrac{\partial\vecEN}{\partial t}\varphi^{\Il}\dxv 
  -\int_{I} \big(\nabla_{\xv}\times\vecBN\big)\varphi^{\Il}\dxv
  +\int_{\partial\Is}\left(\NFLX{\FFE(\Uv)\nv}-\FFE(\Uv)\nv\right)\,\varphi^{\Il}\dS
  +\frac{\ope}{\oce}\int_{I}\vecJN\varphi^{\Il}\dxv = 0,
  \label{eq:semi-discrete:Maxwell:E:00}\\[0.5em]
  &\int_{I}\dfrac{\partial\vecBN}{\partial t}\varphi^{\Il}\dxv 
  +\int_{I}\big(\nabla_{\xv}\times\vecEN\big)\varphi^{\Il}\dxv
  +\int_{\partial\Is}\left(\NFLX{\FFB(\Uv)\nv}-\FFB(\Uv)\nv\right)\,\varphi^{\Il}\dS = 0.
  \label{eq:semi-discrete:Maxwell:B:00}
\end{align}
We take the cross product of equation~\eqref{eq:semi-discrete:Maxwell:E:00}
and $\vecBIl(t)$, and the cross product of equation~\eqref{eq:semi-discrete:Maxwell:B:00} and
$\vecEIl(t)$, we sum over $l$, and in view of the
expansions~\eqref{eq:Ev:expansion} and \eqref{eq:Bv:expansion}, we
obtain the two equations:
\begin{align}
  &\int_{I}\dfrac{\partial\vecEN}{\partial t}\times\vecBN\dxv 
  -\int_{I}\big(\nabla_{\xv}\times\vecBN\big)\times\vecBN\dxv
  +\int_{\partial\Is}\left(\NFLX{\FFE(\Uv)\nv}-\FFE(\Uv)\nv\right)\times\vecBN\dS
  \nonumber\\[0.em]&\qquad
  +\frac{\ope}{\oce}\int_{I}\vecJN\times\vecBN\dxv = 0,
  \label{eq:semi-discrete:Maxwell:E:10}\\[0.5em]
  &\int_{I}\dfrac{\partial\vecBN}{\partial t}\times\vecEN\dxv 
  +\int_{I}\big(\nabla_{\xv}\times\vecEN\big)\times\vecEN\dxv
  +\int_{\partial\Is}\left(\NFLX{\FFB(\Uv)\nv}-\FFB(\Uv)\nv\right)\times\vecEN\dS = 0.
  \label{eq:semi-discrete:Maxwell:B:10}
\end{align}
We subtract~\eqref{eq:semi-discrete:Maxwell:B:10}
from~\eqref{eq:semi-discrete:Maxwell:E:10} and we find that
\begin{align}
  \dfrac{d}{\dt}\int_{I}\big(\vecEN\times\vecBN\big)\dxv
  = &
  -\int_{I} \bigg(\vecBN\times\big(\nabla_{\xv}\times\vecBN\big)+\vecEN\times\big(\nabla_{\xv}\times\vecEN\big)\bigg)\dxv
  \nonumber\\
  & -\int_{\partial\Is}\mathcal{B}_I\dS 
  -\frac{\ope}{\oce}\int_{I}\vecJN\times\vecBN\dxv,
  \label{eq:semi-discrete:Maxwell:ExB}
\end{align}
where
\begin{align}
  \mathcal{B}_I &= \left(\NFLX{\FFE(\Uv)\nv}-\FFE(\Uv)\nv\right)\times\vecBN - \left(\NFLX{\FFB(\Uv)\nv}-\FFB(\Uv)\nv\right)\times\vecEN.\label{eq:BI:def}
\end{align}
Since $-\FFE(\Uv)\nv\times\vecBN=\nv\times\vecBN\times\vecBN=0$ and
$\FFB(\Uv)\nv\times\vecEN=\nv\times\vecEN\times\vecEN=0$, the term
$\mathcal{B}_I$ takes the form given in the statement of Theorem \ref{theorem:semi-discrete:momentum:conservation}.
Finally, the assertion of the theorem follows by
summing~\eqref{eq:semi-discrete:Maxwell:ExB}
to~\eqref{eq:discrete:momentum:20}.


\subsection{Proof of Theorem~\ref{theo:energy} (conservation of total energy)}
\label{ssec:En}

We split the proof of the theorem in three steps.
To ease the notation, we drop the explicit dependence on $\xv$, $\vv$
and $\ts$ of the fields $\fsN$, $\vecEN$, $\vecBN$, and $\vecJN$.
\begin{enumerate}
\item In the first step, we prove that the kinetic energy satisfies
  \begin{align}
    \dfrac{d\calEN_{\KIN}(t)}{\dt}:=
    \dfrac12\sum_{s}\ms\sum_{I}\int_{I}\bigg(\int_{\Ov}\dfrac{\partial\fsN}{\partial t}\ABS{\vv}^2d\vv\bigg)\dxv 
    = \frac{\oce}{\ope}\sum_{I}\int_{I}\vecEN\cdot\vecJN\dxv
    \label{eq:energy:lemma}
  \end{align}
  where $\vecJN$ is the approximate current density
  defined in~\eqref{eq:vecJN:def:b}.
  
  \medskip
\item In the second step, we prove that
  \begin{align}
    \dfrac{d\calEN_{\ELE}(t)}{\dt}:=
    \dfrac12\left(\frac{\oce}{\ope}\right)^2
    \sum_{I}\dfrac{d}{dt}\int_{I}
    \big(|\vecEN|^2 + |\vecBN|^2 \big)\dxv
    = -\frac{\oce}{\ope}\sum_{I}\int_{I}\vecEN \cdot\vecJN\dxv   
    +\left(\frac{\oce}{\ope}\right)^2\widetilde{\Phi},
    \label{eq:energy:lemma2}
  \end{align}
  where 
  \begin{align}
    \widetilde{\Phi} := 
    \sum_{I}\bigg(
    \int_{\partial\Is}
    \big(\FFE\nv-\NFLX{\FFE\nv}\big)\cdot\vecEN\,\dS +
    \int_{\partial\Is}\big(\FFB\nv-\NFLX{\FFB\nv}\big)\cdot\vecBN\,\dS
    - 
    \int_{\partial\Is} \nv\cdot\big( \vecEN\times\vecBN \big) \dS 
    \bigg).
    \label{eq:tildePhi:def}
  \end{align}
  
  \medskip
\item In the third step, we prove that 
  \begin{align}
    \widetilde{\Phi} = -\sum_{\F}\JMPF\leq 0,
    \label{eq:jump:relation}
  \end{align}
  where $\JMPF$ is defined in~\eqref{eq:JF}.
\end{enumerate}

\medskip
\noindent
The assertion of the theorem follows by
substituting~\eqref{eq:energy:lemma2} in~\eqref{eq:energy:lemma}, and
then using definition~\eqref{eq:tildePhi:def} and the jump relation
\eqref{eq:jump:relation}.

\medskip
\PGRAPH{Proof of \eqref{eq:energy:lemma}.}
Since $\ABS{\vv}^2$ is a linear combination of the Hermite polynomials
$\Psiv^{n,m,p}(\xiv)$ for $n+m+p\leq2$, it belongs to the velocity
approximation space.
Therefore, the evolution of the kinetic energy can be computed using
equation~\eqref{eq:semi-discrete:Vlasov} (with $l=0$):
\begin{align}
  &\dfrac{d\calEN_{\KIN}(t)}{\dt}
  = \dfrac12\sum_{s}\ms\sum_{I}\int_{I}\int_{\Ov}\PT{\fsN}\ABS{\vv}^2d\vv\,\dxv         \nonumber\\[0.5em]
  & \quad= 
  -\dfrac12\sum_{s}\ms\sum_{I}\int_{\partial\Is}\NFLX{\nv\cdot\gv^{s,N}}\,\dS
  -\dfrac12\sum_{s}\ms\sum_{I}\int_{I}\frac{\qs}{\ms}\frac{\oce}{\ope}\int_{\Ov}\big(\vecEN+\vv\times\vecBN\big)\cdot\GRADV\fsN\,\ABS{\vv}^2d\vv\,\dxv
  \nonumber\\[0.5em]
  & \quad=: \textrm{T}_1 + \textrm{T}_2,
  \label{eq:Kinenergy}
\end{align}
where $\NFLX{\nv\cdot\gv^{s,N}}$ is the numerical flux at the cell
interfaces $\partial\Is$ of the vector field 
\begin{align}
  \label{eq:aux:gv:velocity}
  \gv^{s,N}(\xv,t) := \int_{\Ov}\vv\,\fsN(\xv,\vv,t)\ABS{\vv}^2d\vv.
\end{align}
Term $\textrm{T}_1$ is zero because of Remark~\eqref{rem:split:flux}.
To compute the term $\textrm{T}_2$ in \eqref{eq:Kinenergy},
we integrate by parts with respect to $\vv$ so that,
\begin{align}
  &\int_{\Ov}\big(\vecEN+\vv\times\vecBN)\cdot\GRADV\fsN\,\ABS{\vv}^2d\vv\nonumber\\[0.5em]
  &\qquad= -\int_{\Ov}\nabla_{\vv}\cdot\Big(\ABS{\vv}^2\big(\vecEN+\vv\times\vecBN)\Big)\fsN\,\dvv
  + \Big[\textrm{zero~boundary~terms~for~$\ABS{\vv}\to\infty$}\Big]\nonumber\\[0.5em]
  &\qquad= -2\vecEN\cdot\int_{\Ov}\vv\fsN\,d\vv,
  \label{eq:energy:T2:05}
\end{align}
since $\GRADV\ABS{\vv}^2=2\vv$ and
$\nabla_{\vv}\cdot\big(\vecEN+\vv\times\vecBN)=0$.
According to the definition of $\textrm{T}_2$
from~\eqref{eq:Kinenergy} and using the definition of $\vecJN$
in~\eqref{eq:vecJN:def:b}, it holds that
\begin{align}
  \textrm{T}_2
  & = -\dfrac12 \sum_{s} \qs\frac{\oce}{\ope}
  \sum_{I}\int_{I}\int_{\Ov}(\vecEN+\vv\times\vecBN)\cdot\GRADV\fsN\,\ABS{\vv}^2d\vv\dxv\nonumber\\
  & = \sum_{s} \qs \frac{\oce}{\ope}
  \sum_{I}\int_{I}\vecEN\cdot\bigg(\int_{\Ov}\vv\fsN\dvv\bigg)\dxv
  = \frac{\oce}{\ope}\sum_{I}\int_{I}\vecEN\cdot\vecJN\dxv.
  \label{eq:energy:T2:15}
\end{align}
Equation~\eqref{eq:energy:lemma} follows by substituting in \eqref{eq:Kinenergy},
$\textrm{T}_1=0$ and the expression of $\textrm{T}_2$
from~\eqref{eq:energy:T2:15}.

\medskip
\PGRAPH{Proof of~\eqref{eq:energy:lemma2}.}
We take the Euclidean product
of~\eqref{eq:semi-discrete:Maxwell:E} with $\vecE^{\Il}$ and sum
over $l=1,\ldots\Nl$,
\begin{align*}
    \sum_{l=1}^{N_l}\bigg(\int_{I} & \dfrac{\partial\vecEN}{\partial t}\cdot\vecE^{\Il}\varphi^{\Il}\dxv
        - \int_{I}(\FFE(\Uv)\nabla_{\xv}\varphi^{\Il})\cdot\vecE^{\Il}\dxv\\
        & + \int_{\partial\Is}\NFLX{\FFE(\Uv)\nv}\cdot\vecE^{\Il}\varphi^{\Il}\dxv 
        + \frac{\ope}{\oce}\int_{I}\vecJN\cdot\vecE^{\Il}\varphi^{\Il}\dxv\bigg) = 0.
\end{align*}
Integrating by parts yields
\begin{align*}
  \int_I \dfrac{\partial \vecEN}{\partial t}\cdot\vecEN\,\dxv
    + \frac{\ope}{\oce}\int_{I}\vecEN\cdot\vecJN\,\dxv
    =  -\int_{I}\big(\nabla_{\xv}\cdot\FFE(\Uv)\big)\cdot\vecEN\dxv
    + \int_{\partial\Is}\big(\FFE\nv-\NFLX{\FFE\nv}\big)\cdot\vecEN\dS.
\end{align*}
Then, we use formula~\eqref{eq:divg:FF} to express the divergence of the flux,
\begin{align}
  \dfrac{d}{dt}\dfrac12\int_{I}|\vecEN|^2\dxv
    + \frac{\ope}{\oce}\int_{I}\vecEN\cdot\vecJN\dxv 
    = \int_{I}\nabla_{\xv}\times\vecBN\,\cdot\vecEN\dxv 
    + \int_{\partial\Is}\big(\FFE\nv-\NFLX{\FFE\nv}\big)\cdot\vecEN\dS.
  \label{eq:energy:10}
\end{align}
Similarly, we multiply
equation~\eqref{eq:semi-discrete:Maxwell:B} by $\vecB^{\Il}$, sum
over $l=1,\ldots\Nl$, integrate by parts, and use
formula~\eqref{eq:divg:FF},
\begin{align}  
  \int_{I}\dfrac{\partial \vecBN}{\partial t}\cdot\vecBN\dxv 
  =
  -\int_{I}\big(\nabla_{\xv}\times\vecEN\big)\cdot\vecBN\dxv
  +\int_{\partial\Is}\big(\FFB\nv - \NFLX{\FFB\nv} \big)\cdot\vecBN\dS.
  \label{eq:energy:15}
\end{align}
Summing~\eqref{eq:energy:10} and~\eqref{eq:energy:15} we obtain,
\begin{align}\label{eq:energy:25}
  \dfrac{d}{dt}\dfrac12\int_{I}\big(|\vecEN|^2+|\vecBN|^2\big)\dxv
    + \frac{\ope}{\oce}\int_{I}\vecEN\cdot\vecJN\dxv
      = \int_{I}\big( \nabla_{\xv}\times\vecBN\cdot\vecEN - \nabla_{\xv}\times\vecEN\cdot\vecBN \big)\dxv
    + \widetilde{\Phi}^{\Is},
\end{align}
where $\widetilde{\Phi}^{\Is}$ is defined as
\begin{align}\label{eq:tildePhi:def:I}
  \widetilde{\Phi}^{\Is} :=
  \int_{\partial\Is}
  \big(\FFE\nv-\NFLX{\FFE\nv}\big)\cdot\vecEN\,\dS +
  \int_{\partial\Is}\big(\FFB\nv-\NFLX{\FFB\nv}\big)\cdot\vecBN\,\dS.
\end{align}
Observe that
\begin{align*}
  \nabla_{\xv}\times\vecBN\,\cdot\vecEN
  -\nabla_{\xv}\times\vecEN\,\cdot\vecBN
  = -\nabla_{\xv}\cdot\big( \vecEN\times\vecBN \big),
\end{align*}
so that \eqref{eq:energy:25} becomes
\begin{align}
  \dfrac{d}{dt}\dfrac12\int_{I}\big(|\vecEN|^2+|\vecBN|^2\big)\dxv
  +\frac{\ope}{\oce}\int_{I}\vecEN\cdot\vecJN\dxv
  = -\int_{I} \nabla_{\xv}\cdot\big( \vecEN\times\vecBN \big) \dxv
  + \widetilde{\Phi}^{\Is}.
  \label{eq:energy:25b}
\end{align}
Using the divergence theorem and summing over the mesh elements
$I\subset\Omega_x$ yields,
\begin{align}
  \dfrac{d}{dt}\dfrac12\sum_{I}\int_{I}\big(|\vecEN|^2+|\vecBN|^2\big)\dxv
  +\frac{\ope}{\oce}\sum_{I}\int_{I}\vecEN\cdot\vecJN\dxv
  = 
  \sum_{I}\bigg(\widetilde{\Phi}^{\Is} -\int_{\partial\Is} \nv\cdot\big( \vecEN\times\vecBN \big) \dxv
  \bigg)
  =: \widetilde{\Phi},
  \label{eq:energy:30}
\end{align}
which gives~\eqref{eq:energy:lemma2}. 

\medskip
\PGRAPH{Proof of\eqref{eq:jump:relation}.}
Let us consider the term
\begin{equation*}
  \widetilde{\Phi}=\sum_{I}
  \bigg(
  \int_{\partial\Is}
  \big(\FFE\nv-\NFLX{\FFE\nv}\big)\cdot\vecEN\,\dS +
  \int_{\partial\Is}\big(\FFB\nv-\NFLX{\FFB\nv}\big)\cdot\vecBN\,\dS
  -\int_{\partial\Is} \nv\cdot\big( \vecEN\times\vecBN \big) \dxv
  \bigg).
\end{equation*}
Note that, by the vector identity \eqref{eq:vecidentity}, it holds
\begin{align*}
  \nv\cdot\big(\vecEN\times\vecBN\big)
  = \dfrac12\big(\FFE\nv\cdot\vecEN + \FFB\nv\cdot\vecBN\big).
\end{align*}
Therefore,
\begin{align*}
  -\nv\cdot\big( \vecEN\times\vecBN \big) + 
  \big(\FFE\nv\cdot\vecEN + \FFB\nv\cdot\vecBN\big)
  = \dfrac12\big(\FFE\nv\cdot\vecEN + \FFB\nv\cdot\vecBN\big),
\end{align*}
and
\begin{align}
  \widetilde{\Phi}
  =
  \sum_{I}\int_{\partial\Is}
  \bigg(
  \dfrac12
    \big(\FFE\nv\cdot\vecEN + \FFB\nv\cdot\vecBN\big)
  -
  \big(\NFLX{\FFE\nv}\cdot\vecEN + \NFLX{\FFB\nv}\cdot\vecBN\big)
  \bigg)\dS.
  \label{eq:Psi:00}
\end{align}
Moreover, it holds
\begin{equation*}
  \FF(\Uv)\nv\cdot\Uv =
  \FFE\nv\cdot\vecEN + \FFB\nv\cdot\vecBN,
  \qquad\qquad
  \NFLX{\FF(\Uv)\nv}\cdot\Uv =
  \NFLX{\FFE\nv}\cdot\vecEN + \NFLX{\FFB\nv}\cdot\vecBN,
\end{equation*}
where the flux $\FF(\Uv)$ is given by~\eqref{eq:divg:FF}.
Moreover, we can reformulate the summation on the boundary cells as a
summation on the cell interfaces by denoting the quantities referring
to the two opposite sides with the superscripts $\pm$, and
rewrite~\eqref{eq:Psi:00} as
\begin{align*}
  \widetilde{\Phi} 
  =
  \sum_{I}\int_{\partial\Is}
  \bigg(
  \frac{1}{2}\FF(\Uv)\nv\cdot\Uv
  -
  \NFLX{\FF(\Uv)\nv}\cdot\Uv
  \bigg)\dS
  = \sum_{\F}\int_{\F}\mathcal{F}_{\F}(\Uvp_{\F},\Uvm_{\F})\dS,
\end{align*}
where
\begin{align*}
  \mathcal{F}_{\F}(\Uvp_{\F},\Uvm_{\F}) :=
  \dfrac12\Big(
  \FF(\Uvp_{\F})\nv^-_{\F}\cdot\Uvp_{\F} +
  \FF(\Uvm_{\F})\nv^+_{\F}\cdot\Uvm_{\F}
  \Big)
  -
  \Big(
  \NFLX{\FF(\Uv_{\F})\nv^-_{\F}}\cdot\Uvp_{\F} +
  \NFLX{\FF(\Uv_{\F})\nv^+_{\F}}\cdot\Uvm_{\F}
  \Big),
\end{align*}
at any given mesh face $\F$.
Now, let $\Fb$ be a mesh face orthogonal to the direction
$\beta$, for $\beta\in\{x,y,z\}$.
Then, the $\beta$-th components of the unit vectors orthogonal to
$\Fb$ are $n^+_{\beta}=-n^-_{\beta}=+1$, while the other components
are zero.
We refer to Figure \ref{fig:numflux:notation} for a schematic
representation of the quantities involved in the definition of the
numerical fluxes.
We recall that for a generic face $\F$ we use the notation
$\FF(\Uv_{\F})\nv_{\F}\cdot\Uv_{\F}=\sum_{\beta\in\{x,y,z\}}n_{\beta}\FF_{\beta}\Uv_{\F}\cdot\Uv_{\F}$,
where the matrices $\FF_{\beta}\in\mathbb{R}^{6\times 6}$ are defined
in Section~\ref{sec:maxwell}.
If the last term is interpreted as the usual matrix-vector product, we can write that
$\FF(\Uv_{\F})\nv_{\F}\cdot\Uv_{\F}=\Uv_{\F}^\top\big(\sum_{\beta\in\{x,y,z\}}n_{\beta}\FF_{\beta}\big)\Uv_{\F}$.
For each direction $\beta$, we denote
$\FFb^{\pm}=\FF^{\pm}\ns^{\pm}=\sum_{\beta}\ns_{\beta}^{\pm}\FFb^{\pm}=\pm\FFb^{\pm}$,
and we recall that $\FFb=\FFb^{+}+\FFb^{-}$ and
$\ABS{\FFb}=\FF^{+}_{\beta}-\FF^{-}_{\beta}$.
A straightforward calculation yields:
\begin{align*}
  \mathcal{F}_{\Fb}(\Uvp_{\Fb},\Uvm_{\Fb}) 
  &= \dfrac12\big((\Uvm_{\Fb})^T\FFb\Uvm_{\Fb} -(\Uvp_{\Fb})^T\FFb\Uvp_{\Fb}\big)-\big(\Uvm_{\Fb}-\Uvp_{\Fb})^T\NFLX{\FFb(\Uv)\nv_{\Fb}}\nonumber\\[0.5em]
  &= -\big(\Uvp_{\Fb}-\Uvm_{\Fb}\big)^T\left(\FFb\dfrac{\big(\Uvp_{\Fb}+\Uvm_{\Fb})}{2}-\NFLX{\FFb(\Uv)\nv_{\Fb}}\right).
\end{align*}
We conclude the proof by considering separately the case of the
\emph{central} and \emph{upwind} numerical flux, and prove that
$\mathcal{F}_{\F}(\Uvp_{\F},\Uvm_{\F})=-\JMPF$ for any face $\F$,
where $\JMPF$ is defined in \eqref{eq:JF}.

\medskip
\noindent
\textbf{Central numerical scheme.}  
The central numerical flux across face $\Fb$ is formulated as:
\begin{align}
  \NFLX{\FFb(\Uv)\nv_{\Fb}} 
  = \FFb\left(\frac{\Uvp_{\Fb}+\Uvm_{\Fb}}{2}\right),
  \label{eq:central:flux:def}
\end{align}
and a direct substitution immediately yields that
$\mathcal{F}_{\Fb}(\Uvp_{\Fb},\Uvm_{\Fb})=0$.

\medskip
\noindent
\textbf{Upwind numerical scheme.} 
The upwind numerical flux across face $\Fb$ is formulated as:
\begin{align*}
  \NFLX{\FFb(\Uv)\nv_{\Fb}} 
  = \FFb^+(\Uvm_{\Fb})+\FFb^-(\Uvp_{\Fb}).
\end{align*}
A direct substitution yields
\begin{align}
  \mathcal{F}_{\Fb}(\Uvp_{\Fb},\Uvm_{\Fb}) 
  &= -\big(\Uvp_{\Fb}-\Uvm_{\Fb}\big)^T
    \left(\matF_{\beta}\dfrac{\big(\Uvp_{\Fb}+\Uvm_{\Fb})}{2}-\FFb^+\Uvm_{\Fb}-\FFb^-\Uvp_{\Fb}\right)                    \nonumber\\[0.5em]
  &= -\dfrac12\big(\Uvp_{\Fb}-\Uvm_{\Fb}\big)^T
    \Big(\big(\FFb^++\FFb^-\big)\big(\Uvp_{\Fb}+\Uvm_{\Fb}\big)-2\FFb^+\Uvm_{\Fb}-2\FFb^-\Uvp_{\Fb})\Big)
    \nonumber\\[0.5em]
  &= -\dfrac12\big(\Uvp_{\Fb}-\Uvm_{\Fb}\big)^T
    \big(\FFb^+-\FFb^-\big)\big(\Uvp_{\Fb}-\Uvm_{\Fb}\big)
    \nonumber\\[0.5em]
  &= -\dfrac12\big(\Uvp_{\Fb}-\Uvm_{\Fb}\big)^T\ABS{\FFb}\big(\Uvp_{\Fb}-\Uvm_{\Fb}\big),
\end{align}
which concludes the proof.
\ENDPROOF

\end{document}